\documentclass[12pt]{article}
\usepackage{amsmath,mathrsfs,amsthm}
\usepackage{caption}
\usepackage{verbatim}
\usepackage{subcaption}
\usepackage{graphicx,psfrag,epsf}
\usepackage{enumerate}
\usepackage{natbib}
\usepackage{bbm}
\usepackage{bm}
\usepackage{amssymb}
\usepackage{lscape}
\usepackage[figuresright]{rotating}
\usepackage[affil-it]{authblk}
\usepackage{verbatim}
\usepackage[figuresright]{rotating}
\usepackage{extarrows}
\usepackage{framed}
\usepackage[colorlinks=true,citecolor=blue,urlcolor  = black]{hyperref}
\usepackage{algorithm}
\usepackage{algorithmic}
\usepackage{url} 
\usepackage{float}
\usepackage{geometry}
\usepackage{lscape} 
\usepackage{multirow}
\usepackage{booktabs}
\usepackage{pifont}
\usepackage{stackengine}
\usepackage{array}
\usepackage{makecell}
\usepackage{multirow}

\newcommand{\blind}{0}
\newcommand\pkg[1]{\texttt{#1}}
\DeclareMathOperator*{\argmax}{arg\,max}
\DeclareMathOperator*{\argmin}{arg\,min}
\newtheorem{definition}{Definition}[section]
\newtheorem{assumption}{Assumption}[section]
\newtheorem{theorem}{Theorem}[section]

\theoremstyle{remark}
\newtheorem{remark}{Remark}[section]
\newtheorem{lemma}{Lemma}[section]
\newtheorem{corollary}{Corollary}[section]

\newcommand{\btheta}{\bm{\theta}}

\begin{document}
\def\spacingset#1{\renewcommand{\baselinestretch}%
{#1}\small\normalsize} \spacingset{1}

\if0\blind
{
   \title{\bf A Variational Spike-and-Slab Approach for Group Variable Selection}
  \author[1]{Buyu Lin\footnote{These authors contribute equally to this work.}}
\author[2]{Changhao Ge{$^\ast$}}
\author[1]{Jun S. Liu}
\affil[1]{Department of Statistics, Harvard University}
\affil[2]{Graduate Group in Applied Mathematics and Computational Science, University of Pennsylvania}  
\date{}

  \maketitle
} \fi

\if1\blind
{
  \bigskip
  \bigskip
  \bigskip
  \begin{center}
    {\LARGE\bf{A Variational Spike-and-Slab Approach for Group Variable Selection}}
\end{center}
  \medskip
} \fi

\bigskip

\begin{abstract}
We introduce a class of generic spike-and-slab priors for high-dimensional linear regression with grouped variables and present a Coordinate-ascent Variational Inference (CAVI) algorithm for obtaining an optimal variational Bayes approximation. Using parameter expansion for a specific, yet comprehensive, family of slab distributions, we obtain a further gain in computational efficiency.
The method can be easily extended to fitting additive models.
Theoretically, we present general conditions on the generic spike-and-slab priors that enable us to derive the contraction rates for both the true posterior and the VB posterior for linear regression and additive models, of which some previous theoretical results can be viewed as special cases. Our simulation studies and real data application demonstrate that the proposed method is superior to existing methods in both variable selection and parameter estimation. 
Our algorithm is implemented in the R package \href{https://github.com/HowardGech/GVSSB}{\pkg{GVSSB}}.

\end{abstract}
\vfill
\newpage
\spacingset{1.5}
\section{Introduction}
Researchers nowadays have the privilege of collecting many explanatory features targeting a specific response variable.
In many applications, these features can be naturally partitioned into disjoint groups. This occurs when dealing with multi-level categorical predictors in regression problems \citep{2006YuanandLin} or when using basis expansions of additive components in high-dimensional additive models \citep{2010HuangAdditive, 2010Meier}. The grouping structure of variables can also be incorporated into a model to leverage domain knowledge. Single nucleotide  variations in the same gene, for example, form a natural group.  The response variable, on the other hand, most likely depends only on a small subset of these groups. Therefore, identifying these relevant groups is crucial for improving the computational efficiency of downstream analysis and enhancing the interpretability of scientific findings.

One way to achieve group sparsity is by imposing a penalty on each group, as implemented in the widely-used Group Lasso method \citep{2006YuanandLin}. The estimator $\bm{\hat\theta}$ is obtained by minimizing the following objective function:
\begin{equation}\nonumber
\frac{1}{2n}\|\bm{Y}-\sum_{i = 1}^G\bm{X}_i\bm{\theta}_i\|^2+\lambda\sum_{i = 1}^G \sqrt{p_i}\|\bm{\theta}_i\|,
\end{equation}
where $p_i$ is the dimension of $\bm{\theta}_i$. The group SCAD method, introduced by \citet{2007WanggroupSCAD}, extends the group Lasso penalty to the SCAD penalty, which also operates at the group level. The benefits of group sparsity have been investigated in several studies, including \citet{2010TongZhang} and \citet{2011Lounici}. For a more comprehensive review of penalization methods for group variable selection, please refer to \citet{2012Huangreview}.

In contrast to penalization methods, a Bayesian method encodes sparsity in its sparsity-promoting prior, such as a continuous-shrinkage prior \citep{2008BayesianLasso,2010Horseshoe} or a spike-and-slab prior \citep{1988Mitchell, 1993George,  1997George}. Identifying relevant groups is often achieved by specifying priors over the whole group of variables \citep{2009Raman, 2010Kyung, 2015xiaofan}. In this paper,  we consider a  discrete spike-and-slab prior for group variable selection, which takes the following form:
\begin{equation}
\begin{aligned}
\label{eq:sas-group}
\sigma^2&\sim g(\sigma^2);\\
S&\sim \pi(S),\ \mbox{where} \  S\subset\{1, \ldots, G\};\\
\bm{\theta}_j&\sim \delta_0(\bm{\theta}_j),\ \  \forall j\notin S, \quad  \mbox{and} \quad
\bm{\theta}_j\sim h(\bm{\theta}_j),\ \ \forall j\in S;
\end{aligned}
\end{equation}
where $g$ is a density function defined on $(0, \infty)$, $S$ indicates all possible $2^G$ models, $\pi(S)$ controls the overall sparsity and enables model selection, $\delta_0$ is the point mass at $0$ (spike distribution) for modeling null features and $h$ is the slab distribution for modeling relevant features (e.g., a multivariate  Gaussian distribution).

To overcome computational challenges associated with the discrete nature of the spike-and-slab prior and the large exploration space in high-dimensional settings, researchers have proposed to replace the exact-spike distribution with a continuous distribution with a small variance. This approach has been successfully applied in linear regression without group structure using normal \citep{2014Rockova} or Laplace distributions \citep{2018sslasso}. In the group case, \citet{2019SSGL}  used a multi-Laplacian distribution as the spike distribution. These works enable us to compute the {\it maximum a posteriori} (MAP) estimates numerically, which converges faster than Markov Chain Monte Carlo (MCMC) methods in general.

A different route for Bayesian to deal with sparsity involves a global-local prior. The horseshoe prior \citep{2009horseshoe, 2010Horseshoe} is a popular example of such a prior in Bayesian linear regression. In the regression scenario without group structures, the horseshoe prior can be expressed as:
\begin{equation}\nonumber
\begin{aligned}
 \theta_j|\lambda_j, \tau&\sim N(0, \tau^2\lambda_j^2),\\
 \lambda_j&\sim C^+(0,1),\ j = 1,\ldots, p,
\end{aligned}
\end{equation}
where $C^+(0,1)$ is the half Cauchy distribution, $\lambda_j$'s and $\tau$ are known as local parameters and global parameter, respectively. The parameter $\tau$ controls the overall sparsity and usually  enforces strong simultaneous shrinkage among coefficients, while the long-tailed distribution imposed on the $\lambda_j$'s enables them to move freely, ensuring that the true signal is not heavily penalized. Recently, \citet{2020horseshoe} extended this approach to subspace shrinkage and introduced the functional horseshoe prior, which can be directly applied to high dimensional additive models.

An alternative strategy for dealing with high-dimensional Bayesian inference problems is to obtain a good analytic approximation to the posterior. One popular method is the variational Bayes (VB) approximation \citep{2017VBreview}, which finds the member in a family of approximation distributions $\mathcal{P}$ that is the closest to  the true posterior distribution in terms of the Kullback-Leibler (KL) divergence. The mean-field family, which assumes that the joint density is a product of the marginals (i.e.,  mutually independent), is a widely adopted choice for $\mathcal{P}$. 
A commonly used algorithm for minimizing the KL divergence between the family $\mathcal{P}$ and the target is the Coordinate Ascent Variational Inference algorithm (CAVI) \citep{bishop:2006:PRML}, which iteratively optimizes each component of the mean-field density while holding all the others fixed. For more applications and a general review of VB, we refer the interested readers to \citet{2017VBreview}.

In this paper, we introduce a scalable method called \emph{Group Variational Spike-and-Slab Bayes} (GVSSB) for grouped linear regression, which uses the mean-field approximation of the true posterior distribution based on the generic spike-and-slab prior in \eqref{eq:sas-group}. 
The mean-field family we considered is similar to the one in \citet{2019rayvilinear}, except that  they rely on an external  estimate of the unknown noise level $\sigma$, whereas we  incorporate $\sigma$ into the mean-field family for simultaneous estimation, which improves the estimation accuracy. 

Many commonly-used slab functions can be written as a scale mixture of multivariate normal distributions, e.g., multi-Laplacian distribution and multivariate  t-distribution. For this type of hierarchically structured distributions, we consider a novel mean-field family, which involves an additional set of parameters to facilitate an analytical formula for each updating step. In doing so, we bypass the time-consuming numerical optimization step that would have appeared in every iteration  had we not used this hierarchical structure. This strategy  significantly reduces the computation time. Furthermore, we also introduce an empirical Bayes based approach to optimize the hyper-parameters, which further improves the performance and stability of the method. In addition to linear models, we extend our method to high-dimensional additive models.
    
A  unified theoretical treatment of spike-and-slab priors in \eqref{eq:sas-group} is still lacking for grouped linear regression. Previous studies are usually restricted to specific choices of spike-and-slab priors and require case-by-case treatments. For example, \citet{2020Boningcontraction} dealt with the Dirac-and-Laplace prior and \citet{2019SSGL} worked with the mixture Laplace prior. 
We generalize their results and provide contraction rates for the large class of spike-and-slab priors characterized by  \eqref{eq:sas-group}. Our theoretical investigation is  similar in spirit to \citet{2019Jiang}, who developed a unified framework to rigorously assess  theoretical properties of spike-and-slab priors in linear regression without group structures. 

    In addition to investigating the  posterior distribution, we also derive theoretical properties of its VB approximation. 
    Under slightly stronger assumptions than those required for proving the  contraction rate for the true posterior, a similar contraction rate holds for the VB posterior. 


The rest of the paper is structured as follows. Section \ref{sec:model} introduces a  class of spike-and-slab priors. Sections \ref{subsec:general-sas} and \ref{subsec:computation-general-sas} discuss variational approximations with general spike-and-slab priors, where detailed updating rules are provided.  Section \ref{subsec:hierarchical-sas} focuses on hierarchical spike-and-slab priors, and Section \ref{subsec:hyperparameter} discusses how to choose hyper-parameters. Section \ref{sec:additive-model} extends the proposed method  to the  high-dimensional additive model. Theoretical properties are investigated in Section \ref{sec:mainresults}. Section \ref{sec:simulation} demonstrates through extensive simulations that our methods achieve the best or near-best performances among all the competitors for both grouped linear models and additive models. Section \ref{sec:realdata} applies our method to the prediction of  ethanol concentration using NIR spectroscopy absorbance values.  Section \ref{sec:conclusion} concludes with a few final remarks. Additional proofs, method derivations, and simulation studies are collected in  Supplementary Materials.

\section{A Class of Spike-and-Slab Priors}\label{sec:model}
Throughout the paper, we assume that each feature
variable $x_j$ has mean 0 and standard deviation 1 marginally, and that the response variable $y$ also has mean 0. In practice, we simply center the response vector $\bm{Y}$ to mean 0, and center each covariate vector $X_j$ and standardize it to $\|X_j\| = \sqrt{n}$.
The error variance $\sigma_\star^2\in (0, \infty)$ is the same for all observations (homogeneous). We let $S_\star$ be the set of groups whose associated coefficients $\bm{\theta}_j^\star$ are nonzero and denote its size by $s_\star=|S_\star|$. Our focus is on the high dimensional regime with $G>n$ and  $\epsilon_n = 
\max\{\sqrt{s_\star\log G/n}, \sqrt{s_\star p_{\max}\log n/n}\}\to 0$, where $p_{\max}$ denotes the maximum group size. The following assumptions characterize a class of generic spike-and-slab priors.
\begin{assumption}
\label{assump:prior}
\text{}
\begin{enumerate}[(a)]
    \item Variance prior: The density function $g(\sigma^2)$ is continuous and positive for any $\sigma^2\in (0,+\infty)$.
    \item Model Selection Prior: $\pi(S)$ satisfies  $\pi(\emptyset)\asymp 1$, and with constants $A_1, A_2>0$, \begin{equation}\nonumber
        \pi(S_\star)\geq e^{-A_1s_\star\log G},\ \ \ \sum_{S: |S|>t}\pi(S)\leq e^{-A_2t\log G}, \forall\ t\geq 1.
    \end{equation} 
    \item Slab Prior: For $z_n = \max_{j\in S_\star}\|\bm{\theta}_j^\star\|+\epsilon_n$ and some constant $A_3>0$, the slab function h(z) satisfies 
    \begin{equation}\nonumber
        \inf_{\bm{z}: \|\bm{z}\|\leq z_n}
        h(\bm{z})\succeq G^{-A_3},
    \end{equation}
    where, $\forall$ positive sequences $a_n$ and $b_n$, $a_n\succeq b_n$ (and $b_n\preceq a_n$) means that $\limsup b_n/a_n<\infty$.
\end{enumerate}
\end{assumption}

Assumption \ref{assump:prior}(a) is easily satisfied if we let $g$ be the inverse-gamma density function. Assumption \ref{assump:prior}(b)
 requires the model selection prior to assign sufficient mass to the true model $S_{\star}$ while down-weighting the large models exponentially fast.   
 The following commonly-used priors satisfy this assumption.
 \begin{enumerate}
     \item (Bernoulli Prior). The i.i.d. Bernoulli prior $\pi(S)$ selects each group $j$ with probability $1/G$. This prior meets the Assumption \ref{assump:prior}(b) with any $A_1>1$ and $A_2 = 1$ \citep{2019Jiang}.
     \item \citet{2020Boningcontraction} assumes $\pi(S) = w(|S|)\binom{G}{|S|}^{-1}$ and
\begin{equation}\nonumber
    \frac{B_1}{(G\lor n^{p_{max}})^{B_3}}<\frac{w(s)}{w(s-1)}<\frac{B_2}{(G\lor n^{p_{max}})^{B_4}},\quad s=1,\cdots,G,
\end{equation}
with constants $B_1, B_2, B_3, B_4>0$. This prior meets the Assumption \ref{assump:prior}(b) with any $A_1>B_3+1$ and any $A_2<B_4$ provided $p_{\max} = O(\log G/\log n)$.
\item (Beta-Binomial Prior). As a special example of the aforementioned prior in \citet{2020Boningcontraction}, the Beta-Binomial prior selects each group $j$ with probability $r$, with $r$ following the distribution Beta$(1, G^u)$. It satisfies Assumption \ref{assump:prior}(b) with $u>1$.
 \end{enumerate}
 
The i.i.d. Bernoulli prior and the Beta-Binomial prior might be the two most well-adopted model selection priors, where both can be described in a uniform way by introducing a set of binary latent variables $(z_i)_{i = 1}^G$ as follows:
\begin{equation}
\label{eq:sas-special}
\begin{aligned}
    z_i\mid w &\sim \text{Bernoulli}(w),\\
    \bm{\theta}_i\mid z_i & \sim z_ih(\bm{\theta}_i)+(1-z_i)\delta_{\bm{0}}.
\end{aligned}  
\end{equation}

 Assumption \ref{assump:prior}(c) requires the tail of the slab distribution to be heavy enough so that the prior can pose enough mass around the true regression coefficients.  Note that Assumption \ref{assump:prior}(c) doesn't prevent us from using the Gaussian slab as long as the variance of the Gaussian slab lies in some interval. Denote the dimension of $\bm{z}$ as $p_j$. For Gaussian slab function $h(\bm{z}) = (2\pi\tau_n^2)^{-p_j/2}\exp[-\|\bm{z}\|^2/(2\tau_n^2)]$, Assumption \ref{assump:prior}(c) can be satisfied with $A_3^\prime z_n^2/\log G\preceq \tau_n^2 \preceq G^{A_3^{\prime\prime}/p_j}$ and $A_3 = 1/A_3^{\prime}+A_3^{\prime\prime}/2$.
 For Laplace slab distribution $h(z)\propto \lambda_n^{p_j}\exp[-\lambda_n\|\bm{z}\|]$, the inverse scale parameter $\lambda_n$ should be chosen such that $G^{-A_3^{\prime}/p_j}\preceq \lambda_n \preceq A_3^{\prime\prime}\log G/z_n $ with $A_3 = A_3^\prime+A_3^{\prime\prime}$ to satisfy Assumption \ref{assump:prior}(c).
 In the case of t slab distribution with degree of freedom $\nu$, $h(\bm{z})\propto (\rho_n^2)^{p_j/2}[1+\rho_n ^2\|\bm{z}\|^2/\nu]^{-(\nu+p_j)/2}$, the choice of $G^{-A_3^{\prime}/p_j}\preceq\rho_n^2\preceq G^{A_3^{\prime\prime}/(\nu+p_j)}/z_n^2$ satisfies Assumption \ref{assump:prior} with $A_3 = A_3^{\prime}/2+A_3^{\prime\prime}/2. $ For the Cauchy slab, its inverse-scale parameter should be chosen in the same way as the t slab with the degree of freedom $\nu = 1$.

\section{Variational Methods for Grouped Variable Selections}
\label{sec:variational}

\subsection{Variational approximations with general spike-and-slab priors}
\label{subsec:general-sas}
The resulting posterior distribution $\Pi(\cdot\mid\bm{X}, \bm{Y})$  under the model selection prior \eqref{eq:sas-special} includes $2^G$ potential models and is thus difficult to evaluate even for moderate $G$. We adopt a Variational Bayes approach to approximate it with the following mean-field family:
\begin{equation}
\label{eq:mean-field-general}
    \mathcal{P}_{MF} = \left\{P_{\bm{\mu}, \bm{\Sigma}, \bm{\gamma}, v} = \prod_{j}q(\bm{\theta}_j)q(\sigma^2)\right\},
\end{equation}
where 
\begin{equation}\label{eq:VB_approx}
q(\sigma^2)\propto e^{-v/(2\sigma^2)}g(\sigma^2)(\sigma^2)^{-n/2}\ \ \ \textnormal{and}\ \ \ q(\bm{\theta}_j) = \gamma_jN(\bm{\mu}_j, \bm{\Sigma}_j)+(1-\gamma_j)\delta_{\bm{0}}.
\end{equation}
The corresponding VB posterior is defined as  the minimizer of the KL divergence between a mean-field density and the true posterior:
\begin{equation}
\label{eq:VB-optimization}
    \widetilde\Pi := \argmin_{P_{\bm{\mu}, \bm{\Sigma}, \bm{\gamma}, v}\in \mathcal{P}_{MF}} \text{KL}(P_{\bm{\mu}, \bm{\Sigma}, \bm{\gamma}, v}\|\Pi(\cdot\mid\bm{X}, \bm{Y})).
\end{equation}
Parameters $v$ and
$\gamma_j, \bm{\mu}_j, \bm{\Sigma}_j$, $j\in [p]$, can be determined via coordinate-ascent variational inference (CAVI) algorithm \citep{bishop:2006:PRML}. The details will be given later.

Similar to the hierarchical representation of the prior \eqref{eq:sas-special}, we let $(z_i)_{i = 1}^G$ be a set of binary random variables, the mean-field density can also be defined equivalently as:
\begin{equation}\nonumber
\begin{aligned}
z_i &\sim Bernoulli(\gamma_i),\\
    \bm{\theta}_i\mid z_i &\sim z_iN(\bm{\mu}_i, \bm{\Sigma}_i)+(1-z_i)\delta_{\bm{0}}.
\end{aligned}
\end{equation}

The mean-field family \eqref{eq:mean-field-general} and \eqref{eq:VB_approx} is inspired by \citet{2019rayvilinear}, and similar mean-field families have been widely used in the literature (\citet{2016huangvariational}, \citet{2012Carbonetto}, \citet{Titsias2011SpikeAS} to name a few).  The mean-field density for each $\bm{\theta}_j$ is designed to capture the  spike-and-slab nature of the prior distribution. While the Gaussian slab in the mean-field family may not be optimal (CAVI suggests that the optimal slab distribution should be proportional to $\exp(-\|\bm{X}_j\bm{\theta}_j-\bm{c}_j\|^2/(2s^2))h(\bm{\theta}_j)$), it still performs well in practice. One reason is that, as the sample size $n$ grows, the effect of the likelihood dominates the prior, leading to Gaussian tails in the posterior \citep{2019rayvilinear}. 

It is important to note that our mean-field family differs from others', such as those in \citet{2019rayvilinear}, in that we directly incorporate $\sigma$ into the family.  
\citet{2019rayvilinear} used an empirical Bayes approach where they first rescale the data using an estimated noise level obtained from external methods, such as scaled Lasso, and treat the noise level as unity, only considering the regression parameter $\bm{\theta}$ in their mean-field family. 
However, if the estimated noise level is inaccurate, which can easily happen when the design matrix is correlated  and the dimension is high \citep{2023Chenguang}, it can lead to an unsatisfactory estimation of the regression parameters.
In contrast, our method directly puts $\sigma^2$ into the mean-field family for simultaneous estimation, which can  help greatly in improving the estimation accuracy of both $\sigma^2$ and regression coefficients.

\subsection{Computational algorithms}\label{subsec:CAVI} 
\label{subsec:computation-general-sas}
In this subsection, we make simplified assumptions that the prior on $\sigma^2$ is Inverse-Gamma with shape parameter $\alpha$ and scale parameter $\beta$, specifically $g(\sigma^2) \propto (\sigma^2)^{-(\alpha+1)}\exp(-\beta/\sigma^2)$\footnote{When $\alpha=\beta=0$, this becomes the noninformative improper prior, i.e. $g(\sigma^2)\propto 1/\sigma^2$.} and that the spike-and-slab prior takes the special, yet well-adopted form \eqref{eq:sas-special}. In this case, $q(\sigma^2)$ becomes an Inverse-Gamma distribution with shape parameter $\alpha+n/2$ and scale parameter $\beta+v/2$.   We now provide a detailed CAVI algorithm to obtain the variational posterior $\widetilde \Pi$ defined as \eqref{eq:VB-optimization}. In CAVI, we iteratively optimize each of the parameters $\bm{\mu}_i$, $\bm{\Sigma}_i$, $\gamma_i$, and $v$, while holding all other parameters fixed.

Specifically, with the latent variable $z_i = 1$ and all other parameters fixed, 
we update the variational parameters $\bm{\mu}_i$ and $\bm{\Sigma}_i$ by minimizing the following objective functions:

\begin{equation}
\label{eq:mu-sigma-update-general}
    \begin{aligned} f_i(\bm{\mu}_i\mid\sim) &:= -2\tilde\sigma^2\mathbbm{E}_{\bm{\mu}_i, \bm{\Sigma}_i}\log h(\bm{\theta}_i)+\bm{\mu}_i^\intercal \bm{X}_i^\intercal\bm{X}_i\bm{\mu}_i-2\bm{Y}^\intercal\bm{X}_i\bm{\mu}_i+2\bm{\mu}_i^\intercal\bm{X}_i^\intercal\sum_{j\neq i}\gamma_j\bm{X}_j\bm{\mu}_j,\\
    g_i(\bm{\Sigma}_i\mid\sim) &:= -\log(|\bm{\Sigma}_i|)-2\mathbbm{E}_{\bm{\mu}_i, \bm{\Sigma}_i}\log h(\bm{\theta}_i)+\frac{1}{\tilde\sigma^2}Tr(\bm{X}_i^\intercal\bm{X}_i\bm{\Sigma}_i),
    \end{aligned}
\end{equation}
where $\tilde\sigma^2$ is defined as the reciprocal of the expectation $\mathbbm{E}_v[1/\sigma^2] = (n/2+\alpha)/(v/2+\beta)$ under the mean-field density, i.e., $\tilde\sigma^2 = (v/2+\beta)/(n/2+\alpha)$. Furthermore, fixing $\bm{\mu}, \bm{\Sigma}$ and $\bm{\gamma}$, we update $v$ as $v = \mathbbm{E}_{\bm{\mu}, \bm{\Sigma}, \bm{\gamma}}\|\bm{Y}-\bm{X}\bm{\theta}\|^2$, whose expansion is equal to 
\begin{equation}
v=\left[
     \bm{Y}^\intercal \bm{Y}-2\sum_{i=1}^G\gamma_i \bm{Y}^\intercal\bm{X}_i\bm{\mu}_i+\sum_{i=1}^G \gamma_i Tr(\bm{X}_i^\intercal\bm{X}_i(\bm{\mu}_i\bm{\mu}_i^\intercal+\bm{\Sigma}_i))+\sum_{i=1}^G\sum_{j\neq i}\gamma_i\gamma_j\bm{\mu}_i^\intercal \bm{X}_i^\intercal \bm{X}_j\bm{\mu}_j\right].
     \label{eq:Gaussian-sigma}
\end{equation}
Using similar calculations, we can also obtain the updating rule for $\gamma_i$:
\begin{equation}
\begin{aligned}
        \gamma_i = \text{logit}^{-1}&\left(\frac{1}{2\tilde\sigma^2}\left[2\bm{\mu}_i^\intercal\bm{X}_i^\intercal\bm{Y}-2\sum_{j\neq i}\gamma_j\bm{\mu}_i^\intercal\bm{X}_i^\intercal \bm{X}_j\bm{\mu}_j-Tr(\bm{X}_i^\intercal\bm{X}_i(\bm{\mu}_i\bm{\mu}_i^\intercal+\bm{\Sigma}_i))\right]\right.\\
        &\left.+\frac{p_i}{2}\log2\pi+\frac{1}{2}\log|\Sigma_i|+\frac{p_i}{2}+\mathbbm{E}_{\bm{\mu}_i, \bm{\Sigma}_i}\log h(\bm{\theta}_i)+\log\frac{w}{1-w}\right).
\end{aligned}
    \label{eq:Gaussian-gamma}
\end{equation}
These calculations can be found in  Supplementary Materials.

Given \eqref{eq:Gaussian-sigma}, we find that the slab function does not affect the updates of $v$, 
but only affects the updates of $\bm{\mu}_i$, $\bm{\Sigma}_i$ and $\gamma_i$ via the expectation $\mathbbm{E}_{\bm{\mu}_i, \bm{\Sigma}_i}\log h(\bm{\theta}_i)$. For the Gaussian slab $h(\bm{\theta}_i)\sim N(0, \lambda^2\bm{I}_{p_i})$, this expectation can be easily computed as
$-(\bm{\mu}_i^\intercal\bm{\mu}_i+Tr(\bm{\Sigma}_i))/\lambda^2$. Thus the updates for $\bm{\mu}_i,\ \bm{\Sigma}_i$ and $\gamma_i$ have nice analytic formulas:
\begin{equation}
    \begin{aligned}
\bm{\Sigma}_i &= \left(\bm{X}_i^\intercal\bm{X}_i/\tilde\sigma^2+\lambda^{2}\bm{I}_{p_i}\right)^{-1},
\quad \bm{\mu}_i = \tilde\sigma^{-2}
\bm{\Sigma}_i \left(\bm{X}_i^\intercal\bm{Y}-\sum_{j\neq i}\gamma_j\bm{X}_i^\intercal \bm{X}_j\bm{\mu}_j\right),\\
    \gamma_i & = \text{logit}^{-1}\bigg(\log\frac{w}{1-w}+\frac{1}{2}\log|\bm{\Sigma}_i|+p_i\log \lambda+\frac{1}{2}\bm{\mu}_i^\intercal\bm{\Sigma}_i^{-1}\bm{\mu}_i\bigg).
    \end{aligned}
    \label{eq:Gaussian-mu,sigma, gamma}
\end{equation}
Note that calculating $\bm{Y}-\sum_{j\neq i}\gamma_j\bm{X}_j\bm{\mu}_j$ for each $i = 1,\ldots, G$ in every iteration can be computationally expensive, with a time complexity of $O(np)$ per calculation, where $p$ is the number of predictors. This results in a total computation complexity of $O(npG)$ for the entire iteration, as we need to calculate a similar term $G$ times. To improve the computation efficiency,  we can define 
\begin{equation}
\label{eq:def-residual}
    \bm{r} = \bm{Y}-\sum_{j}\gamma_j \bm{X}_j\bm{\mu}_j,\ \ \ \bm{r}_i = \bm{r}+\gamma_i\bm{X}_i\bm{\mu}_i.
\end{equation}
 Then, the calculation of $\bm{Y}-\sum_{j\neq i}\gamma_j\bm{X}_j\bm{\mu}_j$ is reduced to the calculation of $\bm{r}_i$, which reduces the computational complexity from $O(np)$ to $O(np_i)$, where $p_i$ is the number of predictors in group $i$. So the time complexity for the whole iteration can be lowered from $O(npG)$ to $O(np)$, which is a significant improvement for large datasets with many groups.

 Other than the Gaussian slab, which benefits from explicit analytical formulas, the expectation $\mathbbm{E}_{\bm{\mu}_i, \bm{\Sigma}_i}\log h(\bm{\theta}_i)$ may not always have a closed-form expression with respect to $\bm{\mu}_i$ and $\bm{\Sigma}_i$. Sometimes, even if such an explicit expression exists, it may be too complicated to be useful. For instance, in the case of multi-Laplacian slab $h(\bm{\theta}_i)\propto \exp[-\lambda\|\bm{\theta}_i\|]$, evaluating $\mathbbm{E}_{\bm{\mu}_i, \bm{\Sigma}_i}\log h(\bm{\theta}_i)$ requires calculating the  $\ell_2$ norm for a general multivariate Gaussian random variable, of which the analytical form is very complicated and  involves special functions such as ``the Lauricella function'' \citep{1992Arak}. For a general slab function, we should resort to numerical or sampling methods such as the reparametrization trick \citep{kingma2013auto} to evaluate and optimize \eqref{eq:mu-sigma-update-general}. Specifically,  we can reparametrize $\bm{\theta}_i = \bm{\Sigma}_i^{1/2}\bm{Z}_i+\bm{\mu}_i$ where $\bm{Z}_i\sim N(0, \bm{I}_{p_i})$ and then deal with $\mathbbm{E}_{\bm{Z}_i}\log h(\bm{\Sigma}_i^{1/2}\bm{Z}_i+\bm{\mu}_i)$. However, this approach involves solving  $G$ such multivariate optimization problems in each iteration, which can be computationally expensive, particularly when the group size is large. Therefore, our method requires modifications for more general slab functions.

\subsection{Hierarchical spike-and-slab priors}
\label{subsec:hierarchical-sas}
By introducing a set of hidden variables $(\alpha_i)_{i = 1}^G$, the multi-Laplacian slab $h(\bm{\theta}_i)\sim \exp[-\lambda\|\bm{\theta}_i\|]$ can be rewritten hierarchically as a scale mixture of multivariate normal distributions, which is easier to handle. Specifically, we can rewrite the density of multi-Laplacian distribution as: 
\begin{equation}\nonumber
\bm{\theta}_i\mid\alpha_i^2\sim N(0, \alpha_i^{-2}\bm{I}_{p_i}),\quad \alpha_i^2\sim \text{Inv-Gamma}\left(\frac{p_i+1}{2}, \frac{\lambda^2}{2}\right).
\end{equation}
This motivates us to study a special family of slab distributions, in which each density can be written as a scale mixture of multivariate normal distributions. Throughout, for any $h(\bm{\theta}_i)$ in this family, we  rewrite it equivalently as:
\begin{equation}
\label{eq:hierarchi-slab}
\bm{\theta}_i\mid\alpha_i^2\sim N(0, \alpha_i^{-2}\bm{I}_{p_i}),\quad \alpha_i^2\sim \tilde h(\alpha_i^2),
\end{equation}
where $\tilde h$ is a one-dimensional density function. This family is quite rich and  includes multivariate Gaussian, multi-Laplacian, and multivariate-t distribution if we choose $\tilde h$ to be constant, inverse Gamma and Gamma distribution, respectively.

As we have introduced a new set of parameters $(\alpha_i)_{i = 1}^G$ into our model, it is natural to include them in the mean-field family of distributions, which can be expressed as:
\begin{equation}
\label{eq:mean-field-augment}
    \mathcal{P}_{MF} = \left\{P_{\bm{\mu}, \bm{\Sigma}, \bm{\gamma}, \bm{\kappa},v} = \prod_{j}q(\bm{\theta}_j, \alpha_j^2)q(\sigma^2)\right\},
\end{equation}
where
\begin{equation}
 q(\bm{\theta}_j, \alpha^2_j) = \gamma_jN(\bm{\mu}_j, \bm{\Sigma}_j)q(\alpha_j^2)+(1-\gamma_j)\delta_{\bm{0}}\tilde h(\alpha_j^2),\quad  q(\alpha_j^2)\propto (\alpha_j^2)^{p_j/2}e^{-\alpha_j^2\kappa_j/2}\tilde h(\alpha^2_j)
 \label{eq:hierarchical-variational}
\end{equation}
and $q(\sigma^2)$ is the same as in \eqref{eq:mean-field-general}. It is worth noting that although we call this family the mean-field family, $\alpha_j$ and $\bm{\theta}_j$ are not independent. Actually, let $z_j$ be the binary latent variable indicating whether $\bm{\theta}_j = 0$, $\alpha_j$ and  $\bm{\theta}_j$ are independent conditional on $z_j$. 

We now present a CAVI algorithm to compute the variational posterior $\tilde\Pi$ under the hierarchical slab distribution \eqref{eq:hierarchi-slab} and the augmented mean-field family \eqref{eq:mean-field-augment}. Similar to the derivation in Section \ref{subsec:computation-general-sas}, $\bm{\mu}_i$ and $\bm{\Sigma}_i$ can be updated as the minimizer of

\begin{equation}
\label{eq:mu-sigma-update-hierarchical}
    \begin{aligned} f_i(\bm{\mu}_i\mid\sim) &:= \tilde\sigma^2\mathbbm{E}_{\kappa_i}\alpha_i^2\bm{\mu}_i^\intercal\bm{\mu}+\bm{\mu}_i^\intercal \bm{X}_i^\intercal\bm{X}_i\bm{\mu}_i-2\bm{Y}^\intercal\bm{X}_i\bm{\mu}_i+2\bm{\mu}_i^\intercal\bm{X}_i^\intercal\sum_{j\neq i}\gamma_j\bm{X}_j\bm{\mu}_j,\\
    g_i(\bm{\Sigma}_i\mid\sim) &:= -\log(|\bm{\Sigma}_i|)+\mathbbm{E}_{\kappa_i}\alpha_i^2\bm{\Sigma}_i+\frac{1}{\tilde\sigma^2}Tr(\bm{X}_i^\intercal\bm{X}_i\bm{\Sigma}_i),
    \end{aligned}
\end{equation}
for which we have explicit solutions:
\begin{equation}\nonumber
    \begin{aligned}    \bm{\Sigma}_i&=\left(\bm{X}_i^\intercal\bm{X}_i/\tilde\sigma^2+\mathbbm{E}_{\kappa_i}\alpha_i^2\bm{I}_{p_i}\right)^{-1},\\
    \bm{\mu}_i&=\bm{\Sigma}_i\frac{\bm{X}_i^\intercal\bm{r}_i}{\tilde\sigma^2},
    \end{aligned}
    \label{eq:mu-signal-hierarchical-update}
\end{equation}
where $\bm{r}_i, \tilde\sigma^2$ is defined the same as \eqref{eq:def-residual} and \eqref{eq:Gaussian-sigma}, and $\mathbbm{E}_{\kappa_i}\alpha_i^2$ is the expectation of $\alpha_i^2$ under $q(\alpha_i^2)$. We can see the updates for $\bm{\Sigma}_i$ and $\bm{\mu}_i$ are rather similar to those for the Gaussian slab \eqref{eq:Gaussian-mu,sigma, gamma}.
However, the update for $\gamma_i$ is slightly different from \eqref{eq:Gaussian-gamma}. Specifically, $\gamma_i$ can be updated as:
\begin{equation}
\gamma_i:=\textnormal{logit}^{-1}\left[\log\frac{w}{1-w}+\frac{1}{2}[\kappa_i\mathbbm{E}_{\kappa_i}\alpha_i^2+\log|\Sigma_i|+\bm{\mu}_i^\intercal\bm{\Sigma}_i^{-1}\bm{\mu}_i]+\log C_i\right],
    \label{eq:GM-gamma}
\end{equation}
where 
\begin{equation}
    C_i=\int(\alpha_i^2)^{p_i/2}e^{-\alpha_i^2\kappa_i/2}\tilde h(\alpha^2_i)d\alpha_i^2
    \label{eq:normalizing-const}
\end{equation}is the normalizing constant of $q(\alpha_i^2)$.

By constraining ourselves to the hierarchical slab functions \eqref{eq:hierarchi-slab}, we are able to convert the original multivariate optimization problem with intractable terms \eqref{eq:mu-sigma-update-general} into a problem of calculating the expectation $\mathbbm{E}_{\kappa_i}\alpha_i^2$ and the normalizing constant $C_i$. These terms are just one dimensional integrals and can be efficiently solved through either numerical integration or Monte Carlo approximation.
Furthermore, for some commonly used slab functions, the closed-form expressions for these terms are available. Specifically, $q(\alpha_i^2)$ is  inverse Gaussian distribution for the multi-Laplacian slab, and Gamma distribution for the multivariate-t slab.  The exact updates for these parameters can be found in Table 1 in Supplementary Materials. Similar ideas of using hierarchical representations to avoid intractable terms in variational approximation have also been  adopted in the high-dimensional logistic regression \citep{2019SSGLneurips}.

Finally, $\kappa_i$ should be updated as the minimizer of the  following loss function:
\begin{equation}
\label{eq:kappa-loss}\nonumber
\ell(\kappa_i\mid\sim) := -\frac{1}{2}\left[(\kappa_i-\bm{\mu}_i^\intercal\bm{\mu}_i-Tr(\bm{\Sigma}_i))\mathbbm{E}_{\kappa_i}\alpha_i^2\right]-\log C_i,
\end{equation}
where $C_i$ is the normalizing constant in \eqref{eq:normalizing-const} and is also related to $\kappa_i$.
This leads to a surprisingly simple updating rule:
\begin{equation}
\label{eq:GM-kappa}
    \kappa_i = \bm{\mu}_i^\intercal\bm{\mu}_i+Tr(\bm{\Sigma}_i),
\end{equation}
which is invariant to the choice of $\tilde h$ and does not involve additional computational cost.

\subsection{Hyper-parameter specification}
\label{subsec:hyperparameter}
Assumption \ref{assump:prior}(c) is essentially an assumption on the hyper-parameter for the slab function $h$. Throughout, we denote this hyper-parameter as $\lambda$, which can be multi-dimensional and interchange the notations $h_{\lambda}$ and $\tilde h_{\lambda}$ with $h$ and $\tilde h$ respectively to highlight their dependences on $\lambda$. Furthermore, Assumption \ref{assump:prior}(b) is actually a requirement on the hyper-parameter $w$ in the special case of \eqref{eq:sas-special}, where $w$ controls the overall model size. Mis-specification of these hyper-parameters can lead to sub-optimal parameter estimation rates \citep{2012needles, 2015lineartheory}. In this section, we  propose a data-adaptive approach to determine $\lambda$ and $w$.

One possible solution to find satisfactory hyper-parameters $\lambda$ and $w$ is by cross-validation. However, since we are dealing with two hyper-parameters, performing two-way cross-validation can be computationally expensive, especially when testing a large number of possible values. 

Another popular approach is the empirical Bayes method, as demonstrated by \citet{2004Johnstone,2005Johnstone} for estimating hyper-parameters in the Gaussian sequence model and wavelet shrinkage. We adopt their idea here by writing out the likelihood with respect to the hyper-parameters and optimizing it. That is, we want to maximize
\begin{equation}
\label{eq:EM-hyper-general}
    \log\Pi(\bm{Y}|\lambda, w) = \log\int_{\bm{\theta}}\int_{\sigma^2}\int_{\bm{z}}\Pi(\bm{Y}, \bm{z}, \sigma^2, \bm{\theta}|\lambda, w)d\bm{z}d\sigma^2d\bm{\theta}
\end{equation}
for \eqref{eq:sas-special} and maximize 
\begin{equation}\nonumber
    \log\Pi(\bm{Y}|\lambda, w) = \log\int_{\bm{\theta}}\int_{\sigma^2}\int_{\bm{z}}\int_{\bm{\alpha}^2}\Pi(\bm{Y}, \bm{\alpha}^2, \bm{z}, \sigma^2, \bm{\theta}|\lambda, w)d\bm{\alpha}^2d\bm{z}d\sigma^2d\bm{\theta}
\end{equation}
for slab distributions with hierarchical representation \eqref{eq:hierarchi-slab}.

Take the optimization under general slab functions \eqref{eq:EM-hyper-general} as an example. By treating $\bm{z}, \bm{\theta}, \sigma^2$ as hidden variables, the optimization falls into the EM framework. However, the traditional EM algorithm requires to compute the expected log-likelihood under the true conditional posterior $P(\bm{z}, \bm{\theta}, \sigma^2|\bm{Y}, \lambda, w)$ in the E-step. This involves intractable calculations since there are $2^G$ possible combinations of $\bm{z}$. The variational EM algorithm, on the other hand, sidesteps this issue by using a simpler, tractable distribution $q(\bm{z}, \bm{\theta}, \sigma^2)$ to approximate the true posterior distribution $P(\bm{z}, \bm{\theta}, \sigma^2|\bm{Y}, \lambda, w)$. Here, we choose $q$ from the mean-field family \eqref{eq:mean-field-general}. By Jensen's inequality, we have
\begin{equation}
\begin{aligned}
    \log\Pi(\bm{Y}|\lambda,w)&=\log\int_{\bm{\theta}}\int_{\sigma^2}\int_{\bm{z}}\Pi(\bm{Y}, \bm{z}, \sigma^2,
    \bm{\theta}|\lambda, w)\\
    &\geq\int_{\bm{\theta}}\int_{\sigma^2}\int_{\bm{z}}q(\bm{z},\bm{\theta},\sigma^2)\log\frac{\Pi(\bm{Y,z,\sigma^2, \theta}|\lambda,w)}{q(\bm{z},\bm{\theta}, \sigma^2)}:=\mathcal{L}(q,\lambda, w),
\end{aligned}
\label{eq:elbo}
\end{equation}
where the surrogate function $\mathcal{L}(q, \lambda, w)$ is usually called Evidence Lower Bound or ELBO.

Then the optimization can be cast into two steps. In the E-step, the mean-field distribution $q$ is optimized to close the gap between the ELBO and the true log likelihood, which is equivalent to minimizing the KL divergence between the mean-field family and the true posterior and can be achieved via CAVI algorithm.
In the M-step, the hyper-parameters $\lambda$ and $w$ are optimized by maximizing the ELBO. The detailed algorithm can be found in Algorithm \ref{alg:update} in Supplementary Materials.

The ELBO $\mathcal{L}(q, \lambda, w)$ can be rewritten as a sum of the expected log-likelihood of the data and the KL divergence between the mean-field density $q$ and the prior $\Pi(\bm{z}, \sigma^2, \bm{\theta}\mid \lambda, w)$:
\begin{equation}\nonumber
    \mathcal{L}(q, \lambda, w) = \mathbbm{E}_{q}\left[\log\mathbbm{P}(\bm{Y}\mid\bm{X}, \bm{z}, \bm{\theta}, \sigma^2)\right]-\text{KL}\left(q(\bm{z}, \bm{\theta}, \sigma^2)\|\Pi(\bm{z}, \bm{\theta}, \sigma^2\mid \lambda, w)\right),
\end{equation}
where the first term is a constant with respect to hyper-parameters $\lambda$ and $w$. Hence, maximizing the ELBO with respect to $\lambda$ and $w$ is equivalent to minimizing the KL divergence between $q$ and $\Pi$. For the general slab function $h$, this optimization gives rise to the updates:
\begin{equation}
    \begin{aligned}
    w&=\frac{\sum_{i=1}^G\gamma_i}{G}, \ \quad 
    \lambda = \argmax \sum_{i = 1}^G\gamma_i\mathbbm{E}_{\bm{\mu}_i, \bm{\Sigma}_i}\log h_\lambda(\bm{\theta}_i).
    \end{aligned}
    \label{eq:general-hyper}
\end{equation}
We can see the update for $w$ does not depend on the slab function $h_{\lambda}$. For the Gaussian slab, where $h_{\lambda}(\bm{\theta}_i)\sim \exp(-\lambda\|\bm{\theta}\|_i^2)$, the update for $\lambda$ in \eqref{eq:general-hyper} is reduced to :
\begin{equation}\nonumber
\lambda=\frac{\sum_{i=1}^G\gamma_ip_i}{\sum_{i=1}^G\gamma_iTr(\bm{\Sigma}_i+\bm{\mu}_i\bm{\mu}_i^\intercal)}.
\end{equation}

For the hierarchical spike-and-slab prior, the updates are:
\begin{equation}
    \begin{aligned}
    &w=\frac{\sum_{i=1}^{G}\gamma_i}{G}, \ \quad 
\lambda=\argmax \bigg[\sum_{i=1}^G \gamma_i\int_{\alpha_i^2}q(\alpha_i^2)\log\tilde h_{\lambda}(\alpha_i^2)d\alpha_i^2\bigg],
    \end{aligned}
    \label{eq:GM-hyper}
\end{equation} 
where the closed-form solution is available for multi-Laplacian  and multivariate-t slabs (see Table 1 in Supplementary Materials). Similar idea of hyper-parameter specification has been adopted in \citet{2012Carbonetto} for individual variable selection and in \citet{2018bivas} for bi-level variable selection. Our method differs from these approaches as we include the noise variance $\sigma^2$ in our mean-field family, whereas they regard $\sigma^2$ as a hyper-parameter and update it in the M-step.

\subsection{Extensions to additive models}\label{sec:additive-model}
 
Let $\bm{Y}\in\mathbbm{R}^n$ be the response vector and let $\bm{X}\in\mathbbm{R}^{n\times p}$ be the design matrix. We consider the following nonparametric additive model:
\begin{equation}
\label{eq:additive-model-1}
Y_i=\mu+\sum_{j=1}^pf_j(X_{ij})+\epsilon_i,\quad \epsilon_i\sim N(0,\sigma^2)
\end{equation}
where $\mu$ is a constant and the $f_j$'s are smooth univariate functions. For the identification purpose, we assume that all $f_j$'s are centered, i.e.
$    \sum_{i = 1}^nf_j(X_{ij}) = 0. $
A basic idea for dealing with the additive model is to approximate each $f_j$ by a linear combination of a set of pre-defined basis functions $\{\phi_{jk}\}_{1\leq k\leq d}$:
\begin{equation}\nonumber
    f_j(x)\approx \sum_{k=1}^d\theta_{jk}\phi_{jk}(x),\quad j=1,\cdots, p,
\end{equation}
where $\theta_{jk}$ are unknown parameters. One commonly-used basis function is B-splines \citep{2010HuangAdditive}. Let $ \bm{\widetilde X}_{j} = \{\phi_{jk}(X_{ij})\}_{1\leq i\leq n, 1\leq k\leq d}$ be the $n\times d$ matrix where the $(i, k)$th element $\bm{\widetilde X}_{j, ik} =\phi_{jk}(X_{ij})$, then the additive model \eqref{eq:additive-model-1} can be approximated as:
\begin{equation}\nonumber
\label{eq:linear-additive}
\bm{Y}\approx\mu+\sum_{j=1}^p  \bm{\widetilde X}_{j}\bm{\theta}_j+\epsilon_i,\quad \epsilon_i\sim N(0,\sigma^2).
\end{equation}
To simplify the model, we can standardize $\bm{Y}$ and discard the intercept $\mu$.

  When the number of predictors $p$ is large, it is often the case that only a small fraction of the $f_j$'s are non-zero. Therefore, our goal is to estimate a sparse $(\bm{\theta}_j)_{j = 1}^p$, where most of $\bm{\theta}_j$'s are zero vectors.  This problem fits well into the framework of group variable selection, and   can be addressed using the Variational Bayes approach developed before.  More specifically, the design matrices for each group are the matrices of basis functions, $\bm{\widetilde X}_j, j = 1,\ldots, p$. We employ the spike-and-slab prior \eqref{eq:sas-special} as our working prior to induce the group sparsity. The slab function can be set to a general $h_{\lambda}$ or the hierarchical slab \eqref{eq:hierarchi-slab}. To approximate the resulting posterior distribution, we use Variational approximations with the mean-field family \eqref{eq:mean-field-general} or \eqref{eq:mean-field-augment}. The optimization details are summarized in Algorithm 1 in Supplementary Materials.

Similar additive models have been considered by a number of authors. 
In order to select variables as well as estimate the unknown functions, there are two main approaches. The first one uses the penalization method to identify important components, with methods such as group lasso \citep{2009Ravikumar} 
 and adaptive group lasso \citep{2010HuangAdditive}.  \citet{2010Meier} introduced the sparsity-smoothness penalty, which not only shrinks coefficients to zero but also controls the smoothness of the estimated functions. The second approach is a Bayesian one, where different priors are assigned to the parameters. Some examples include the multivariate extension of the Dirichlet-Laplace prior
\citep{Wei2019SparseBA} and the functional horseshoe prior \citep{2020horseshoe}. \citet{2019SSGL} generalized the spike-and-slab group Lasso method to the additive model. These methods typically use either MCMC to sample from the posterior or a block-descent algorithm to find the posterior mode. In contrast, our method utilizes variational inference to approximate the posterior distribution.

\section{Theoretical Results} \label{sec:mainresults}

We establish the contraction rates for both the exact posterior distribution under the generic spike-and-slab prior \eqref{eq:sas-group} with Assumption \ref{assump:prior} and  the VB posterior $\widetilde\Pi$  resulting from  \eqref{eq:VB-optimization} with mean field family \eqref{eq:mean-field-general} and \eqref{eq:mean-field-augment}.  Proofs for all theorems can be found in Supplementary Materials. 

For a  vector $\bm{\theta}=(\btheta_1^\top,\ldots,\btheta_G^\top)^\top$, we let $S_{\theta} = \{j\in\{1,\ldots, G\}: \bm{\theta}_j\neq \bm{0}\}$ be the index set of nonzero groups in $\bm{\theta}$ with $s_\theta = |S_\theta|$.  We denote the $\ell_1$, $\ell_2$, and $\ell_{2,1}$  norms of $\bm{\theta}$ as $\|\bm{\theta}\|_1$, $\|\bm{\theta}\|_2$, and $\|\bm{\theta}\|_{2,1}$, respectively, where $\|\bm{\theta}\|_{2,1} = \sum_{i = 1}^G\|\bm{\theta}_i\|_2$
is the sum of the within-group $\ell_2$ norms. 
For the design matrix $\bm{X}$, we define  $\|\bm{X}\|_o = \max\{\|\bm{X}_j\|:1\leq j\leq G\}$, which is the maximum matrix $\ell_2$ norm among all sub-matrices $\bm{X}_j$. For two positive sequences $a_n$ and $b_n$, we write $a_n\asymp b_n$ if both $a_n\preceq b_n$ and $a_n\succeq b_n$, where  definitions of ``$\preceq$ '' and ``$\succeq$'' can be found in Assumption \ref{assump:prior}. Moreover, we write $a_n\lesssim b_n$  (or $a_n\gtrsim b_n$) to mean  $a_n <b_n$  (or $a_n>b_n$) for sufficiently large $n$.

\subsection{Grouped linear regression}

We assume that the true generative model is:
\begin{equation}
\label{eq: true-model}
\bm{Y} = \bm{X}\bm{\theta}^{\star}+\bm{\epsilon},\  \bm{\epsilon}\sim N(\bm{0}, \sigma_\star^{2}\bm{I}_{n})
\end{equation}
where ${\bm{\theta}^\star}^\intercal = ({\bm{\theta}^\star_1}^\intercal,\ldots, {\bm{\theta}^\star_G}^\intercal)$, $\bm{X} = [\bm{X}_1,\ldots, \bm{X}_G]$ and $\sigma_\star^2\in(0,\infty).$ Define $S_\star = \{j\in \{1,\ldots, G\}: \bm{\theta}_j^\star\neq\bm{0}\}$ to be the index set of nonzero groups with  $s_\star = |S_\star|$.  
Let $p_i$ be the dimension of $\bm{\theta}_i^\star$ and let $p_{\max} = \max_{1\leq i\leq G} p_i$. We first introduce the following definitions and assumptions.

\begin{definition}[Full-rank models] Let $\mathcal{F}$ be the set of all full-rank models, i.e.,
\begin{equation}\nonumber
    \mathcal{F} := \{S\subset\{1,\ldots, p\}: \bm{X}_{S}^\top\bm{X}_{S}\ \textnormal{is invertable}\}.
\end{equation}
\end{definition}

\begin{definition}[Minimum United Eigenvalue (MUEV)] The minimum united eigenvalue of order $t$ for the design matrix X is defined as 
\begin{equation}\nonumber
    \textnormal{MUEV}(t) :=  \min_{S_\theta\in\mathcal{F}:|S_\theta\cup S|\leq t}\inf_{\bm{\theta}\neq\bm{0}} \left\{\frac{\|\bm{X}\bm{\theta}\|^2}{\|\bm{X}\|_o^2\|\bm{\theta}\|_2^2}\right\}
\end{equation}
\end{definition}

\begin{assumption}[MUEV condition]
\label{assump:MUEV}
There exists a constant $K>0$ such that 
\begin{equation}\nonumber
    \lambda = \lambda(K) := \textnormal{MUEV}((K+1)s)>0.
\end{equation}
\end{assumption}

The definition of MUEV can be seen as a generalization of that in the non-grouped linear regression \citep{2019Jiang}. Another well-adopted local eigenvalue used in the literature \citep{2020Boningcontraction,2019SSGL} is:

\begin{definition}
\label{def:sparse-eigenvalue}
(Minimum Sparse eigenvalue (MSEV)). The smallest scaled singular value  of dimension $t$ is defined as:
\begin{equation}\nonumber
    \textnormal{MSEV}(t) = \inf\left\{\frac{\|\bm{X}\bm{\theta}\|^2}{\|\bm{X}\|_o^2\|\bm{\theta}\|_2^2},\ \  0\leq s_{\bm{\theta}}\leq t\right\}.
\end{equation}
\end{definition}
In their papers, they need $\textnormal{MSEV}(T)$ to be larger than zero for some constant $T>0$. Comparing MUEV with MSEV, the following lemma essentially states that our MUEV condition is weaker than the condition required by MSEV. The proof of this Lemma is the same as the one in \citet{2019Jiang} and is thus omitted.
\begin{lemma}
For any $t>s_\star$,
\begin{equation}\nonumber
    \textnormal{MUEV}(t)\geq \textnormal{MSEV}(t).
\end{equation}
\end{lemma}

We further make an assumption on the configuration of the problem,  a similar assumption has also appeared in \citet{2019SSGL}. It allows the number of group $G$ to grow near exponentially with $n$. The maximum group size can also grow with $n$, but should grow at most in the order of $\log G/\log n$. It also specifies the growth rate of the true model size $s_\star$.

\begin{assumption}
\label{assump:configuration}
Assume $(G, s_\star, p_{\max})$ satisfies: 
\begin{equation}\nonumber
    G\gg n,\ \ \log G = o(n),\ \ s_\star = o(n/\log G),\ \  p_{\max}  = O(\log G/\log n).
\end{equation}
\end{assumption}

Let $\mathbbm{P}_\star$ denote the measure associated with the true model \eqref{eq: true-model}. Now we are ready to present the theorem regarding the contraction rate for the exact posterior distribution.

\begin{theorem}[Posterior Contraction]
\label{thm:posterior} If for some constants $C_1>0$ and $C_2>0$, Assumptions \ref{assump:prior} and \ref{assump:MUEV} hold with $A_1+A_3+1<(A_2-C_1-C_2)K$, and Assumption \ref{assump:configuration} holds. Then, for any constants $M_1, M_2>\sqrt{8(\max\{A_2, 1\}+C_2+1)K}$, the posterior distribution $\Pi(\sigma, S, \bm{\theta}\mid \bm{X}, \bm{Y})$ concentrates on 
\begin{equation}
\nonumber
    \widehat\Omega= \left\{(\sigma, S, \bm{\theta}): \ \ 
    \begin{aligned}
    & \frac{\sigma^2}{\sigma_\star^2}\in\bigg[\frac{1-M_1\epsilon_n}{1+M_1\epsilon_n}, \frac{1+M_1\epsilon_n}{1-M_1\epsilon_n}\bigg],\\
    &|S\backslash S_\star|\leq Ks_\star,\  S\in\mathcal{F},\\
    & \|\bm{\theta}-\bm{\theta^\star}\|\leq M_2\sqrt{n}\sigma_\star\epsilon_n/(\sqrt{\lambda}\|\bm{X}\|_o)
    \end{aligned}\right\}
\end{equation}
in the sense that
\begin{equation}
\label{eq:posterior-contraction}
    \mathbbm{P}_\star\left(\Pi(\widehat\Omega\mid \bm{X}, \bm{Y})\geq 1-e^{-C_1Ks_\star\log G}\right)\gtrsim 1-e^{-C_2Ks_\star\log G/2}.
\end{equation}
\end{theorem}
\begin{remark}
The key lemma in our proof is Lemma  C.1 in Supplementary Materials. This Lemma was first introduced in \citet{bernardo1998information} to establish the consistency of the posterior distribution and was later used to establish the contraction rate for the Bayesian high dimensional regression with shrinkage priors and spike-and-slab priors, respectively \citep{2023Song, 2019Jiang}.
\end{remark}

This theorem provides the posterior contraction rate for estimating both the regression coefficients $\bm{\theta}^\star$ and the noise level $\sigma_\star$ with general slab functions.  The contraction rates match those derived by \citet{2020Boningcontraction}, in which they only dealt with the multi-Laplacian slab function. When $p_{\max} = 1$, the posterior contraction rate for estimating $\bm{\theta}^\star$ simplifies to $M_2\sigma_\star\epsilon_n/\sqrt{\lambda}$, which is consistent with the contraction rate obtained when sparsity is imposed at the individual level \citep{2019Jiang}. Furthermore, the theorem  implies that the selection set is unlikely to substantially overshoot the true relevant group $S_\star$, and the posterior distribution will concentrate most of its mass on a set with a bounded number of false positives.

Compared with the results in \citet{2019Jiang}, our theorem provides a more detailed characterization of the relationship between the exponential terms in the contraction rate and constants $A_1, A_2, A_3, K, M_1, M_2$ in Assumptions \ref{assump:prior} and \ref{assump:MUEV}.
This is useful in  proving the subsequent contraction rate for the VB posterior. For the VB posterior, we need an additional assumption on the smoothness of the slab function $h(\bm{\theta}_j)$ and a more stringent MUEV condition on the design matrix. 
\begin{assumption}[Local Log Lipschitz]
\label{assump:lipschitz}
For any $(S,\bm{\theta})\in \widehat\Omega$ and for any $j\in S$
\begin{equation}\nonumber
    |\log h(\bm{\theta}_j)-\log h(\bm{\theta}^\star_j)|\leq L\|\bm{\theta}_j-\bm{\theta}^\star_j\|,
\end{equation}
where $L$ can depend on $n$ and $p$.
\end{assumption}

Then for spike-and-slab prior with both general slab function and hierarchical slab function, the VB posterior possesses the following contraction rate:
\begin{theorem}[VB posterior contraction]
\label{thm:VB-contraction}
Suppose Assumptions \ref{assump:prior} and \ref{assump:MUEV} hold with $A_1+A_3+1<A_2K$, and Assumption \ref{assump:configuration} holds.
Further assume for some $\rho_n\gg \max\{s_\star\epsilon_n, 1\}$, there exists $K$, s.t. $\lambda(\rho_nK)>0$.  Then for any constants $M_1, M_2>\sqrt{8(\max\{A_2, 1\}+1)K}$ and 
\begin{equation}
\label{eq:assump-constant-group}
   L = O\bigg(\max\left\{\min\bigg\{\frac{\sqrt{n\log G}}{s_\star}, \|\bm{X}\|_o\sqrt{\log n}\bigg\},\min\bigg\{\sqrt{n\log G}\epsilon_n, \|\bm{X}\|_o\sqrt{\log n}s_\star\epsilon_n\bigg\}\right\}\bigg),
\end{equation}
the VB posterior distribution $\widetilde\Pi(\sigma, S, \bm{\theta})$ concentrates on 
\begin{equation}\nonumber
    \widehat\Omega_{\rho_n}= \left\{(\sigma, S, \bm{\theta}):
    \begin{aligned}
    & \frac{\sigma^2}{\sigma_\star^2}\in\bigg[\frac{1-M_1\rho_n^{1/2}\epsilon_n}{1+M_1\rho_n^{1/2}\epsilon_n}, \frac{1+M_1\rho_n^{1/2}\epsilon_n}{1-M_1\rho_n^{1/2}\epsilon_n}\bigg],\\
    &|S\backslash S_\star|\leq K\rho_ns_\star,\  S\in\mathcal{F},\\
    & \|\bm{\theta}-\bm{\theta^\star}\|\leq M_2\rho_n^{1/2}\sqrt{n}\sigma_\star\epsilon_n/(\sqrt{\lambda(\rho_nK)}\|\bm{X}\|_o)
    \end{aligned}\right\}
\end{equation}
in the sense that
\begin{equation}\nonumber
\mathbbm{P}_\star\left(\widetilde\Pi(\widehat\Omega_{\rho_n})\right)\to 1.
\end{equation}
\end{theorem}
\begin{remark}
The main ingredient in our proof is to bound the KL divergence between VB posterior $\widetilde\Pi$ and the exact posterior $\Pi(\cdot\mid\bm{X}, \bm{Y})$. We adopt the technical framework developed in \citet{2019rayvilinear}, with an additional treatment for the unknown variance. Once we obtain the KL divergence bound, we can apply Theorem 5 in \citet{2019rayvilinear} to establish the VB contraction rate. Similar to their results, our theorem also holds for other mean-field families, such as
\begin{equation}
\label{eq:MF-family-2}
    \mathcal{P}_{MF}^{\prime} = \left\{P_S=q_S(\bm{\theta})q(\sigma^2)\right\},
\end{equation}
where $q_S(\bm{\theta})$ indicates a distribution with a single fixed support set $S\subset\{1,\ldots, G\}$. To be more specific, we can choose $q_S$ to be the product of a multivariate normal distribution with a Dirac distribution at zero, i.e., $q_S = N(\bm{\mu}_S, \bm{\Sigma}_S)\otimes \delta_{S^c}$, where $\bm{\mu}_S\in\mathbbm{R}^{p_S}$ and $\bm{\Sigma}_S\in \mathbbm{R}^{p_S\times p_S}$ is a positive definite matrix with either general form or block-wise diagonal form.
\end{remark}

Our theorem suggests that the VB posterior $\widetilde\Pi(\sigma, S, \bm{\theta})$ has a high probability of concentrating on a small neighborhood of the true parameter values. This provides a theoretical guarantee for the accuracy and stability of the VB posterior inference. The contraction rate for $\sigma_\star$ may be of independent interests since \citet{2019rayvilinear} only considers the contraction rates for $\bm{\theta}^\star$ and  treat $\sigma_\star$ as known.

Note that the VB contraction rate for both  $\sigma_\star$ and $\bm{\theta}^\star$ differs from its posterior counterpart by a factor of $\rho_n^{1/2}$. This factor also appears in \citet{2019rayvilinear}, where they only need $\rho_n\to\infty$ (can be arbitrarily slow). In contrast, our theorem has an additional requirement of $\rho_n\gg s_\star\epsilon_n$. We found that this extra requirement mainly arises from the VB estimation of $\sigma_\star$.   If   $\sigma_\star$ is known \textit{a priori} or $s_\star\epsilon_n = O(1)$, i.e., the signal is relatively sparse,  we can drop this condition and our VB contraction rate becomes similar to the one given by \citet{2019rayvilinear}. However, we believe this is merely a technical assumption and our method performs well when the noise level is unknown and the number of relevant groups increases.

We are left with the question that whether the restriction on the Lipschitz constant $L$ \eqref{eq:assump-constant-group} contradicts with Assumption \ref{assump:prior}(c).  Take the multivariate Gaussian slab  as an example, where $h(\bm{\theta}_j) = (2\pi\tau_n^2)^{-p_j/2}\exp[-\|\bm{\theta}_j\|^2/(2\tau_n^2)]$, we can choose
$L = 2z_n/\tau_n^2$. In this case, according to \eqref{eq:assump-constant}, $\tau_n^2$  has a lower bound, but this does not violate the upper bound given by Assumption \ref{assump:prior}(c). A similar argument applies to other commonly-used slab functions including the multi-Laplacian slab and the multivariate-t slab.

\subsection{Sparse generalized additive models}

We assume that the true generative model is:
\begin{equation}
\label{eq:additive-model}\nonumber
Y_i = \sum_{j = 1}^p f_{j}^\star(X_{ij})+\epsilon_i,\ \ \epsilon_i\sim N(0, \sigma_\star^2),
\end{equation}
where each $X_{ij}$ takes value in $[a,b]$, with $a<b$ being finite numbers, and  each $f_{j}^\star\in C^{\kappa}[0,1]$ for some $\kappa\in \mathbbm{N}$, i.e., each $f_j^\star$ is continuous up to the $\kappa$-th derivative. As we discussed in Section \ref{sec:additive-model}, the additive model \eqref{eq:additive-model} can be written as:
\begin{equation}\nonumber
    Y_i = \sum_{j = 1}^p  \bm{\widetilde X}_{ij}\bm{\theta}_{j}^\star+\delta_i+\epsilon_i,\ \epsilon_i\sim N(0, \sigma_\star^2),
\end{equation}
where 
$\bm{\theta}_{j}^\star$ is a $d$-dimensional vector of basis coefficients and 
$\delta_i$ is the bias. 

Denote $\bm{\widetilde X}_j = (\bm{\widetilde X}_{1j},\ldots,\bm{\widetilde X}_{nj})^\intercal$, $\bm{\widetilde X} = [\bm{\widetilde X}_1,\ldots,\bm{\widetilde X}_p]$, $\bm{\delta} = (\delta_1,\ldots, \delta_n)^\intercal$ and ${\bm{\theta}^\star}^\intercal = ({\bm{\theta}_1^\star}^\intercal,\ldots, {\bm{\theta}_p}^\intercal)$. Let $s_\star$ be the number of nonzero function components. Define $\epsilon_n^2 = \max\{s_\star\log p/n, s_\star d\log n/n\}$, we make the following assumptions:
\begin{assumption}
\label{assump:additive}
\text{}
\begin{enumerate}[(a)]
    \item Assume that $p\gg n$, $\log p = o(n)$ and $d \asymp n^{1/(2\kappa+1)}$.
    \item The true number of nonzero functions satisfies $s_\star = o(n/\log p\wedge n^{2\kappa/(2\kappa+1)}/\log n)$.
\item $\widetilde{\bm{X}}$ satisfies Assumption \ref{assump:MUEV} and $\lambda_{\max}(\widetilde{\bm{X}}^\intercal \widetilde{\bm{X}})\leq k_1n$ for some constant $k_1>0$.
\item The bias $\bm{\delta}$ satisfies $\|\bm{\delta}\|\lesssim \sqrt{s_\star n}d^{-\kappa}$.
\end{enumerate}    
\end{assumption}
Assumption \ref{assump:additive}(a) assumes that $d$, the size of truncated basis expansions, is of the same order as $n^{1/(2\kappa+1)}$, which is a commonly assumed for additive models \citep{highdstatbook, 2019SSGL}. Assumption \ref{assump:additive}(b) ensures $n\epsilon_n^2\to 0$. Assumption \ref{assump:additive}(c) and (d) are hard to verify but have been shown to hold with appropriate basis functions, e.g., cubic B-splines \citep{2016Yoo, 1998shen}. Similar sets of assumptions have also appeared in \citet{2019SSGL} and \citet{Wei2019SparseBA}. Let $\mathbbm{P}_\star$ denote the probability with respect to the true data generating distribution \eqref{eq:additive-model}.

\begin{theorem}[VB posterior contraction$^\prime$]
\label{thm:VB-contraction-additive}
Suppose Assumptions \ref{assump:prior} and \ref{assump:MUEV} hold with $A_1+A_3+1<A_2K$, and Assumption \ref{assump:configuration} holds.
Further assume for some $\rho_n\gg \max\{s_\star\epsilon_n, 1\}$, there exists $K$, s.t. $\lambda(\rho_nK)>0$.  Then for any constants $M_1, M_2>\sqrt{8(\max\{A_2, 1\}+1)K}$ and 
\begin{equation}
\label{eq:assump-constant}
   L = O\bigg(\max\left\{\min\bigg\{\frac{\sqrt{n\log p}}{s_\star}, \|\bm{\widetilde X}\|_o\sqrt{\log n}\bigg\},\min\bigg\{\sqrt{n\log p}\epsilon_n, \|\bm{\widetilde X}\|_o\sqrt{\log n}s_\star\epsilon_n\bigg\}\right\}\bigg),
\end{equation}
the VB posterior distribution $\widetilde\Pi(\sigma, S, \bm{\theta})$ concentrates on 
\begin{equation}\nonumber
    \widehat\Omega_{\rho_n}= \left\{(\sigma, S, \bm{\theta}):
    \begin{aligned}
    &\frac{\sigma^2}{\sigma_\star^2}\in\bigg[\frac{1-M_1\rho_n^{1/2}\epsilon_n}{1+M_1\rho_n^{1/2}\epsilon_n}, \frac{1+M_1\rho_n^{1/2}\epsilon_n}{1-M_1\rho_n^{1/2}\epsilon_n}\bigg],\\
    &|S\backslash S_\star|\leq K\rho_ns_\star,\  S\in\mathcal{F},\\
    & \frac{1}{\sqrt{n}}\|\widetilde{\bm{X}}\bm{\theta}-\sum_{j = 1}^pf_j^\star(\bm{X}_j)\|\leq M_2\rho_n^{1/2}\sqrt{n}\sigma_\star\epsilon_n/(\sqrt{\lambda(\rho_nK)}\|\bm{\widetilde X}\|_o)
    \end{aligned}\right\}
\end{equation}
in the sense that
\begin{equation}
\label{eq:VB-additive-posterior-contraction}
\mathbbm{P}_\star\left(\widetilde\Pi(\widehat\Omega_{\rho_n})\right)\to 1.
\end{equation}
\end{theorem}
\begin{remark}
    For the exact posterior distribution, we can get rid of the $\rho_n$ in the above theorem so that  the contraction rate for estimating  $\sum_{j = 1}^{p}f_j^\star(\bm{X_j})$  is $n\epsilon_n^2 \asymp s\log p+sn^{1/(2\kappa+1)}\log n$. The minimax rate for estimating $\sum_{j = 1}^{p}f_j^\star(\bm{X_j})$ is $s\log p+sn^{1/(2\kappa+1)}$ \citep{raskutti2010minimax}, which is only better than  our  posterior concentration rate  by a $\log n$ term.
\end{remark}

\section{Simulation Study}
\label{sec:simulation}

We consider three special slab functions: multivariate Gaussian , multi-Laplacian, and multivariate-t.  We set the initial value of $\bm{\gamma}$ in the VB approximation \eqref{eq:VB_approx} to $(1/G, \ldots, 1/G)$ and initialize $\bm{\mu}$ using ridge regression, where we select the penalization parameter through 10-fold cross-validation. We adopt a prioritized updating strategy as in \citet{2019rayvilinear}, where we start by updating groups with the strongest signal coefficients and finish with those with weak or no signals. 

We use a stopping criterion based on the difference between the binary entropy of the updated $(\gamma_i)_{i=1}^G$ and the original $ (\gamma_{i,\ old})_{i = 1}^G$, as well as the difference between the updated $\tilde\sigma$ and the original $\tilde\sigma_{old}$. Specifically, after each iteration, we compute the binary entropy $H(\gamma_i):=-\gamma_i\log \gamma_i-(1-\gamma_i)\log(1-\gamma_i)$ for each group $i=1,\cdots,G$, and terminate the algorithm if $\max\{|H(\gamma_i)-H(\gamma_{i,old})|: i=1,\cdots,G\}<\epsilon_H$ and $|\tilde\sigma-\tilde\sigma_{old}|<\epsilon_\sigma$, where $\epsilon_H$ and $\epsilon_\sigma$ are pre-determined thresholds. This stopping criterion is similar to that used by \citet{2016huangvariational}  and \citet{2019rayvilinear}. The proposed algorithm is summarized in Algorithm 1 in Supplementary Materials, and specific update rules are provided in Table 2 in Supplementary Materials.

For evaluation purposes, we use TP, TN, FP, and FN to denote the numbers of true positives, true negatives, false positives, and false negatives, respectively. We evaluate the model selection performance via precision, recall and  Matthews correlation coefficient (MCC), where  $\textnormal{precision} = \textnormal{TP}/(\textnormal{TP}+\textnormal{FP})$, $\textnormal{recall} = \textnormal{TP}/(\textnormal{TP}+\textnormal{FN})$ and 
\begin{equation}\nonumber
\text{MCC}=\frac{\text{TP}\times\text{TN}-\text{FP}\times\text{FN}}{\sqrt{(\text{TP+FP})(\text{TP+FN})(\text{TN+FP})(\text{TN+FN})}}.
\label{eq:MCC}
\end{equation}
MCC is a reliable and informative score in binary classification tasks \citep{2020MCC} and higher values indicate better model selection performances. Throughout, each dot in the figure represents the average from 200 independent runs.

\subsection{Grouped linear regression}
\label{subsec:simu-grouplinear}

We focus on the comparison between our method (GVSSB) and three popular methods in high-dimensional grouped linear regression: Group Lasso \citep{2006YuanandLin}, Adaptive Group Lasso \citep{2008Wang}, and Spike-and-Slab Group Lasso \citep{2019SSGL} (short-handed as Glasso, AdGlasso, and SSGL, respectively), which are implemented as R packages \pkg{grpreg}, \pkg{glmnet}, and \pkg{SSGL}. We also consider three variants of our method by selecting different slab functions: multivariate Gaussian (GVSSB-G), multi-Laplacian (GVSSB-L) and multivariate Cauchy (GVSSB-C). To initialize our algorithms, we set $\lambda=1$ and $w=1/G$.  For group lasso and adaptive group lasso, we determine the penalization parameters through 10-fold cross-validation. In the following examples, we summarize the resulting  precision, recall, MCC, and logarithm of the mean squared error (MSE) based on 200 independent runs.

We simulate the response vector $\bm{y}$ from the linear model $\bm{y} = \bm{X}\bm{\theta}^\star+\bm{\epsilon}$ with $\bm{\epsilon}\sim N(\bm{0}, \sigma_\star^2I_n)$ and randomly located $k$ nonzero groups. We generate feature i.i.d. vectors from $N(0, \Sigma)$ where $\Sigma_{ij} = 0.6$ for $i,j$ in the same group, and $\Sigma_{ij} = \rho$ for $i,j$ not in the same group, where $\rho$ varies case by case. The signal-to-noise ratio is defined as $\textnormal{SNR} = \textnormal{Var}(\bm{X}\bm{\theta}^\star)/\sigma_\star^2$.

\medskip

\begin{figure}[t]
    \centering
    \includegraphics[width=0.92\columnwidth]{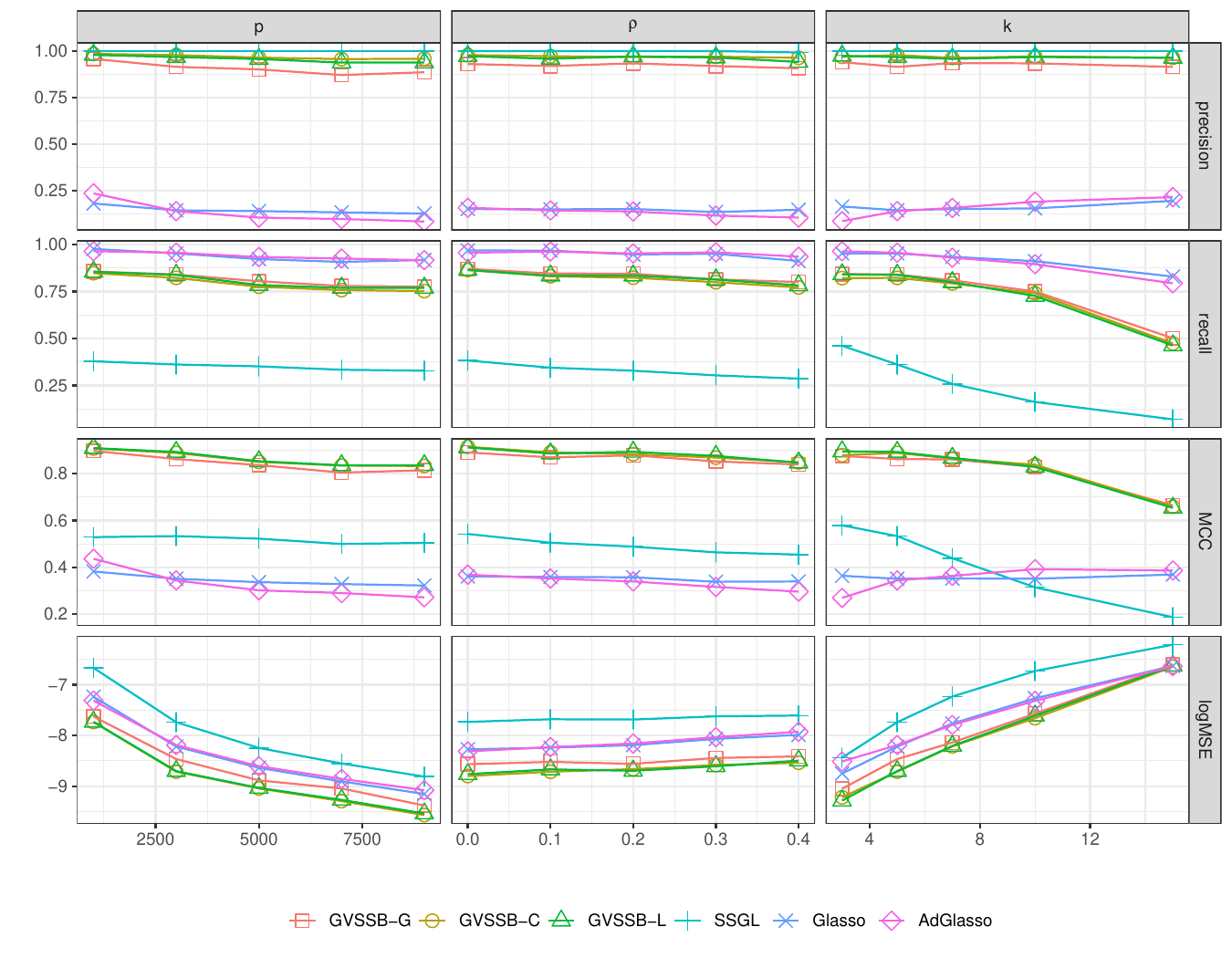}
    \caption{Example 1 of sparse grouped linear regression. $(\rho, k) = (0.2 , 5)$, $p$ is varied from $1000$ to $9000$ in the left column. $( p, k) = (3000 , 5)$, $\rho$ is varied from $0$ to $0.4$ in the middle column. $( p, \rho) = ( 3000, 0.2)$, $k$ is varied from $3$ to $15$ in the right column.}
    \label{fig:r}
\end{figure}

\textbf{Example 1}: We consider a scenario where the sample size $n=200$, the group size $p_i = 5$ across all groups and the noise level $\sigma_\star =1$.
The coefficients in every nonzero group are sampled uniformly from the interval $[-0.5, 0.5]$. Feature size $p = Gp_i$, between group correlation $\rho$ and the nonzero group number $k$ are varied. The results are summarized in Figure \ref{fig:r}. Our method consistently outperforms the other three methods in terms of  precision, MCC, and MSEs across all settings. Among the three variants of our method, multi-Laplacian slab and multivariate Cauchy slab perform similarly, while the multivariate Gaussian slab performs worse in terms of precision and MSE. GLasso and AdGlasso sacrifice some precision for high recall rates, resulting in the selection of  an excessive number of null groups \citep{2013FengConsistentCF}.
In contrast, SSGL is overly conservative, selecting very few groups and performing the worst in terms of MSE.

\medskip

\begin{figure}[t]
    \centering
    \includegraphics[width = 0.92\columnwidth]{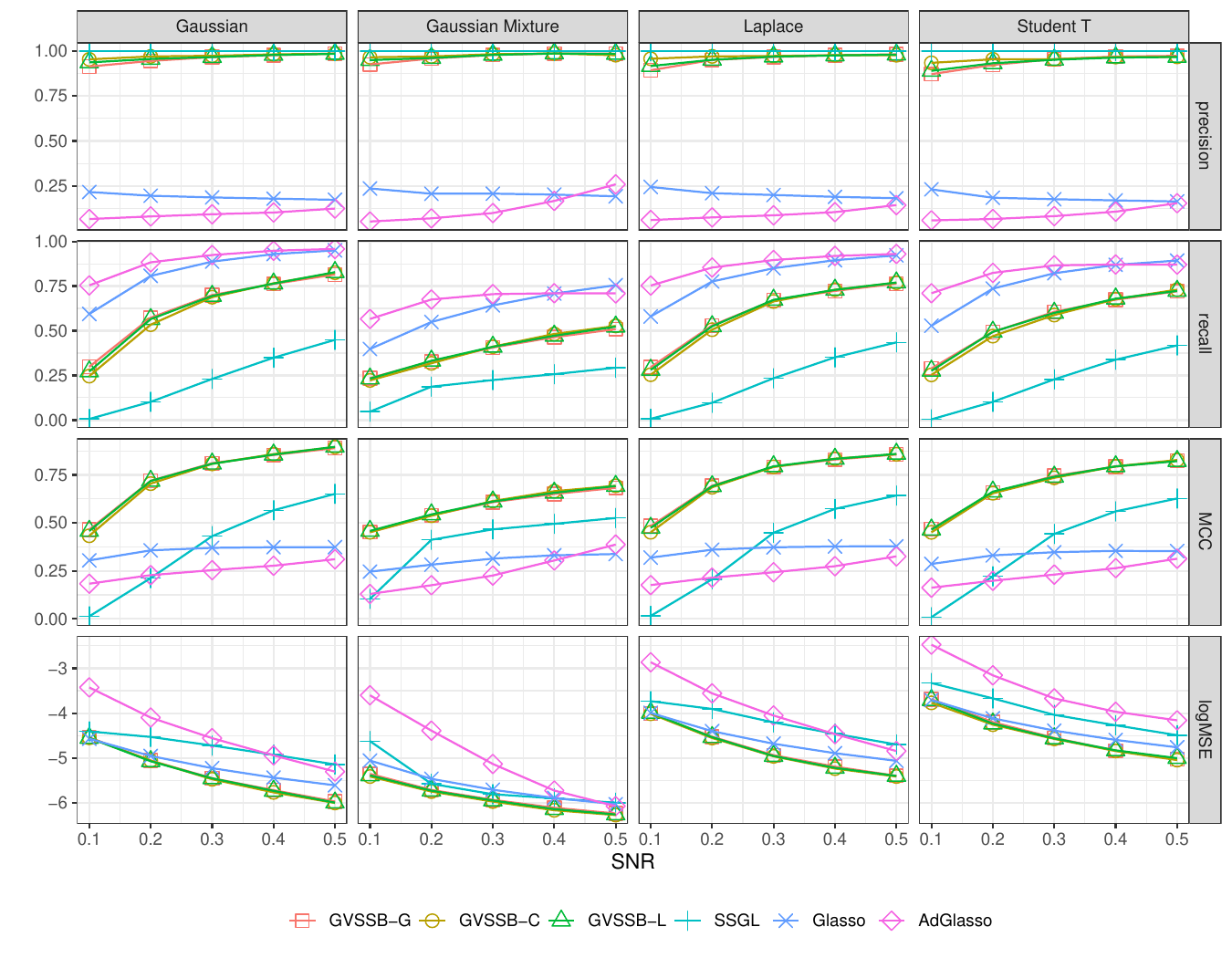}
    \caption{Example 2 of sparse grouped linear regression. $(n, p, p_i, \rho, k)=(800, 2000, 5, 0.2, 5)$ across all settings. The SNR varies from 0.1 to 0.5.}
    \label{fig:snr}
\end{figure}

\textbf{Example 2:}  With $n=800$, $G=400$, $p_i=5$, $k = 5$ and $\rho=0.2$ fixed, the coefficients in every nonzero group are sampled randomly from four distributions: Laplace, Gaussian,  Gaussian Mixture, and t distribution with the degree of freedom equals to $3$. We vary $\sigma_\star$ to obtain the signal-to-noise ratios ranging from $0.1$ to $0.5$. Results are summarized in Figure \ref{fig:snr}. We observe that all three variants of our method demonstrate robust performances across different true signal specifications. Our method outperforms Group Lasso and Adaptive Group Lasso in terms of precision, MCC, and MSE, while these methods have significantly higher recall rates.  This means that our method can capture and only capture those groups that contribute the most to the response vector $\bm{y}$. On the other hand, SSGL remains very parsimonious in selecting groups, as in the previous example.

Section B in Supplementary Materials provides additional simulation results. In Section B.1, we compare the proposed method to the
``Oracle'' Gibbs sampler in a moderate-scale problem. In Section B.2, we assess the estimation accuracy of the unknown noise level $\sigma_\star$ using the proposed method, comparing it with SSGL and the Gibbs sampler.  In Section B.3, we examine the performance of the proposed method with and without the EM optimization of the hyper-parameters. We also investigate the sensitivity of the results with respect to the initialization of hyper-parameters.

\subsection{Sparse generalized additive model }\label{subsec:simu-additive}

In addition to the six methods we compared in Section \ref{subsec:simu-grouplinear}, we also include \textit{Sparse Additive Models} (SpAM) of  \citet{2009Ravikumar} as implemented in the R package \pkg{SAM}, for comparisons. 
The penalization parameter for SpAM is chosen via 10-fold cross-validation. Two examples are considered in this section. Three variants of our proposed method perform rather similarly in these two settings, thus we only report the results associated with multivariate Cauchy slab. Since there is no true underlying coefficient $\bm{\theta}^\star$ in the additive model, we examine the mean-squared prediction error using $500$ leave-out test data points.

\medskip

\begin{figure}[t]
   \centering
    \includegraphics[width=0.92\columnwidth]{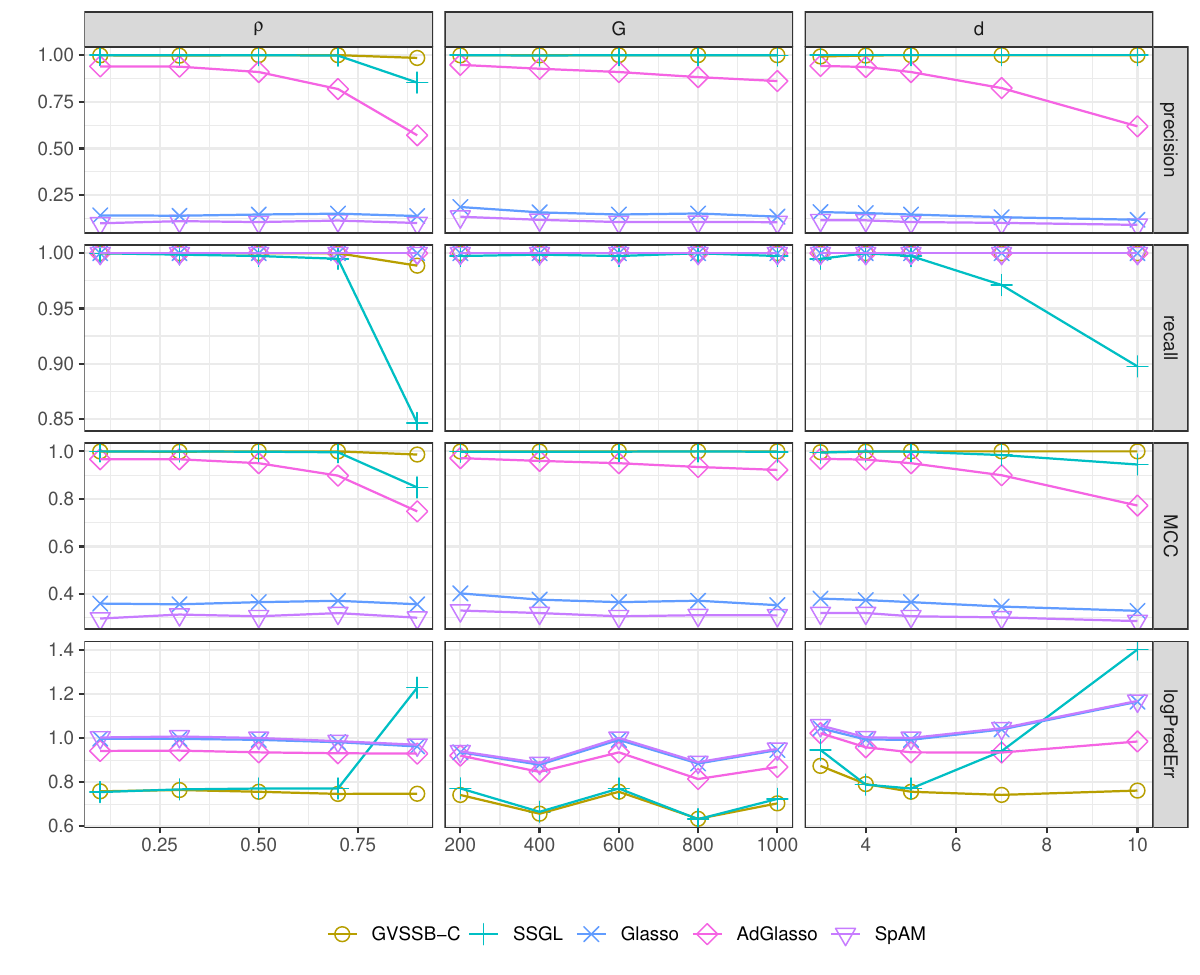}
    \caption{Example 1 of additive model. $(n, G, d) = (200, 600, 5)$, $\rho$ varies from $0.1$ to $0.9$ in the left column. $(n, d, \rho) = (200, 5, 0.5)$, $G$ varies from $200$ to $1000$ in the middle column. $(n, G, \rho) = (200, 600, 0.5)$, $d$ varies from $3$ to $10$ in the right column.}
    \label{fig:additive}
\end{figure}

\textbf{Example 1.} The true model is $Y_i = f_1(x_{i1})+f_2(x_{i2})+f_3(x_{i3})+f_4(x_{i4})+\epsilon_i$, where $\epsilon_i\overset{i.i.d.}{\sim}N(0, 1)$ for $i = 1,\ldots, n$, with $f_1(x) = 5\sin(x)$, $f_2(x) = 2(x^2-0.5)$, $f_3(x) = e^x$ and $f_4(x) = 3x$. The covariates are  generated independently from $N(0,\Sigma)$, where $\Sigma$ has an auto-regressive structure with $\Sigma_{ij}=\rho^{|i-j|}$. The experiment is carried out with a fixed sample size of $n = 200$, the correlation parameter $\rho$ and the number of features are varied. Furthermore, we expand each additive component using $d$ basis functions and $d $ is also varied to test the robustness of each method. The results are summarized in Figure \ref{fig:additive}.

\medskip

\begin{figure}[t]
    \centering
    \includegraphics[width = 0.85\columnwidth]{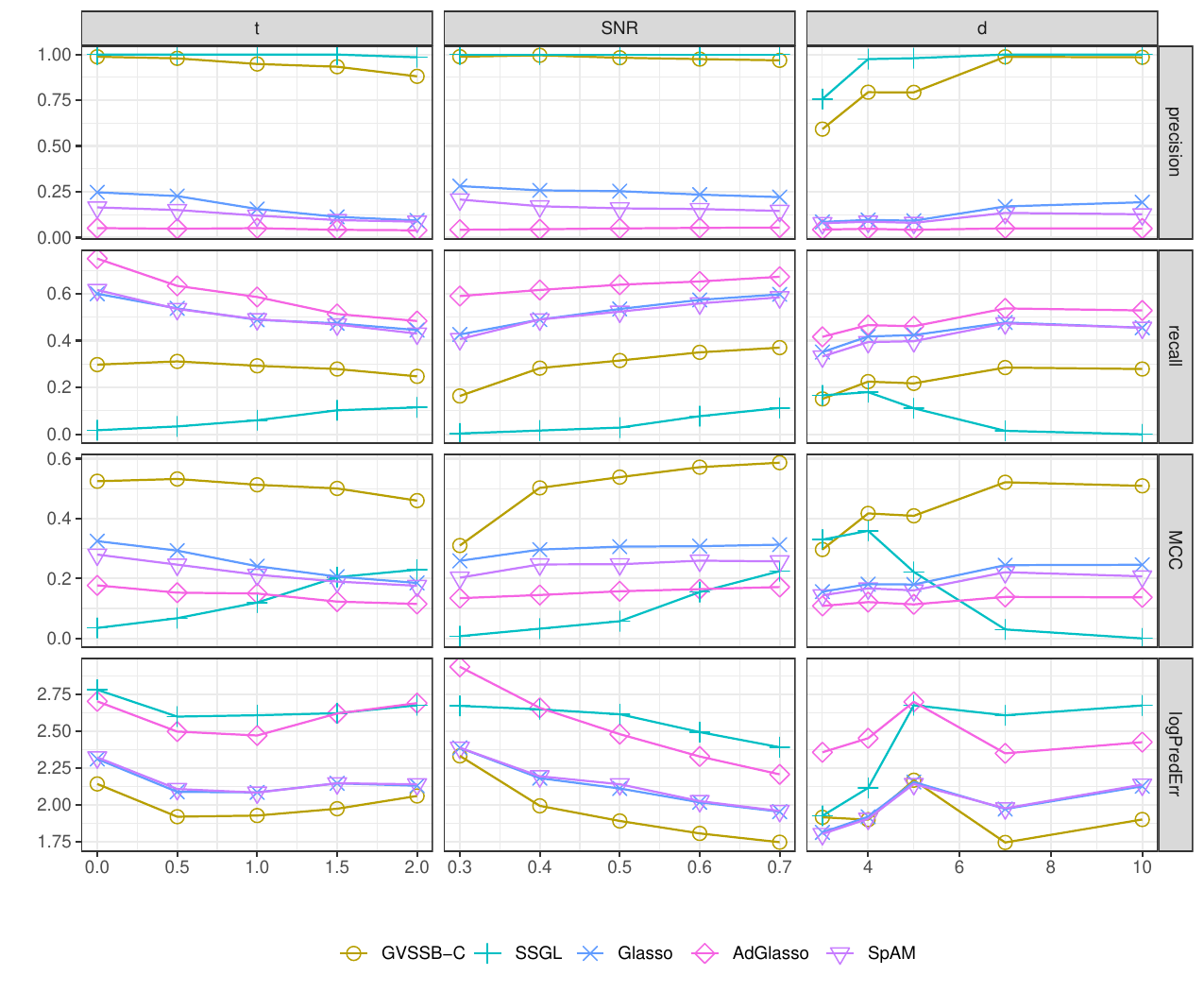}
    \caption{Example 2 of Additive Model. $(n, \text{SNR}, d) = (200, 0.5, 5)$, $t$ varies from $0$ to $2$ in the left column. $(n, d, t) = (200, 5 , 0.5)$, $\text{SNR}$ varies from $0.3$ to $0.7$ in the middle column. $(n, t, \text{SNR}) = (200, 0.5, 0.5)$, $d$ varies from $3$ to $10$ in the right column.}
    \label{fig:compound}
\end{figure}

\textbf{Examples 2:} This is equivalent to Example 1 in \citet{2006Lin}. The true model is $Y_i = f_1(x_{i1})+f_2(x_{i2})+f_3(x_{i3})+f_4(x_{i4})+\epsilon_i$, where $\epsilon_i\overset{i.i.d.}{\sim}N(0, \sigma_\star^2)$ for $i = 1,\ldots, n$, with $f_1(x) = 5x$, $f_2(x) = 3(2x-1)^2$, $f_3(x) = 4\sin(2\pi x)/(2-\sin(2\pi x))$ and $f_4(x) = 6(0.1\sin(2\pi x)+0.2\cos(2\pi x)+0.3 \sin^2(2\pi x)+0.4\cos^3(2\pi x)+0.5\sin^3(2\pi x))$. The standard deviation of the noise $\sigma_\star$ is chosen to vary the signal-to-noise ratios from $0.3$ to $0.7$. The covariates are generated via $X_{ij} = (W_{ij}+tU_i)/(1+t)$, for $i = 1,\ldots, n$ and $j = 1,\ldots, p$ where $W_{i1},\ldots, W_{ip}$ and $U_i$ are i.i.d. from $\textnormal{Uniform}(0,1)$. Therefore, $\textnormal{corr}(X_{ij}, X_{ik}) = t^2/(1+t^2)$ for any $j\neq k$. We fix the sample size $n = 200$ and the feature size $p = 600$. The correlation $t$ and the number of basis functions $d$ are varied. The results are summarized in Figure \ref{fig:compound}.

In Example 1, the proposed method outperforms other methods in all four criteria and across all settings, except when the correlation $\rho = 0.9$. This shows the effectiveness of our method in both coefficient estimates and model selection. Adaptive group Lasso performs better than group Lasso and SpAM in terms of the precision and the  prediction error. SSGL performs second to best when the correlation  $\rho$ is small and when the size of basis expansions $d$ is small. However, its performance deteriorates substantially when correlation $\rho = 0.9$ and when the size of basis functions $d$ increases, indicating that SSGL is not as robust as others.

In Example 2, while group Lasso, adaptive group Lasso, and SpAM have higher recall rates than SSGL and GVSSB, the precision of the latter two methods are close to $1$, which suggests that they rarely select any false positives. In terms of the prediction error, GVSSB performs slightly better than other methods. The performances of all methods become better when the SNR increases and worse when the correlation $t$ becomes large or when the size of basis functions $d$ is larger than needed.

\section{A Real Data Example}
\label{sec:realdata}

In this section, we investigate the performances of various methods for  fitting the  Near Infrared (NIR) Spectroscopy data, which is available in the R package \pkg{chemometrics}. This dataset was initially analyzed by \cite{2009Liebmann} and subsequently examined by \cite{2014McKay} and \cite{2020horseshoe}. The data consists of glucose and ethanol concentrations (in g/L) for $166$ alcoholic fermentation mashes made from various feedstocks, such as rye, wheat, and corn. Our objective is to predict the ethanol concentrations using $235$ variables that contain NIR spectroscopy absorbance values collected from a transflectance probe within the wavelength range of $1115$-$2285$ nanometers (nm). 

We apply the additive model in our analysis to capture the relationship between ethanol concentrations and NIR spectroscopy absorbance values. The absorbance values are expanded by $d$ basis functions, where $d$ varies from 3 to 15. The sample size for this model is $n=166$, and the number of potential groups is $p=235$. To assess the model's performance, we report the ten-fold  prediction error. Specifically, the dataset is partitioned into ten equal folds. In each run, we train our model using none folds and use the remaining  one to calculate the prediction error. This process is then repeated ten times, with each fold serving as the validation set once. Finally, we compute the average prediction error across all ten runs. On each training set, we further use $10$-fold cross-validation to choose the slab function which yields the best prediction error among three options: multivariate Gaussian slab, multi-Laplacian slab, and multivariate Cauchy slab. The penalization parameters for other methods are also chosen in this way.

Results are summarized in Table \ref{table:real-data} and show that our method achieves the smallest prediction error among considered methods except for $d = 3$. When $d = 3$, SpAM enjoys the best prediction error. The smallest prediction error is achieved by the proposed method  when $d$ equals $4$. When $d$ grows, the difference in prediction error between our method and the compared methods becomes more noticeable, indicating that our method is less susceptible to overfitting. When $d$ is relatively small, e.g., $d\leq 8$, Adaptive Group Lasso is the second-best method after ours. However, it fails to perform well when $d$ is equal to $10$ and $15$. SpAM and Group Lasso both perform robustly across different values of $d$. On the other hand, SSGL shows unstable performance across all settings.

\begin{center}
\begin{table}[h]
\centering
\begin{tabular}{cccccc}
\toprule
\multicolumn{1}{c}{$d$} & \multicolumn{1}{c}{GVSSB} & \multicolumn{1}{c}{Glasso} & \multicolumn{1}{c}{AdGlasso} & \multicolumn{1}{c}{SSGL} & \multicolumn{1}{c}{SpAM} \\ \hline
3                       & 6.18               & 7.95               & 6.15                  & 14.9              &\textbf{4.84}               \\
4                       & \textbf{4.53}                & 6.71                 & 5.69                   & 23.96             & 6.03               \\
5                       & \textbf{4.57}               & 5.63                 & 5.84                   & 37.34             & 6.42               \\
6                       & \textbf{5.60}               & 7.59                & 6.13                   & 84.62            & 8.21           \\
7                       & \textbf{5.96}                & 6.53                & 6.74                  & 102.59           & 8.02              \\
8                       & \textbf{5.88}               & 7.48                & 6.39                  & 83.9             & 7.54              \\
10                      & \textbf{6.06}               & 9.17              & 11.38             & 108.01           & 11.33            \\
15                      & \textbf{7.68}              & 11.79              & 27.91                 & 104.21           & 14.42     \\
\bottomrule
\end{tabular}
\caption{Average prediction errors  for NIR data analysis. The lowest prediction error for each $d$ is marked in bold. }
\label{table:real-data}
\end{table}
\end{center}
\section{Conclusion}
\label{sec:conclusion}
 We have introduced a large class of spike-and-slab priors for variable selection and linear regression involving grouped variables. To approximate the true posterior, we have established a variational Bayes framework that can handle both generic slab functions and hierarchical slab functions. The proposed method is then extended to the additive model. Our method is shown to be particularly attractive as it outperformed empirically all the published methods we considered and exhibits near optimal contraction rate theoretically. The convergence of the proposed algorithm is much faster than traditional Bayesian approaches, which rely heavily on MCMC sampling techniques under such settings.

 Several directions for further developments are worthwhile to consider. Our method is currently designed for group sparsity, it would be interesting to extend our method to bi-level variable selection, i.e., the variable selection at both the group levels and the individual variables within the relevant groups. Going beyond the linear model, it is also of interest to investigate the applicability of our method for dealing with generalized linear models and other nonlinear models in view of more complex data. Theoretically, our theorems  only apply pointwisely to pre-fixed hyper-parameters, how to make the theorems hold uniformly over a range of hyper-parameters and determine if the proposed hyper-parameter specification can correctly pin down the specified hyper-parameter into this range is of immediate interest.

\bibliographystyle{chicago}
\bibliography{vi.bib}

\newpage
\appendix
\section{Algorithm and Implementation Details}
Here we present the algorithm for  GVSSB with hierarchical slab functions.  The algorithm for  general slab functions can be easily modified.  We summarize the parameter updates for two specific slab functions, namely the multi-Laplacian and multivariate t distributions, in Table \ref{table:update-rule}.

\begin{algorithm}[h]
\caption{Component-wise Grouped Variational Bayes}
\begin{algorithmic}[1]
\renewcommand{\algorithmicrequire}{ \textbf{Initialize:}}
\REQUIRE $\Delta_H,\Delta_\sigma,\lambda,w,\tilde\sigma,\bm{\Sigma},\bm{\mu},\bm{\gamma},\bm{\kappa}$
\WHILE{$\max\{\Delta_H-\epsilon_H,\Delta_\sigma-\epsilon_\sigma\}>0$}
\STATE $\bm{a}:=\text{order}(\{\Vert\bm{\mu}_i\Vert\}_{i=1}^G)$
\STATE $\bm{\gamma}_{\text{old}}:=\bm{\gamma}$
\FOR{$k=1 \text{ to } G$}
\STATE $i:=a_k$

\STATE update $\bm{\Sigma}_i,\bm{\mu}_i,\bm{\gamma}_i$ \hfill $\backslash\backslash$ see equation \eqref{eq:mu-sigma-update-hierarchical} and equation \eqref{eq:GM-gamma}
\STATE update ${\kappa}_i$ \hfill $\backslash\backslash$ see equation \eqref{eq:GM-kappa}
\ENDFOR
\STATE update $\lambda,w$ \hfill $\backslash\backslash$ see equation \eqref{eq:GM-hyper}
\STATE$\Delta_H:= \max\{|H(\gamma_i)-H(\gamma_{i, old})|, i = 1,\ldots, G\}$
\STATE $\tilde\sigma_{old}=\tilde\sigma$
\IF{$\Delta_H\leq\epsilon_H$}
\STATE $\quad\tilde\sigma^2 := \mathbbm{E}[\|\bm{Y}-\sum_{i = 1}^{G}\bm{X}_i\bm{\theta}_i\|_2^2]/n$ \hfill $\backslash\backslash$ see equation \eqref{eq:Gaussian-sigma}
\ENDIF
\STATE $\Delta_\sigma:=|\tilde\sigma-\tilde\sigma_{old}|$

\ENDWHILE
\end{algorithmic}
\label{alg:update}
\end{algorithm}

\begin{table}[thp]
\centering
\begin{tabular}{lcc}
\hline                                                                            & Multi-Laplacian                                                                                                   & Multivariate T with degree of freedom $\nu$                                                                                                      \\ \hline
$\bm{\Sigma}_i$             &$\left(\bm{X}_i^\intercal\bm{X}_i/\tilde{\sigma}^2+\frac{\lambda}{\sqrt{\kappa_i}}\bm{I}_{p_i}\right)^{-1}$&$\left(\bm{X}_i^\intercal\bm{X}_i/\tilde{\sigma}^2+\frac{\nu+p_i}{\nu\lambda^2+\kappa_i}\bm{I}_{p_i}\right)^{-1}$ \\
$\bm{\mu}_i$                                        & $\bm{\Sigma}_i{\bm{X}_i^\intercal\bm{r}_i}/{\tilde{\sigma}^2}$                                              & $\bm{\Sigma}_i{\bm{X}_i^\intercal\bm{r}_i}/{\tilde{\sigma}^2}$                                                    \\
\multirow{3}{*}{$\gamma_i$} & $\text{logit}^{-1}(\log\frac{w}{1-w}+\frac{1}{2}(\log|\bm{\Sigma}_i|$                                     & $\text{logit}^{-1}(\log\frac{w}{1-w}+\frac{1}{2}(\log|\bm{\Sigma}_i|+\nu\log\frac{\nu\lambda^2}{2}$                                            \\
                            & $+\log\pi+p_i\log\frac{\lambda^2}{2}-\lambda\sqrt{\kappa_i}$                           & $-(\nu+p_i)\log\frac{\nu\lambda^2+\kappa_i}{2}+\kappa_i\frac{\nu+p_i}{\nu\lambda^2+\kappa_i}$                                 \\
                            &$-2\log\Gamma(\frac{p_i+1}{2})+\bm{\mu}_i^\intercal\bm{\Sigma}_i^{-1}\bm{\mu}_i))$
                            &                                       $+\bm{\mu}_i^\intercal\bm{\Sigma}_i^{-1}\bm{\mu}_i))+\log\Gamma(\frac{\nu+p_i}{2})-\log\Gamma(\frac{\nu}{2})$                                                                                                     \\
$\kappa_i$                  & $\bm{\mu}_i^\intercal\bm{\mu}_i+Tr(\bm{\Sigma}_i)$                                                        & $\bm{\mu}_i^\intercal\bm{\mu}_i+Tr(\bm{\Sigma}_i)$                                                              \\
$w$                            & ${(\sum_{i=1}^G\gamma_i)}/{G}$                                                                            & ${(\sum_{i=1}^G\gamma_i)}/{G}$                                                                                  \\
$\lambda^2$&    $\sum_{i=1}^G\gamma_i(p_i+1)/\sum_{i=1}^G\gamma_i(\frac{\sqrt{\kappa_i}}{\lambda}+\frac{1}{\lambda^2})$                                                                                                       &      $\sum_{i=1}^G\gamma_i/\sum_{i=1}^G\gamma_i\frac{\nu+p_i}{\nu\lambda^2+\kappa_i}$                                                                                                           \\ \hline
\end{tabular}
\caption{Update rules for multi-Laplacian distirbution and multivariate t distribution with degree of freedom $\nu$. }
\label{table:update-rule}
\end{table}

\section{Additional Numerical Results}\label{sec:additional_results}
\subsection{Comparison with Gibbs sampler}\label{subsec:simu-MCMC}
In this section, we compare our method with Bayesian Group Lasso with Spike-and-Slab prior (BGLSS), which is a Gibbs Sampler with Dirac-and-Laplace prior available in the R package \pkg{MBSGS}. Unlike our method, they use the Monte Carlo EM algorithm to automatically calibrate hyper-parameters.

We consider a scenario where  $(n, G, p_i, k) = (200, 200, 5, 10)$. We generate features independently from $N(0, \Sigma)$, where $\Sigma_{ij}=0.6$ for $i\neq j$ in the same group and $\Sigma_{ij}=0.2$ for $i$, $j$ not in the same group. The coefficients in relevant groups are sampled i.i.d. from $Unif[-0.5, 0.5]$. We vary the signal-to-noise ratio (SNR) from $0.5$ to $2.5$. For BGLSS, we set 5000 burnin steps and use the median value of the coefficients estimated from the following 5000 steps as the final estimation \citep{2015xiaofan}.

The results for the log mean squared error (MSE) and Matthews correlation coefficient (MCC) are summarized in Table \ref{table:MCMC}. We can see that GVSSB consistently performs on par with or surpasses BGLSS in terms of MSE, regardless of the choice of slab functions. Moreover, GVSSB achieves higher MCC values, indicating better variable selection performance. These results demonstrate that the variational approach is not inferior to the Gibbs sampler, while also offering the advantage of shorter running times.

\begin{table}[thp]
    \centering
\begin{tabular}{crrrrrrrrrr}
\toprule
SNR                  & \multicolumn{2}{c}{0.5} & \multicolumn{2}{c}{1}  & \multicolumn{2}{c}{1.5} & \multicolumn{2}{c}{2}      & \multicolumn{2}{c}{2.5}                               \\ \hline
\multicolumn{1}{r}{} & MSE  & MCC  & MSE     & MCC & MSE  & MCC     & MSE  & MCC  & MSE     & MCC\\ \hline
GVSSB-G              & -5.42   & 0.21 & -5.56    & 0.49  & -5.80 & 0.63    & -6.06   & 0.72 & -6.33    & 0.79   \\
GVSSB-C              & -5.42   & 0.18 & -5.56    & 0.43  & -5.77 & 0.58    & -6.03   & 0.69 & -6.28    & 0.76   \\
GVSSB-L              & -5.40   & 0.22 & -5.56    & 0.47  & -5.80 & 0.61    & -6.04   & 0.70 & -6.29    & 0.77\\
BGLSS                & -5.48   & 0.13  & -5.53  & 0.26  & -5.61   & 0.37 & -5.72 & 0.49  & -5.93                    & 0.61                 \\ 
\bottomrule
\end{tabular}
\caption{The simulation results of GVSSB and BGLSS. $n=200,\ G=200,\ p_i=5,\ k=10,\ \Sigma_{ii}=1,\  \Sigma_{ij}=0.6$ \text{ for } $i\neq j$ in the same group and $\Sigma_{ij}=0.2\text{ for } i,j$ not in the same group. Each row of the design matrix is generated independently from $N(0,\Sigma)$.}
\label{table:MCMC}
\end{table}

\subsection{Estimation of the noise level}

In Section \ref{sec:variational}, we highlighted one of the advantages of GVSSB, which is its ability to incorporate the unknown noise level into the mean-field family. This allows for simultaneous estimations of both the noise level $\sigma_\star^2$ and the regression coefficients.  We compare our methods with SSGL and BGLSS and evaluate the performance using $|\hat\sigma^2/\sigma_\star^2-1|$. The experimental setup here is the same as in Section \ref{subsec:simu-MCMC}, with the exception $k=5$. The signal-to-noise ratio (SNR) ranges from 0.5 to 1.5.

For the Gibbs sampler (BGLSS), we discovered that the choice of the initial value for $\sigma^2$ has a significant impact on the performance. If we simply set $\sigma^2 = 1$ as the starting point, which is done in the R package \pkg{MBSGS}, the MCMC algorithm fails to converge. To mitigate the mixing problem, we compute the initial value of $\sigma^2$ using the scaled Lasso. We set the iteration number to $10000$ and discarded the first $5000$ samples as burn-in. The estimation is obtained using the sample mean of $\sigma^2$ from the remaining $5000$ samples. 
    \begin{table}[h]
    \centering
    
\begin{tabular}{lccccc}
\toprule
         SNR    & 0.5 & 0.7 & 0.9 & 1.2 & 1.5\\
             \hline

GVSSB-G & 0.15    & 0.15     &0.14    &0.12      & 0.11 \\

GVSSB-C  & 0.19     & 0.18      & 0.16     & 0.13     & 0.12 \\
GVSSB-L  &0.19    & 0.18      & 0.16   & 0.13    & 0.12 \\
SSGL   &  0.44  & 0.53     & 0.60&0.60    & 0.56\\
BGLSS &0.33 &0.36  &0.32  & 0.26& 0.20 \\
\bottomrule
\end{tabular}
\caption{Simulation results of $|\hat{\sigma}^2/\sigma_\star^2-1|$. $n = 200,\  G = 200,\  p_i = 5,\ k = 5$. The covariance structure of features is the same as in Section \ref{subsec:simu-MCMC}. SNR ranges from 0.5 to 1.5.}
\label{table:sigma}
\end{table}

The results presented in Table \ref{table:sigma} clearly show that our  methods consistently outperform their Bayesian counterparts. Among the three variants of GVSSB, the multivariate Gaussian slab performs slightly better, while the multi-Laplacian slab and multivariate Cauchy slab exhibit similar performances. BGLSS performs better than SSGL but does not achieve the same level of accuracy as the GVSSB methods. SSGL tends to favor parsimony in variable selection, which can lead to the exclusion of relevant groups and consequently an overestimation of $\sigma_\star^2$.  

One possible reason that BGLSS performs worse than our method may be that their prior for $\bm{\theta}_j$ depends on the unknown noise level $\sigma^2$. Specifically, for nonzero groups, they set $\bm{\theta}_j\mid\sigma^2\propto \exp(-\lambda\|\bm{\theta}_j\|/\sigma)$. Thus, the Gibbs updates for $\bm{\theta}_j$'s and $\sigma^2$ are coupled together, potentially leading to a slower  convergence. This is validated by our observation that the mixing for $\sigma^2$ can be poor at times (auto-correlation exceeding 0.5 even after gaps) and also potentially explains why a bad initialization can cause the algorithm to fail to converge.

\subsection{The effect of updating hyper-parameters}\label{subsec:simu-hyperparameter}

In Section \ref{sec:mainresults}, we have proved the robustness of the VB approach to the choice of hyper-parameters, provided that the hyper-parameters fall within certain wide intervals. However, in finite sample cases, the accuracy can still be affected by the choice of hyper-parameters. Improper choices may lead to poor performances (\citet{2019rayvilinear}). In this section,  we show that the EM hyper-parameter updates can be helpful in mitigating the sensitivity to hyper-parameters. 

\begin{figure}[t]
    \centering
    \includegraphics[width =0.9 \columnwidth]{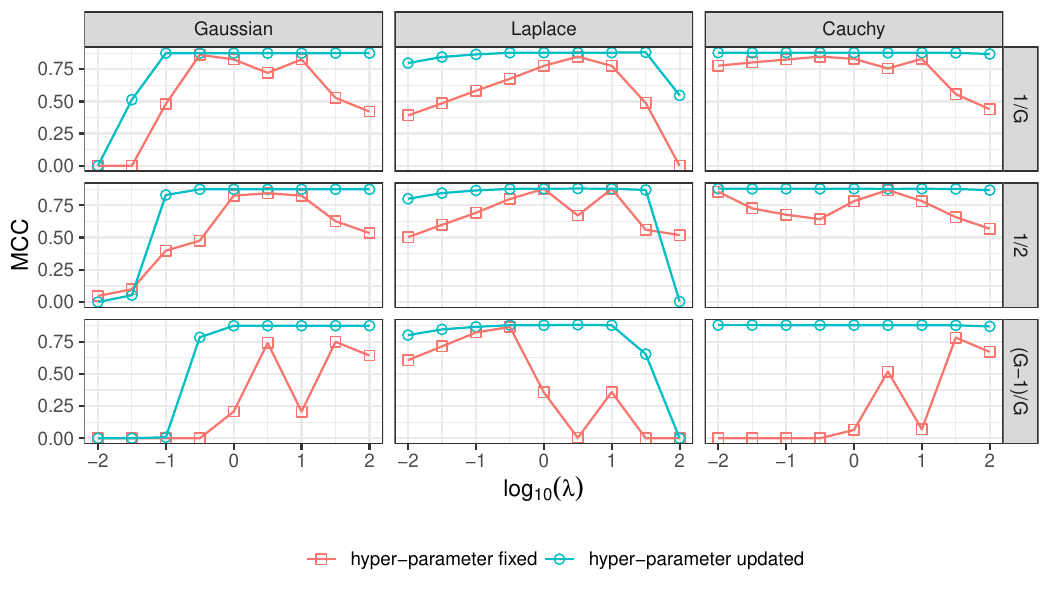}
    \caption{MCC results for GVSSB with and without EM updates. $n=600,\ G=600,\ p_i=5,\ k=5$. The initialization of $\lambda$ ranges from 0.01 to 100, with $w\in\{\frac{1}{G},\frac{1}{2},\frac{G-1}{G}\}$. }
    \label{fig:mcc-em}
\end{figure}

\begin{figure}[t]
    \centering
    \includegraphics[width=0.9\columnwidth]{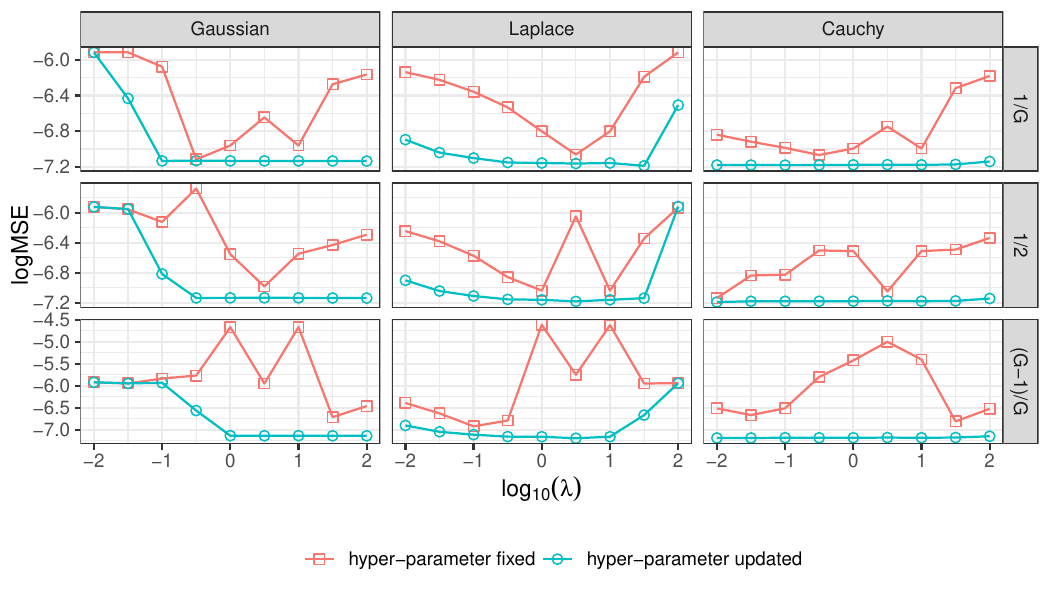}
    \caption{MSE results for GVSSB with and without EM updates. 
 $n=600,\  G=600,\  p_i=5,\  k=5$. The initialization of $\lambda$ ranges from 0.01 to 100, with $w\in\{\frac{1}{G},\frac{1}{2},\frac{G-1}{G}\}$.}
    \label{fig:mse-em}
\end{figure}

We set $n = 600,\ G = 600,\ p_i = 5,\ k = 5$. Each sample is generated independently from a normal distribution $N(0,\Sigma)$ with $\Sigma_{ij}=0.6$ if variable $i$ and $j$ belong to the same group and $\Sigma_{ij}=0.2$ for $i$ and $j$ belong to different groups. The true coefficients are generated from a uniform distribution $Unif[-1,1]$. The SNR is set to $0.5$.  To initialize the hyper-parameter $\lambda$, we consider values ranging from 0.01 to 100, with $w$ taking on three different values: $1/G$, $1/2$, and $(G-1)/G$. We then summarize the MCC and the log(MSE) in Figure \ref{fig:mcc-em} and \ref{fig:mse-em} respectively.  In both figures, we compare the performance of our algorithm with and without the EM update of hyper-parameters.

From Figure \ref{fig:mcc-em} and Figure \ref{fig:mse-em}, we found that the EM updates improve the performance of our algorithm in most cases and  also stabilize the algorithm across different initializations. When the hyper-parameters are not properly chosen, the EM updates can regain some power even if the original algorithm has no variable selection ability.
Hence, we recommend using the EM update unless one is highly confident about the value of the hyper-parameters in practical scenarios.

\section{Proofs}
\subsection{Proof of Theorem \ref{thm:posterior}}
Equation \eqref{eq:posterior-contraction} is equivalent to 
\begin{equation}
\label{eq:posterior-similar}
    P_\star\left(\Pi(\widehat\Omega^c\mid \bm{X}, \bm{Y})\geq e^{-C_1Ks_{\star}\log G}\right)\lesssim e^{-C_2Ks_{\star}\log G/2}.
\end{equation}

Similarly to \citet{2019Jiang}, the main recipe to prove \eqref{eq:posterior-similar} is the following lemma.

\begin{lemma}
\label{lemma:posterior-key}
Consider a parametric model $\{P_{\bm{\beta}}\}_{\bm{\beta}\in \mathcal{B}}$, and a data generation $\bm{D}$ from the true parameter $\bm{\beta}_\star\in\mathcal{B}$. Let
$\Pi(\bm{\beta})$ be a prior distribution over $\mathcal{B}$. If
\begin{enumerate}[(a)]
    \item $\Pi(\mathcal{B}_0)\leq \delta_0$;
    \item There exists a test function $\phi(\bm{D})$ such that \begin{equation}\nonumber
        \sup_{\bm{\beta}\in \mathcal{B}_{\textnormal{test}}}\mathbbm{E}_{\bm{\beta}}[1-\phi(\bm{D})]\leq \delta_1,\ \ \ \mathbbm{E}_{\bm{\beta}_\star}[\phi(\bm{D})]\leq \delta_1^\prime;
    \end{equation}
    \item \begin{equation}\nonumber
        \mathbbm{P}_{\bm{\beta}_\star}\left(\int\frac{P_{\bm{\beta}}(\bm{D})}{P_{\bm{\beta}_\star}(\bm{D})}d\Pi(\bm{\beta})\leq \delta_2\right)\leq\delta_2^\prime;
    \end{equation}
\end{enumerate}
Then for any $\delta_3$, 
\begin{equation}\nonumber
    \mathbbm{P}_{\bm{\beta}_\star}\left(\Pi(\mathcal{B}_0\cup \mathcal{B}_{\textnormal{test}}\mid \bm{D})\geq \frac{\delta_0+\delta_1}{\delta_2\delta_3}\right)\leq \delta_1^\prime+\delta_2^\prime+\delta_3.
\end{equation}
\end{lemma}

In our case, we let

\begin{equation}\nonumber
    \begin{aligned}
    &\bm{D} = (\bm{X}, \bm{Y}),\ \  \bm{\beta} = (\sigma, S, \bm{\theta}),\ \ P_{\bm{\theta}}(\bm{D}) = N(\bm{Y}\mid \bm{X}\bm{\theta}, \sigma^2\bm{I}),\\
    &\mathcal{B}_0 = \{(\sigma, S, \bm{\theta}): |S\backslash S_{\star}|> Ks_\star)\}\cup \{(\sigma, S, \bm{\theta}): S\notin \mathcal{F})\},\\
     &\mathcal{B}_{\textnormal{test}} = \mathcal{B}_1\cup\mathcal{B}_2,\\
     &\mathcal{B}_1 = \left\{(\sigma, S, \bm{\theta})\in\mathcal{B}_0^c: \frac{\sigma^2}{\sigma_\star^2}\notin\bigg[\frac{1-M_1\epsilon_n}{1+M_1\epsilon_n}, \frac{1+M_1\epsilon_n}{1-M_1\epsilon_n}\bigg]\right\},\\
     &\mathcal{B}_2 = \left\{(\sigma, S, \bm{\theta})\in\mathcal{B}_0^c\cap \mathcal{B}_1^c: \|\bm{\theta}-\bm{\theta}^\star\|> M_2\sigma_\star\sqrt{n}\epsilon_n/(\|\bm{X}\|_o\sqrt{\lambda})\right\}.
    \end{aligned}
\end{equation}

We take the test function as 
\begin{equation}\nonumber
    \phi = \max\{\phi_1,\phi_2\},
\end{equation}
with 
\begin{equation}\nonumber
\begin{aligned}
    &\phi_1 = \mathbbm{1}\left\{\max_{S\in\mathcal{F}:|S\backslash S_{\star}|\leq Ks_\star}\bigg|\frac{\bm{Y}^\intercal(\bm{I}-\bm{P}_{S\cup S_\star})\bm{Y}}{n\sigma_\star^2}-1\bigg|>M_1\epsilon_n\right\},\\
    & \phi_2 = \mathbbm{1}\left\{\max_{S\in\mathcal{F}:|S\backslash S_{\star}|\leq Ks_\star}\|\bm{X}_{S\cup S_{\star}}^\dagger\bm{Y}-\bm{\theta}_{S\cup S_{\star}}^\star\|>M_2\sigma_\star\sqrt{n}\epsilon_n/(2\|\bm{X}\|_o\sqrt{\lambda})\right\},
\end{aligned}
\end{equation}
where for any matrix $A$, $A^\dagger = (A^\intercal A)^{-1}A^\intercal$. 
Then we prove the following lemmas:
\begin{lemma}
\label{lemma:posterior-step-a}
Suppose Assumption \ref{assump:prior} hold. Then
\begin{equation}\nonumber
    \Pi(\mathcal{B}_0)\leq e^{-[A_2K-o(1)]s_\star\log G}.
\end{equation}
\end{lemma}
The proof of this lemma follows similarly to the proof of Lemma 5.2 in \citet{2019Jiang} and is thus omitted.

\begin{lemma}
\label{lemma:posterior-step-b1}
\begin{equation}\nonumber
    \mathbbm{E}_\star\phi_1\lesssim e^{-[M_1^2/8-K-o(1)]n\epsilon_n^2},\ \ \ \sup_{(\sigma, S, \bm{\theta})\in \mathcal{B}_1}\mathbbm{E}_{(\sigma, S, \bm{\theta})}(1-\phi_1)\lesssim e^{-[M_1^2/8-o(1)]n\epsilon_n^2}
\end{equation}
\end{lemma}
\textit{Proof of Lemma \ref{lemma:posterior-step-b1}.} We can rewrite $\phi_1$ as 
\begin{equation}\nonumber
\begin{aligned}
    \phi_1 &= \mathbbm{1}\left\{\max_{S\in\mathcal{F}:|S\backslash S_{\star}|\leq Ks_\star}\bigg|\frac{\bm{\epsilon}^\intercal(\bm{I}-\bm{P}_{S\cup S_\star})\bm{\epsilon}}{n}-1\bigg|>M_1\epsilon_n\right\}\\
    &\leq \mathbbm{1}\left\{\max_{|S\backslash S_{\star}|\leq Ks_\star}\bigg|\frac{\bm{\epsilon}^\intercal(\bm{I}-\bm{P}_{S\cup S_\star})\bm{\epsilon}}{n}-1\bigg|>M_1\epsilon_n\right\}.
\end{aligned}
\end{equation}
Note that the projection matrices $P_{S_1\cup S_{\star}}\leq P_{S_2\cup S_{\star}}$ for nested models $S_1\subseteq S_2$, we know $\bm{\epsilon}^\intercal(\bm{I}-\bm{P}_{S\cup S_\star})\bm{\epsilon}$ achieves its maximum and minimum value when $S\backslash S_\star = \emptyset$ and $|S\backslash S_\star| = Ks_\star$. For each $|S\backslash S_\star| = Ks_\star\ \textnormal{or}\ 0$, we have:
\begin{equation}\nonumber
\begin{aligned}
&\mathbbm{P}_\star\left(\bigg|\frac{\bm{\epsilon}^\intercal(\bm{I}-\bm{P}_{S\cup S_\star})\bm{\epsilon}}{n}-1\bigg|>M_1\epsilon_n\right)\\
&\leq \mathbbm{P}\left(\frac{\chi^2_{n-s_\star}}{n}>1+M_1\epsilon_n\right)+\mathbbm{P}\left(\frac{\chi^2_{n-(K+1)s_\star p_{\max}}}{n}<1-M_1\epsilon_n\right).
\end{aligned}
\end{equation}
It's easy to show $n\epsilon_n \gg (K+1)s_\star p_{\max}$ by Assumption \ref{assump:configuration}. Thus, by Lemma S.3.2 in \citet{2019Jiang} (concentration inequality for the  Chi-square random variable), we have:
\begin{equation}\nonumber
\begin{aligned}
&\mathbbm{P}\left(\frac{\chi^2_{n-s_\star}}{n}>1+M_1\epsilon_n\right)\preceq e^{-(1/8-o(1))n\epsilon_n^2},\\
&\mathbbm{P}\left(\frac{\chi^2_{n-(K+1)s_\star p_{\max}}}{n}<1-M_1\epsilon_n\right)\preceq e^{-(1/8-o(1))n\epsilon_n^2}.
\end{aligned}
\end{equation}
Then we know $$\mathbbm{E}_\star \phi_1\leq \left[1+\binom{G}{Ks_\star}\right]\times e^{-(1/8-o(1))n\epsilon_n^2}\leq e^{-[M_1^2/8-K-o(1)]n\epsilon_n^2}.$$

Then we show the second part of the lemma. Define 
\begin{equation}\nonumber
    \phi_{1, S} = \mathbbm{1}\left\{\bigg|\frac{\bm{Y}^\intercal(\bm{I}-\bm{P}_{S\cup S_\star})\bm{Y}}{n\sigma_\star^2}-1\bigg|\leq M_1\epsilon_n\right\}.
\end{equation}
Then we can rewrite $\phi_1 = \max_{S\in\mathcal{F}:|S\backslash S_\star|\leq Ks_\star} \phi_{1, S}$. Thus $\phi_1\geq \phi_{1, S}$ for any $\{S\in\mathcal{F}:|S\backslash S_\star|\leq Ks_\star\}$. So for any $S\in\mathcal{F},\ |S\backslash S_{\star}|\leq Ks_{\star}$, we have:
\begin{equation}\nonumber
    \sup_{(\sigma, S, \bm{\theta})\in \mathcal{B}_1}\mathbbm{E}_{(\sigma, S, \bm{\theta})}(1-\phi_1)\leq \sup_{(\sigma, S, \bm{\theta})\in \mathcal{B}_1}\mathbbm{E}_{(\sigma, S, \bm{\theta})}(1-\phi_{1, S}).
\end{equation}

We then provide a uniform bound for  each $\mathbbm{E}_{(\sigma, S, \bm{\theta})}(1-\phi_{1, S})$.
\begin{equation}\nonumber
    \mathbbm{E}_{(\sigma, S, \bm{\theta})}(1-\phi_{1, S}) = \mathbbm{P}_{(\sigma, S, \bm{\theta})}\left(\bigg|\frac{\bm{\epsilon}^\intercal(\bm{I}-\bm{P}_{S\cup S_\star})\bm{\epsilon}}{n\sigma_\star^2/\sigma^2}-1\bigg|\leq M_1\epsilon_n\right)
\end{equation}
The restriction $\sigma^2/\sigma_\star^2\notin[\frac{1-M_1\epsilon_n}{1+M_1\epsilon_n}, \frac{1+M_1\epsilon_n}{1-M_1\epsilon_n}]$ of $\mathcal{B}_1$ implies that 
\begin{equation}\nonumber
    \bigg|\frac{\bm{\epsilon}^\intercal(\bm{I}-\bm{P}_{S\cup S_\star})\bm{\epsilon}}{n\sigma_\star^2/\sigma^2}-1\bigg|\leq M_1\epsilon_n\Longrightarrow \bigg|\frac{\bm{\epsilon}^\intercal(\bm{I}-\bm{P}_{S\cup S_\star})\bm{\epsilon}}{n}-1\bigg|\geq M_1\epsilon_n.
\end{equation}
Then 
\begin{equation}\nonumber
\begin{aligned}
&\mathbbm{P}_{(\sigma, S, \bm{\theta})}(1-\phi_{1, S})\leq \mathbbm{P}\left( \bigg|\frac{\bm{\epsilon}^\intercal(\bm{I}-\bm{P}_{S\cup S_\star})\bm{\epsilon}}{n}-1\bigg|\geq M_1\epsilon_n\right)\\
&\leq\mathbbm{P}\left(\frac{\chi_{n-s_\star}^2}{n}>1+M_1\epsilon_n\right)+\mathbbm{P}\left(\frac{\chi_{n-(K+1)s_\star p_{\max}}^2}{n}<1-M_1\epsilon_n\right)
&\lesssim e^{-[1/8-o(1)]n\epsilon_n^2}.
\end{aligned}
\end{equation}
Note this bound holds uniformly for all $(\sigma, S, \bm{\theta})\in\Theta_1$, we have proved the second part of the lemma.

\begin{lemma}
\label{lemma:posterior-step-b2}
\begin{equation}\nonumber
    \mathbbm{E}_\star\phi_2\lesssim e^{-[M_2^2/8-K-o(1)]n\epsilon_n^2},\ \ \ \sup_{(\sigma, S, \bm{\theta})\in \mathcal{B}_2}\mathbbm{E}_{(\sigma, S, \bm{\theta})}(1-\phi_2)\lesssim e^{-[M_2^2/8-o(1)]n\epsilon_n^2}
\end{equation}
\end{lemma}
\textit{Proof of Lemma \ref{lemma:posterior-step-b2}.} We begin to prove the first part. For any $S\in\mathcal{F}$ such that $|S\backslash S_\star|\leq Ks_\star$, we have:
\begin{equation}\nonumber
\begin{aligned}
&\frac{\|\bm{X}_{S\cup S_{\star}}^\dagger\bm{Y}-\bm{\theta}_{S\cup S_{\star}}^\star\|^2}{\sigma_\star^2} =  \bm{\epsilon}^\intercal\bm{X}_{S\cup S_{\star}}(\bm{X}_{S\cup S_{\star}}^\intercal \bm{X}_{S\cup S_{\star}})^{-2}\bm{X}_{S\cup S_{\star}}\bm{\epsilon}\\
&\leq \frac{\bm{\epsilon}^\intercal\bm{X}_{S\cup S_{\star}}(\bm{X}_{S\cup S_{\star}}^\intercal \bm{X}_{S\cup S_{\star}})^{-1}\bm{X}_{S\cup S_{\star}}\bm{\epsilon}}{\lambda_{\min}(\bm{X}_{S\cup S_{\star}}^\intercal \bm{X}_{S\cup S_{\star}})}\leq \frac{\bm{\epsilon}^\intercal\bm{P}_{S\cup S_{\star}}\bm{\epsilon}}{\|X\|_o^2\lambda},
\end{aligned}
\end{equation}
where $\bm{P}_{S\cup S_{\star}} = \bm{X}_{S\cup S_{\star}}(\bm{X}_{S\cup S_{\star}}^\intercal \bm{X}_{S\cup S_{\star}})^{-1}\bm{X}_{S\cup S_{\star}}$.
This implies 
\begin{equation}\nonumber
    \phi_2\leq \mathbbm{1}\left\{\max_{S\in\mathcal{F}: |S\backslash S_\star|\leq Ks_\star}\bm{\epsilon}^\intercal \bm{P}_{S\cup S_\star}\bm{\epsilon}>M_2^2n\epsilon_n^2/4\right\}
\end{equation}
Similar to the proof of Lemma \ref{lemma:posterior-step-b1}, we have:
\begin{equation}\nonumber
\begin{aligned}
\mathbbm{E}_\star\phi_2&\leq \binom{G}{Ks_\star}\times\mathbbm{P}\left(\chi^2_{(K+1)s_\star p_{\max}}>M_2^2n\epsilon_n^2/4\right)\\
&\leq \binom{G}{Ks_\star}\times e^{-[M_2^2/8-o(1)]n\epsilon_n^2}\leq e^{-[M_2^2/8-K-o(1)]n\epsilon_n^2}.
\end{aligned}
\end{equation}
Where the second inequality comes from Lemma S.3.2., part (b) in \citet{2019Jiang}.

Then we prove the second part. Similar to the proof of the second part in Lemma \ref{lemma:posterior-step-b2}, define 
\begin{equation}\nonumber
    \phi_{2, S}  = \mathbbm{1}\left\{\|\bm{X}_{S\cup S_{\star}}^\dagger\bm{Y}-\bm{\theta}_{S\cup S_{\star}}^\star\|>M_2\sqrt{n}\sigma_\star\epsilon_n/(2\|\bm{X}\|_o\sqrt{\lambda})\right\},
\end{equation}
then $\phi_2 = \max_{S\in\mathcal{F}:|S\cup S_\star|\leq Ks_\star} \phi_{2, S}$. We then have:
\begin{equation}\nonumber
    \sup_{(\sigma, S, \bm{\theta})\in\mathcal{B}_2}\mathbbm{E}_{(\sigma, S, \bm{\theta})}(1-\phi_2)\leq\sup_{(\sigma, S, \bm{\theta})\in\mathcal{B}_2}\mathbbm{E}_{(\sigma, S, \bm{\theta})}(1-\phi_{2, S})
\end{equation}
We proceed to bound
\begin{equation}\nonumber
    \mathbbm{E}_{(\sigma, S, \bm{\theta})}(1-\phi_{2, S}) = \mathbbm{P}_{(\sigma, S, \bm{\theta})}\left(\|\bm{X}^\dagger_{S\cup S_{\star}}\bm{Y}-\bm{\theta}^\star_{S\cup S_\star}\|\leq \frac{M_2\sigma_\star\sqrt{n}\epsilon_n}{2\|X\|_o\sqrt{\lambda}}\right)
\end{equation}
In $\mathcal{B}_2$, we know:
\begin{equation}\nonumber
\begin{aligned}
\|\bm{X}^\dagger_{S\cup S_{\star}}\bm{Y}-\bm{\theta}^\star_{S\cup S_\star}\| & = \|\bm{X}^\dagger_{S\cup S_{\star}}(\bm{X}_{S\cup S_{\star}}\bm{\theta}_{S\cup S_{\star}}+\bm{X}_{S^c\cap S^c_{\star}}\bm{\theta}_{S^c\cup S^c_{\star}}+\sigma\bm{\epsilon})-\bm{\theta}_{S\cup S_{\star}}^\star\|\\
&= \|\bm{\theta}_{S\cup S_{\star}}- \bm{\theta}_{S\cup S_{\star}}^\star+\sigma \bm{X}^\dagger_{S\cup S_{\star}}\bm{\epsilon}\|\geq \|\bm{\theta}_{S\cup S_{\star}}- \bm{\theta}_{S\cup S_{\star}}^\star\|- \sigma\| \bm{X}^\dagger_{S\cup S_{\star}}\bm{\epsilon}\|
\end{aligned}
\end{equation}
Then we have:
\begin{equation}\nonumber
\begin{aligned}
\mathbbm{E}_{(\sigma, S, \bm{\theta})}(1-\phi_{2, S})&\leq \mathbbm{P}_{(\sigma, S, \bm{\theta})}\left(\|\bm{\theta}_{S\cup S_{\star}}- \bm{\theta}_{S\cup S_{\star}}^\star\|- \sigma\| \bm{X}^\dagger_{S\cup S_{\star}}\bm{\epsilon}\|\leq \frac{M_2\sigma_\star\sqrt{n}\epsilon_n}{2\|\bm{X}\|_o\sqrt{\lambda}}\right)\\
      & = \mathbbm{P}_{(\sigma, S, \bm{\theta})}\left(\sigma\| \bm{X}^\dagger_{S\cup S_{\star}}\bm{\epsilon}\|\geq \|\bm{\theta}_{S\cup S_{\star}}- \bm{\theta}_{S\cup S_{\star}}^\star\|- \frac{M_2\sigma_\star\sqrt{n}\epsilon_n}{2\|\bm{X}\|_o\sqrt{\lambda}}\right)\\
      &\leq \mathbbm{P}_{(\sigma, S, \bm{\theta})}\left( \sigma\| \bm{X}^\dagger_{S\cup S_{\star}}\bm{\epsilon}\|>\frac{M_2\sigma_\star\sqrt{n}\epsilon_n}{2\|\bm{X}\|_o\sqrt{\lambda}}  \right)\\
      &\leq \mathbbm{P}_{(\sigma, S, \bm{\theta})}\left(\|\bm{P}_{S\cup S_\star}\epsilon\|> \frac{M_2}{2}\sqrt{\frac{1-M_1\epsilon_n}{1+M_1\epsilon_n}}\sqrt{n}\epsilon_n\right)\\
      &\leq \mathbbm{P}\left(\chi_{(K+1)s_\star p_{\max}}^2> \bigg(\frac{M_2}{2}\sqrt{\frac{1-M_1\epsilon_n}{1+M_1\epsilon_n}}\bigg)^2n\epsilon_n^2\right)\\
      &\leq e^{-[M_2^2/8-o(1)]n\epsilon_n^2},
\end{aligned}
\end{equation}
where the second and the third inequality follow from the definition of $\mathcal{B}_2$ and the last inequality follows from the part (b) in Lemma S.3.2 in \citet{2019Jiang}. We have thus completed the proof of Lemma \ref{lemma:posterior-step-b2}.

\begin{lemma}
\label{lemma:posterior-step-c}
Suppose Assumption \ref{assump:prior} holds. For any constant $\eta>0$, 
\begin{equation}\nonumber
\mathbbm{P}_\star\left(\int\frac{N(\bm{Y}\mid \bm{X\theta}, \sigma^2\bm{I})}{N(\bm{Y}\mid \bm{X\theta}^\star, \sigma_\star^2\bm{I})}d\Pi(\sigma, S, \bm{\theta})\leq e^{-(A_1+A_3+1+\eta)s_\star\log G}\right)\lesssim e^{-C_\eta s_\star\log G},
\end{equation}
where the constant $C_\eta>0$ depends on $\eta$.
\end{lemma}

\textit{Proof of Lemma \ref{lemma:posterior-step-c}.} Similar to the proof of Lemma 5.5 in \citet{2019Jiang}, the proof consists of four steps.
\begin{enumerate}[(1)]
    \item Let
    \begin{equation}
    \label{eq:definition-Omega-star}
    \Omega_\star= \left\{(\sigma, S, \bm{\theta}):
    \begin{aligned}
    & \sigma^2/\sigma_\star^2\in[1, 1+\eta_1\epsilon_n^2],\\
    &S = S_\star,\\
    & \|\bm{\theta}_j-\bm{\theta}^\star_j\|\leq \eta_2\sigma_\star\epsilon_n/(s_\star\sqrt{p_{\max}}),\  j\in S_\star
    \end{aligned}\right\}.
\end{equation}
Then it is obvious that 
\begin{equation}\nonumber
    \int\frac{N(\bm{Y}\mid \bm{X\theta}, \sigma^2\bm{I})}{N(\bm{Y}\mid \bm{X\theta}^\star, \sigma_\star^2\bm{I})}d\Pi\geq \Pi(\Omega_\star)\inf_{(\sigma, S, \bm{\theta})\in\Omega_\star}\frac{N(\bm{Y}\mid \bm{X\theta}, \sigma^2\bm{I})}{N(\bm{Y}\mid \bm{X\theta}^\star, \sigma_\star^2\bm{I})};
\end{equation}
\item Prove if $\eta_2<1$ then
\begin{equation}\nonumber
    \Pi(\Omega_\star)\gtrsim e^{-[A_1+A_3+1+o(1)]s_\star\log G};
\end{equation}
\item Prove that if
\begin{equation}\nonumber
\|\bm{P}_{S_\star}\bm{\epsilon}\|^2\leq Cn\epsilon_n^2\ \textnormal{and}\ \|\bm{\epsilon}\|^2< 2Cn,
\end{equation}
then for any small $\eta_3>0$:
\begin{equation}\nonumber
    \inf_{(\sigma, S, \bm{\theta})\in\Omega_\star}\frac{N(\bm{Y}\mid \bm{X\theta}, \sigma^2\bm{I})}{N(\bm{Y}\mid \bm{X\theta}^\star, \sigma_\star^2\bm{I})}\gtrsim e^{-[C\eta_1+\eta_2^2/2+\sqrt{C}\eta_2+\eta_3+o(1)]s_\star\log G};
\end{equation}
\item Prove
\begin{equation}\nonumber
    \mathbbm{P}_\star\left(\|\bm{P}_{S_\star}\bm{\epsilon}\|^2\leq Cn\epsilon_n^2\ \textnormal{and}\ \|\bm{\epsilon}\|^2< 2Cn\right)\gtrsim 1-e^{-[C/2-o(1)]n\epsilon_n^2}.
\end{equation}
\end{enumerate}

Once we established the aforementioned $4$ steps, we then choose sufficiently small $\eta_1, \eta_2, \eta_3$ and suitable $C>0$ such that
\begin{equation}
\label{eq:eta-define}
    \eta>C\eta_1+\eta_2^2/2+\sqrt{C}\eta_2+\eta_3.
\end{equation}
Then we have:
\begin{equation}\nonumber
\begin{aligned}
&\ \ \ \ \mathbbm{P}_\star\left(\int\frac{N(\bm{Y}\mid \bm{X\theta}, \sigma^2\bm{I})}{N(\bm{Y}\mid \bm{X\theta}^\star, \sigma_\star^2\bm{I})}d\Pi(\sigma, S, \bm{\theta})\leq e^{-(A_1+A_3+1+\eta)s_\star\log G}\right)\\
&\leq \mathbbm{P}_{\star}\left(\Pi(\Omega_\star)\inf_{(\sigma, S, \bm{\theta})\in\Omega_\star}\frac{N(\bm{Y}\mid \bm{X\theta}, \sigma^2\bm{I})}{N(\bm{Y}\mid \bm{X\theta}^\star, \sigma_\star^2\bm{I})}\leq e^{-(A_1+A_3+1+\eta)s_\star\log G}\right)\\
&\leq \mathbbm{P}_{\star}\left(\inf_{(\sigma, S, \bm{\theta})\in\Omega_\star}\frac{N(\bm{Y}\mid \bm{X\theta}, \sigma^2\bm{I})}{N(\bm{Y}\mid \bm{X\theta}^\star, \sigma_\star^2\bm{I})}\leq e^{-(\eta-o(1))s_\star\log G}\right)\\
&\leq \mathbbm{P}_\star\left(\|\bm{P}_{S_\star}\bm{\epsilon}\|^2> Cn\epsilon^2\ \textnormal{or}\ \|\bm{\epsilon}\|^2> 2Cn\right)\lesssim e^{-[C/2-o(1)]n\epsilon_n^2}.
\end{aligned}
\end{equation}
Set $C_\eta = C/2$, we complete the proof. Thus it remains to prove Steps (2), (3) and (4).

\textit{Proof of (2):}
\begin{equation}\nonumber
    \Pi(\Omega_\star) = \int_{\sigma_\star^2}^{(1+\eta_1\epsilon_n^2)\sigma_\star^2}g(\sigma^2)d\sigma^2\times \pi(S_\star)\times\prod_{j\in S_\star}\int_{\|\bm{\theta}_j-\bm{\theta}^\star_j\|\leq \eta_2\sigma_\star\epsilon_n/(s_\star\sqrt{p_{\max}})}h(\bm{\theta}_j)d\bm{\theta}_j.
\end{equation}
By Assumption \ref{assump:prior}, the first term is bounded from below as:
\begin{equation}
\label{eq:ratio-2-1}
    \int_{\sigma_\star^2}^{(1+\eta_1\epsilon_n^2)\sigma_\star^2}g(\sigma^2)d\sigma^2\geq \eta_1\sigma_\star^2\epsilon_n^2g(\sigma_\star^2)/2\gtrsim e^{-cs_\star\log G}
\end{equation}
for any $c>0$.

The third term is bounded below by 
\begin{equation}
\label{eq:posterior-likelihood-ratio}
    \begin{aligned}
    &\quad \prod_{j\in S_\star}\int_{\|\bm{\theta}_j-\bm{\theta}^\star_j\|\leq \eta_2\sigma_\star\epsilon_n/(s_\star\sqrt{p_{\max}})}h(\bm{\theta}_j)d\bm{\theta}_j\\
    &\geq \prod_{j\in S_\star}\int_{\|\bm{\theta}_j-\bm{\theta}^\star_j\|\leq \eta_2\sigma_\star\epsilon_n/(s_\star\sqrt{p_{\max}})}1d\bm{\theta}_j\times \left(\inf_{\|\bm{z}\|\leq \max_{j\in S_\star}\|\bm{\theta}^\star_j\|+\eta_2\sigma_\star\epsilon_n/(s_\star\sqrt{p_{\max}})}h(\bm{z})\right)^{s_\star}\\
    &\geq \prod_{j\in S_\star}V_{p_j}(\eta_2\sigma_\star\epsilon_n/(s_\star\sqrt{p_{\max}}))\times G^{-A_3s_\star} \\
    &\gtrsim \prod_{j\in S_\star} p_j^{-p_j/2}(\eta_2\sigma_\star\epsilon_n/(s_\star\sqrt{p_{\max}}))^{p_j}\times G^{-A_3s_\star}\\
    & \gtrsim p_{\max}^{-s_\star p_{\max}/2} e^{s_\star p_{\max}\log\epsilon_n-s_\star p_{\max}\log(s_\star\sqrt{p_{\max}})}G^{-A_3s_\star}
    \end{aligned}
\end{equation}
where $V_{p_j}(\eta_2\sigma_\star\epsilon_n/(s_\star\sqrt{p_{\max}}))$ denotes the volume of a $p_j$ dimension ball with radius $\eta_2\sigma_\star\epsilon_n/(s_\star\sqrt{p_{\max}})$. The second inequality of \eqref{eq:posterior-likelihood-ratio} follows from Assumption \ref{assump:prior} and the last two inequalities follow from the fact that:
\begin{equation}\nonumber
    V_{p_j}(R)\gtrsim   p_j^{-p_j/2}R^{p_j}\geq p_{\max}^{-p_{\max}/2} R^{p_{\max}}
\end{equation}
for any $R<1$.

Then it's easy to show that 
\begin{equation}
\label{eq:ratio-2-3}
    \prod_{j\in S_\star}\int_{\|\bm{\theta}_j-\bm{\theta}^\star_j\|\leq \eta_2\sigma_\star\epsilon_n/(s_\star\sqrt{p_{\max}})}h(\bm{\theta}_j)d\bm{\theta}_j\gtrsim e^{-(A_3+1+o(1))s_\star\log G}
\end{equation}
Combine \eqref{eq:ratio-2-1}, \eqref{eq:ratio-2-3} and the fact that  $\pi(S_\star)\gtrsim e^{-A_1s_\star\log G}$, the proof of (2) is accomplished.

\textit{Proof of (3):}
\begin{equation}
\label{eq:step-c3-1}
\begin{aligned}
-2\log \Delta & = \|\sigma_\star\bm{\epsilon}+\bm{X}(\bm{\theta}^\star-\bm{\theta})\|^2/\sigma^2-\|\bm{\epsilon}\|^2+2n\log(\sigma^2/\sigma_\star^2)\\
& = (\sigma_\star^2/\sigma^2-1)\|\bm{\epsilon}\|^2+\|\bm{X}(\bm{\theta}^\star-\bm{\theta})/\sigma\|^2+2\sigma_\star\bm{\epsilon}^\intercal\bm{X}(\bm{\theta}^\star-\bm{\theta})/\sigma^2+2n\log(\sigma^2/\sigma_\star^2).
\end{aligned}
\end{equation}

By the definition of \eqref{eq:definition-Omega-star}, we have:
\begin{equation}
\label{eq:step-c3-2}
    \log(\sigma^2/\sigma_\star^2)\leq \eta_1\epsilon_n^2, \ \ \|\bm{\theta}^\star-\bm{\theta}\|\leq \eta_2\sigma_\star\epsilon_n/\sqrt{s_\star p_{\max}},\ \ \|\bm{X}(\bm{\theta}^\star-\bm{\theta})\|\leq \sqrt{ns_\star p_{\max}}\times\frac{\eta_2\sigma_\star\epsilon_n}{\sqrt{s_\star p_{\max}}}.
\end{equation}

Furthermore, we have:
\begin{equation}
\label{eq:step-c3-3}
    \bm{\epsilon}^\intercal\bm{X}(\bm{\theta}^\star-\bm{\theta}) = \bm{\epsilon}^\intercal\bm{X}_{S_\star}(\bm{\theta}_{S_\star}^\star-\bm{\theta}_{S_\star})\leq\|\bm{X}_{S_\star}^\intercal\bm{\epsilon}\|\|\bm{\theta}^\star-\bm{\theta}\|.
\end{equation}
Note 
\begin{equation}
\label{eq:step-c3-4}
    \|\bm{X}_{S_\star}^\intercal\bm{\epsilon}\| = \|\bm{X}_{S_\star}^\intercal\bm{P}_{S_\star}\bm{\epsilon}\|\leq \sqrt{ns_\star p_{\max}}\sqrt{nC}\epsilon_n.
\end{equation}

Combining \eqref{eq:step-c3-1}, \eqref{eq:step-c3-2}, \eqref{eq:step-c3-3} and \eqref{eq:step-c3-4}, we have proved the part (3).

\textit{Proof of Part (4).} It follows from the fact that:
\begin{equation}\nonumber
\begin{aligned}
&\mathbbm{P}(\|\bm{P}_{S_\star}\bm{\epsilon}\|^2> Cn\epsilon_n^2) \leq \mathbbm{P}(\chi^2_{s_\star p_{\max}}>Cn\epsilon_n^2),\\
&\mathbbm{P}(\|\bm{\epsilon}\|^2> 4n) \leq \mathbbm{P}(\chi^2_{n}>4n).
\end{aligned}
\end{equation}
Then by the concentration of the chi-squared  random variable (Lemma S.3.2 in \citet{2019Jiang}), we have concluded the proof.

We set
\begin{equation}\nonumber
    \begin{aligned}
    \delta_0 = e^{-[A_2K-o(1)]s_\star\log G},\ \ &\delta_1 = e^{-[\min\{M_1, M_2\}^2/8-o(1)]s_\star\log G},\ \ \delta^{\prime}_1 = e^{-[\min\{M_1, M_2\}^2/8-K-o(1)]s_\star\log G},\\
    & \ \ \delta_2 = e^{-[A_1+A_3+1+\eta]s_\star\log G},\ \ \delta^{\prime}_2 = e^{-C_\eta s_\star\log G}.
    \end{aligned}
\end{equation}

Let $\eta = C_2K/2$, then we choose sufficiently small $\eta_1, \eta_2$ and $\eta_3$ in \eqref{eq:eta-define}, which allows us to set $C = C_2K$. Further define $\delta_3 = e^{-C_2Ks_\star\log G/2}$, then we have:
\begin{equation}\nonumber
\begin{aligned}
&\frac{\delta_0+\delta_1}{\delta_2\delta_3}\asymp e^{-[(A_2-C_2)K-(A_1+A_3+1)]s_\star\log G}\lesssim e^{-C_1Ks_\star\log G}\\
&\delta_1^{\prime}+\delta_2^{\prime}+\delta_3\asymp e^{-C_2Ks_\star\log G/2}.
\end{aligned}
\end{equation}
Thus we have completed the proof.

\subsection{Proof of Theorem \ref{thm:VB-contraction}}
We first prove Theorem \ref{thm:VB-contraction} for general slab functions and the proof for hierarchical  slab functions  can be derived similarly and we comment on some of the differences at the end. We first upper bound the KL divergence between the VB posterior $\widetilde\Pi$ and the true posterior $\Pi(\cdot\mid\bm{X}, \bm{Y})$ and then use Theorem 5 in \citet{2019rayvilinear} to complete the proof.

The posterior distribution can be represented as 
\begin{equation}\nonumber
    \Pi(\cdot|\bm{X},\bm{Y}) = \sum_{S\in\mathcal{F}}q_S\Pi_S(\cdot|\bm{X}, \bm{Y})\otimes\delta_{S^c}
\end{equation}
where $\sum_{S\in\mathcal{F}}q_S = 1$ and $\Pi_S(\cdot|Y)$ is the posterior for $\bm{\theta}_S\in \mathbbm{R}^{p_S}$ in the restricted model $\bm{Y} = \bm{X}_S\bm{\theta}_S+\sigma\bm{\epsilon}$.

Define the event $\Gamma = \Gamma_0\cap\Gamma_1\cap\Gamma_2\cap\Gamma_3\cap\Gamma_4\cap\Gamma_5$, as
\begin{equation}\nonumber
\begin{aligned}
&\Gamma_0 = \{\Pi\left(S\in\mathcal{F}:|S\backslash S_\star|\leq Ks_\star\mid \bm{X},\bm{Y}\right)\geq 1-e^{-C_1Ks_\star\log G}\}\\
&\Gamma_1 = \bigg\{\Pi\left(\frac{\sigma^2}{\sigma_\star^2}\in\bigg[\frac{1-M_1^{\prime}\epsilon_n}{1+M_1^{\prime}\epsilon_n}, \frac{1+M_1^{\prime}\epsilon_n}{1-M_1^{\prime}\epsilon_n}\bigg]\mid \bm{X},\bm{Y}\right)\geq 1-e^{-C_1K^{\prime}s_\star\log G}\bigg\}\\
&\Gamma_2 = \bigg\{\Pi\left(\|\bm{\theta}-\bm{\theta^\star}\|\leq M_2^{\prime}\sqrt{n}\sigma_\star\epsilon_n/(\sqrt{\lambda(K^\prime)}\|\bm{X}\|_o)\mid \bm{X},\bm{Y}\right)\geq 1-e^{-C_1K^{\prime}s_\star\log G}\bigg\}\\
&\Gamma_3 = \left\{\bigg|\frac{\|(\bm{I}-\bm{P}_{\tilde S})\bm{\epsilon}\|^2}{n}-1\bigg|<C\epsilon_n\right\},\\
&\Gamma_4 = \left\{\bigg|\frac{\|\bm{P}_{ \tilde S}\bm{\epsilon}\|^2}{n}\bigg|<C\epsilon_n^2\right\}\\
&\Gamma_5 = \left\{\max_{i\in [G]}\|\bm{X}_i^\intercal \bm{\epsilon}\|\leq 4\|\bm{X}\|_o\sqrt{p_{\max}\log G}\right\},
\end{aligned}
\end{equation}
where $K'$ is chosen so that $C_1K^{\prime}\geq K+2$,  $\min\{M_1^{\prime}, M_2^{
\prime}\}>\sqrt{8(\max\{A_2, 1\}+C_2+1)K^{\prime}}$ and $\tilde S$ is defined later in \eqref{eq:condition-tilde-S}, which satisfies $|\tilde S\backslash S_\star|\leq Ks_\star$.

By Theorem \ref{thm:posterior}, we have $\mathbbm{P}_\star(\Gamma_0\cap \Gamma_1\cap \Gamma_2)\to 1$.
Note that
\begin{equation}\nonumber
\|(\bm{I}-\bm{P}_{\tilde S})\bm{\epsilon}\|^2  \overset{d}{=} \chi_{n-p_{\tilde S}}^2,
\end{equation}
then for any constant $C>0$,
\begin{equation}\nonumber
    \begin{aligned}
    \mathbbm{P}\left(\frac{\|(\bm{I}-\bm{P}_{\tilde S})\bm{\epsilon}\|^2}{n}>1+C\epsilon_n\right)\leq e^{-[C^2/8-o(1)]n\epsilon_n^2}.
    \end{aligned}
\end{equation}
Similarly,
\begin{equation}\nonumber
\|\bm{P}_{\tilde S}\bm{\epsilon}\|^2 \overset{d}{=} \chi_{p_{\tilde S}}^2
\end{equation}
Then for any constant $C>0$
\begin{equation}\nonumber
    \begin{aligned}
    \mathbbm{P}\left(\frac{\|\bm{P}_{\tilde S}\bm{\epsilon}\|^2}{n}>C\epsilon_n^2\right)\leq e^{-[C^2/8-o(1)]n\epsilon_n^2}.
    \end{aligned}
\end{equation}
For $\Gamma_5$, let $\bm{V}_i = \bm{X}_i^\intercal\bm{\epsilon}$, then $\bm{V}_i\sim N(\bm{0}, \bm{X}_i^\intercal \bm{X}_i)$. Define $\bm{U}_i = (\bm{X}_i^\intercal\bm{X}_i)^{-1/2}\bm{V}_i$, then $\|\bm{U}_i\|^2\sim\chi_{p_i}^2$. Using Lemma 8.1 in \citet{highdstatbook}, we have:
\begin{equation}\nonumber
    \mathbbm{P}\left(\max_{1\leq i\leq G}\|\bm{U}_i\|^2/p_i\geq 16\log G\right)\leq \exp[-\log G]\to 0.
\end{equation}
Then we have, with probability converging to 1, 
\begin{equation}\nonumber
    \|\bm{V}_i\|\leq \|\bm{X}\|_o\|\bm{U}_i\|\leq 4\|\bm{X}\|_o\sqrt{p_{\max}\log G}.
\end{equation}
Thus, we have $\mathbbm{P}(\Gamma) \to 1$.

Define $\vartheta = M_2^{\prime}\sqrt{n}\sigma_\star\epsilon_n/(\sqrt{\lambda(K^{\prime})}\|\bm{X}\|_o)$.
Under event $\Gamma$, since
\begin{equation}\nonumber
    \Pi(\bm{\theta}:\|\bm{\theta}^\star_{ S_\theta^c}\|>\vartheta\mid \bm{X}, \bm{Y})\leq \Pi(\bm{\theta}:\|\bm{\theta}-\bm{\theta}^\star\|>\vartheta\mid \bm{X}, \bm{Y}),
\end{equation}
it follows that:
\begin{equation}\nonumber
    \sum_{S\in\mathcal{F}:|S\backslash S_\star|\leq Ks_\star,
    \|\bm{\theta}^\star_{S^c}\|\leq \vartheta}q_s\geq  1-e^{-C_1K^\prime s_\star\log G}-e^{-C_1K s_\star\log G}\geq \frac{1}{2}.
\end{equation}
Furthermore, we have:
\begin{equation}\nonumber
    \left|\{S\in\mathcal{F}:|S
    \backslash S_\star|\leq Ks_\star\}\right|\leq \sum_{t = 0}^{(K+1)s_\star}\binom{G}{t}\leq eG^{(K+1)s_\star}.
\end{equation}
Thus there exists a set $\tilde S$ such that:
\begin{equation}
\label{eq:condition-tilde-S}
    |\tilde S\backslash S_\star|\leq Ks_\star,\ \ \ \|\bm{\theta}^\star_{\tilde S^c} \|\leq \vartheta,\ \ \ q_{\tilde S}\gtrsim (2e)^{-1}G^{-(K+1)s_\star}.
\end{equation}
We first upper bound the KL divergence between the VB posterior $\widehat Q$ arising form the family $\mathcal{P}^\prime_{MF}$ and the true posterior $\Pi(\cdot\mid \bm{X},\bm{Y})$ where $\mathcal{P}^\prime_{MF}$ is defined as \eqref{eq:MF-family-2}.
\begin{lemma}
\label{lemma:KL-bound-Q}
For the VB posterior $\widehat Q$ arising from the family $\mathcal{P}^\prime_{MF}$, it satisfies:
\begin{equation}\nonumber
\begin{aligned}
\text{KL}(\widehat Q\|\Pi(\cdot\mid\bm{X}, \bm{Y}))\mathbbm{1}_{\Gamma}&\preceq (K+1)s_\star\log G+p_{\max}|\tilde S|+L|\tilde S|\left(\frac{s_\star^{1/2}\vartheta}{\lambda(K)}+\frac{\sqrt{p_{\max}\log G}}{\lambda(K)\|\bm{X}\|_o}\right)\\
& +\epsilon_n\left(|\tilde S|^{1/2}\|X\|_o\sqrt{n}\epsilon_n\vartheta+\|\bm{X}\|_o^2|\tilde S|\vartheta^2\right).
\end{aligned}
\end{equation}
\end{lemma}
\textit{Proof of Lemma \ref{lemma:KL-bound-Q}.\text{}}
Note an $N_S(\bm{\mu}_S, \bm{\Sigma}_S)\otimes \delta_{S^c}$ distribution is only absolutely continuous with respect to the $q_S\Pi_S(\cdot\mid\bm{X},\bm{Y})\otimes \delta_{S^c}$, then we have:
\begin{equation}\nonumber
\begin{aligned}
  \inf_{\widehat Q\in \mathcal{Q}}\text{KL}(\widehat Q\|\Pi(\cdot|\bm{X}, \bm{Y}))& = \inf_{S, \bm{\mu}_S, \bm{\Sigma}_S, v} \mathbbm{E}_{\bm{\mu}_S, \bm{\Sigma}_S, v}\frac{\log dN(\bm{\mu}_S, \bm{\Sigma}_S)\otimes q(\sigma^2)\otimes\delta_{S^c}}{q_Sd\Pi_S(\cdot|\bm{X}, \bm{Y})\otimes\delta_{S^c}}\\
  &\leq \log \frac{1}{q_{\tilde S}}+ \inf_{\bm{\mu}_{\tilde S}, \bm{\Sigma}_{\tilde S}, v} \mathbbm{E}_{\bm{\mu}_{\tilde S}, \bm{\Sigma}_{\tilde S}, v}\frac{\log dN(\bm{\mu}_{\tilde S}, \bm{\Sigma}_{\tilde S})\otimes \pi(\sigma^2)}{d\Pi_{\tilde S}(\cdot|\bm{X}, \bm{Y})}
\end{aligned}
\end{equation}
We can calculate $\Pi_{\tilde S}(\cdot\mid\bm{X}, \bm{Y})$ as follows:
\begin{equation}\nonumber
    \frac{1}{D_\pi}e^{-\frac{1}{2\sigma^2}\|\bm{Y}-\bm{X}_{\tilde S}\bm{\theta}_{\tilde S}\|^2+\frac{v}{2\sigma_\star^2}-\frac{v}{2\sigma_\star^2}}(\sigma^2)^{-n/2}g(\sigma^2)\prod_{j\in\tilde S}h(\bm{\theta}_j)
\end{equation}
where $D_{\pi}$ is the normalizing constant. We add $\frac{v}{2\sigma_\star^2}$ and subtract $\frac{v}{2\sigma_\star^2}$ in the exponent to conveniently compare it with the VB posterior. Now let
\begin{equation}\nonumber
   \bm{\mu}_{\tilde S} = (\bm{X}_{\tilde S}^\intercal \bm{X}_{\tilde S})^{-1}\bm{X}_{\tilde S}^\intercal \bm{Y}\ \ \text{and}\ \  \bm{\Sigma}_{\tilde S} = \sigma_\star^2(\bm{X}_{\tilde S}^\intercal \bm{X}_{\tilde S})^{-1},
\end{equation}
the VB posterior $\widehat Q$ is
\begin{equation}
\label{eq:define-mu-Sigma}
    \frac{1}{D_{N}}e^{-\frac{1}{2\sigma_\star^2}\|\bm{Y}-\bm{X}_{\tilde S}\bm{\theta}_{\tilde S }\|^2+\frac{1}{2\sigma_\star^2}\|(\bm{I}-\bm{P}_{\tilde S})\bm{Y}\|^2-\frac{v}{2\sigma^2}}(\sigma^2)^{-n/2}g(\sigma^2)\prod_{j\in \tilde S}h(\bm{\theta}_j^\star),
\end{equation}
where $D_{N}$ is the normalizing constant. Note that the product of $h(\bm{\theta}_j^\star)$ is injected to match the form of $\Pi(\cdot\mid\bm{X}, \bm{Y})$.

Choose $v = \|(\bm{I}-\bm{P}_{\tilde S})\bm{Y}\|^2$, then we have:
\begin{equation}
\label{eq:KL-decomposition}
\begin{aligned}
    &\ \ \ \ \mathbbm{E}_{\bm{\mu}_{\tilde S}, \bm{\Sigma}_{\tilde S}, v}\frac{\log dN(\bm{\mu}_{\tilde S}, \bm{\Sigma}_{\tilde S})\otimes \pi(\sigma^2)}{d\Pi_{\tilde S}(\cdot|\bm{X},\bm{Y})}\\
    &= \log\frac{D_\pi}{D_N}+\mathbbm{E}_{v}\bigg[\frac{\sigma_\star^2}{\sigma^2}-1\bigg]\bigg(\frac{\mathbbm{E}_{\bm{\mu}_{\tilde S}, \bm{\Sigma}_{\tilde S}}\|\bm{Y}-\bm{X}_{\tilde S}\bm{\theta}_{\tilde S}\|^2-v}{2\sigma_\star^2}\bigg)\\
    &+\frac{\|(\bm{I}-\bm{P}_{\tilde S})\bm{Y}\|^2-v}{2\sigma_\star^2}+\sum_{j\in \tilde S}\mathbbm{E}_{\bm{\mu}_{\tilde S}, \bm{\Sigma}_{\tilde S}}[\log h(
    \bm{\theta}^\star_j)-\log  h(\bm{\theta}_j)]\\
    & \leq \log\frac{D_\pi}{D_N}+\mathbbm{E}_{v}\bigg[\frac{\sigma_\star^2}{\sigma^2}-1\bigg]\bigg(\frac{ \|(\bm{I}-\bm{P}_{\tilde S})\bm{Y}\|^2+tr(\bm{P}_{\tilde S})-v}{2\sigma_\star^2}\bigg)\\
    &+\frac{\|(\bm{I}-\bm{P}_{\tilde S})\bm{Y}\|^2-v}{2\sigma_\star^2}+L\mathbbm{E}_{\bm{\mu}_{\tilde S}, \bm{\Sigma}_{\tilde S}}\left[\sum_{j\in\tilde S}\|\bm{\theta}^\star_{j}-\bm{\theta}_{j}\|\right]\\
    &\leq\log \frac{D_\pi}{D_N}+\mathbbm{E}_{v}\bigg[\frac{\sigma_\star^2}{\sigma^2}-1\bigg]\frac{p_{\tilde S}}{2\sigma_\star^2}+L|\tilde S|^{1/2}\mathbbm{E}_{\bm{\mu}_{\tilde S}, \bm{\Sigma}_{\tilde S}}\left[\|\bm{\theta}^\star_{\tilde S}-\bm{\theta}_{\tilde S}\|\right].
    \end{aligned}
\end{equation}
We first upper bound $\log (D_\pi/D_N)$.
\begin{equation}\nonumber
\frac{D_\pi}{D_N} = \frac{\int\int \exp\left(-\frac{1}{2\sigma^2}\|\bm{Y}-\bm{X}_{\tilde S}\bm{\theta}_{\tilde S}\|^2+\frac{v}{2\sigma_\star^2}-\frac{v}{2\sigma_\star^2}\right)(\sigma^2)^{-(n/2)}g(\sigma^2)\prod_{j\in\tilde S}h(\bm{\theta}_j)d\sigma^2d\bm{\theta}_{\tilde S}}{\int\int \exp\left(-\frac{1}{2\sigma_\star^2}\|\bm{Y}-\bm{X}_{\tilde S}\bm{\theta}_{\tilde S}\|^2+\frac{1}{2\sigma_\star^2}\|(\bm{I}-\bm{P}_{\tilde S})\bm{Y}\|^2-\frac{v}{2\sigma^2}\right)(\sigma^2)^{-(n/2)}g(\sigma^2)\prod_{j\in\tilde S}h(\bm{\theta}^\star_j)d\sigma^2d\bm{\theta}_{\tilde S}}.
\end{equation}

Let
\begin{equation}\nonumber
B_{\tilde S} = \left\{(\bm{\theta}_{\tilde S}, \sigma^2)\in \mathbbm{R}^{p_{\tilde S}}\times \mathbbm{R}: \|\bm{\theta}_{\tilde S}-\bm{\theta}^{\star}_{\tilde S}\|\leq 2\vartheta, \frac{\sigma^2}{\sigma_\star^2}\in\bigg[\frac{1-M_1^{\prime}\epsilon_n}{1+M_1^{\prime}\epsilon_n}, \frac{1+M_1^{\prime}\epsilon_n}{1-M_1^{\prime}\epsilon_n}\bigg]\right\}.
\end{equation}
Let $\bm{\bar\theta}_{\tilde S}$ denote the extension of a vector $\bm{\theta}_{\tilde S}\in \mathbbm{R}^{p_{\tilde S}}$ to $\mathbbm{R}^p$ with $\bm{\bar\theta}_{\tilde S, j} = \bm{\bar\theta}_{\tilde S, j}$ for $j\in \tilde S$ and $\bm{\bar\theta}_{\tilde S, j} = 0$ for $j\notin\tilde S$. On $\Gamma$, we have:
\begin{equation}\nonumber
\begin{aligned}
    \Pi_{\tilde S}(B^c_{\tilde S}) & \leq \Pi_{\tilde S}(\bm{\theta}_{\tilde S}\in \mathbbm{R}^{p_{\tilde S}}: \|\bm{\theta}_{\tilde S}-\bm{\theta}^\star_{\tilde S}\|> 2\vartheta|\bm{X},\bm{Y})+\Pi_{\tilde S}\left(\sigma^2: \frac{\sigma^2}{\sigma_\star^2}\notin\bigg[\frac{1-M_1^{\prime}\epsilon_n}{1+M_1^{\prime}\epsilon_n}, \frac{1+M_1^{\prime}\epsilon_n}{1-M_1^{\prime}\epsilon_n}\bigg]\mid\bm{X},\bm{Y}\right)\\
    &= \frac{q_{\tilde S}}{q_{\tilde S}}     \bigg(\Pi_{\tilde S}(\bm{\theta}_{\tilde S}\in \mathbbm{R}^{p_{\tilde S}}:\|\bm{\bar\theta}_{\tilde S}-\bm{\theta}^\star\|>2\vartheta-\|\bm{\theta}^\star_{\tilde S^c}\|\mid \bm{X},\bm{Y})\\
    &\ \ \ \ \ \ \ \ \ \ \ \ \ \ \ \ \ \ \ \ \ \ \ \ \ \ \ +\Pi_{\tilde S}\bigg(\sigma^2: \frac{\sigma^2}{\sigma_\star^2}\notin\bigg[\frac{1-M_1^{\prime}\epsilon_n}{1+M_1^{\prime}\epsilon_n}, \frac{1+M_1^{\prime}\epsilon_n}{1-M_1^{\prime}\epsilon_n}\bigg]\mid\bm{X},\bm{Y}\bigg)\bigg)\\
    & \leq q_{\tilde S}^{-1}\left(\Pi(\bm{\theta}\in \mathbbm{R}^p:\|\bm{\theta}-\bm{\theta}^\star\|>\vartheta\mid\bm{X},\bm{Y})+\Pi\bigg(\sigma^2:\frac{\sigma^2}{\sigma_\star^2}\notin\bigg[\frac{1-M_1^{\prime}\epsilon_n}{1+M_1^{\prime}\epsilon_n}, \frac{1+M_1^{\prime}\epsilon_n}{1-M_1^{\prime}\epsilon_n}\bigg]\mid\bm{X},\bm{Y}\bigg)\right)\\
    &\leq 2e^{1+(K+1)s_\star\log G-(K+2)s_\star\log G}\leq 1/2
\end{aligned}
\end{equation}

Thus, we have:
\begin{equation}\nonumber
\begin{aligned}
    \Pi_{\tilde S}(B_{\tilde S}|Y)\mathbbm{1}_{\Gamma} &= \frac{\int\int_{B_{\tilde S}} \exp\left(-\frac{1}{2\sigma^2}\|\bm{Y}-\bm{X}_{\tilde S}\bm{\theta}_{\tilde S}\|^2\right)(\sigma^2)^{-n/2}g(\sigma^2)\prod_{j\in\tilde S}h(\bm{\theta}_j)d\sigma^2d\bm{\theta}_{\tilde S}}{\int\int \exp\left(-\frac{1}{2\sigma^2}\|\bm{Y}-\bm{X}_{\tilde S}\bm{\theta}_{\tilde S}\|^2\right)(\sigma^2)^{-n/2}g(\sigma^2)\prod_{j\in\tilde S}h(\bm{\theta}_j)d\sigma^2d\bm{\theta}_{\tilde S}}\mathbbm{1}_{\Gamma_1}\\
    &\geq \frac{1}{2}\mathbbm{1}_{\Gamma}.
\end{aligned}
\end{equation}
Therefore on $\Gamma$, we can upper bound $\log\frac{D_\pi}{D_N}$ as 
\begin{equation}
\begin{aligned}
\label{eq:KL-constant-decomposition}
    &\log\frac{2\int\int_{B_{\tilde S}} \exp\left(-\frac{1}{2\sigma^2}\|\bm{Y}-\bm{X}_{\tilde S}\bm{\theta}_{\tilde S}\|^2+\frac{v}{2\sigma_\star^2}-\frac{v}{2\sigma_\star^2}\right)(\sigma^2)^{-(n/2)}g(\sigma^2)\prod_{j\in\tilde S}h(\bm{\theta}_j)d\sigma^2d\bm{\theta}_{\tilde S}}{\int\int_{B_{\tilde S}} \exp\left(-\frac{1}{2\sigma_\star^2}\|\bm{Y}-\bm{X}_{\tilde S}\bm{\theta}_{\tilde S}\|^2+\frac{1}{2\sigma_\star^2}\|(\bm{I}-\bm{P}_{\tilde S})\bm{Y}\|^2-\frac{v}{2\sigma^2}\right)(\sigma^2)^{-(n/2)}g(\sigma^2)\prod_{j\in\tilde S}h(\bm{\theta}^\star_j)d\sigma^2d\bm{\theta}_{\tilde S}}\\
    &\leq \sup_{(\bm{\theta}_{\tilde S},\ \sigma^2)\in B_{\tilde S}} \bigg(1-\frac{\sigma_\star^2}{\sigma^2}\bigg)\frac{\|\bm{Y}-\bm{X}_{\tilde S}\bm{\theta}_{\tilde S}\|^2-v}{2\sigma_\star^2}-\frac{\|(\bm{I}-\bm{P}_{\tilde S})\bm{Y}\|^2-v}{2\sigma_\star^2}+\left(\sum_{j\in\tilde S}h(\bm{\theta}_j)-\sum_{j\in \tilde S}h(\bm{\theta}^\star_j)\right)+\log 2\\
&\leq \sup_{(\bm{\theta}_{\tilde S},\ \sigma^2)\in B_{\tilde S}} \bigg(1-\frac{\sigma_\star^2}{\sigma^2}\bigg)\frac{\|\bm{Y}-\bm{X}_{\tilde S}\bm{\theta}_{\tilde S}\|^2-v}{2\sigma_\star^2}+L|\tilde S|^{1/2}\|\bm{\theta}_{\tilde S}-\bm{\theta}^\star_{\tilde S}\|+\log 2
\end{aligned}
\end{equation}
Combing \eqref{eq:KL-decomposition} and \eqref{eq:KL-constant-decomposition}, it suffices to upper bound 
\begin{equation}
\label{eq:expectation-and-difference}
    \mathbbm{E}_{\bm{\mu}_{\tilde S}, \bm{\Sigma}_{\tilde S}}\left[\|\bm{\theta}^\star_{\tilde S}-\bm{\theta}_{\tilde S}\|\right]\ \ \ \textnormal{and}\ \ \ \sup_{\bm{\theta}_{\tilde S}\in B_{\tilde S}}\|\bm{Y}-\bm{X}_{\tilde S}\bm{\theta}_{\tilde S}\|^2-v.
\end{equation}
For the first term in \eqref{eq:expectation-and-difference}, we have:
\begin{equation}
    \mathbbm{E}_{\bm{\mu}_{\tilde S}, \bm{\Sigma}_{\tilde S}}\left[\|\bm{\theta}^\star_{\tilde S}-\bm{\theta}_{\tilde S}\|\right]\leq \sqrt{ \mathbbm{E}_{\bm{\mu}_{\tilde S}, \bm{\Sigma}_{\tilde S}}\left[\|\bm{\theta}^\star_{\tilde S}-\bm{\theta}_{\tilde S}\|^2\right]} = \sqrt{\|\bm{\mu}_{\tilde S}-\bm{\theta}^{\star}_{\tilde S}\|^2+Tr(\bm{\Sigma}_{\tilde S})}\leq \|\bm{\mu}_{\tilde S}-\bm{\theta}^{\star}_{\tilde S}\|+\sqrt{Tr(\bm{\Sigma}_{\tilde S})}.
    \label{eq:ineq-l2-bound}
\end{equation}
Note $\bm{\Sigma}_{\tilde S} = \sigma_\star^2(\bm{X}_{\tilde S}^\intercal \bm{X}_{\tilde S})^{-1}$, then we have:
\begin{equation}
\label{eq:Tr-bound1}
    Tr(\bm{\Sigma}_{\tilde S})\leq \sigma_\star^2|\tilde S|p_{\max}\Lambda_{\max}((\bm{X}_{\tilde S}^\intercal \bm{X}_{\tilde S})^{-1}) = \frac{\sigma_\star^2|\tilde S|p_{\max}}{\Lambda_{\min}(\bm{X}_{\tilde S}^\intercal \bm{X}_{\tilde S})}.
\end{equation}
Note we have:
\begin{equation}
\label{eq:Tr-bound2}
    \Lambda_{\min}(\bm{X}_{\tilde S}^\intercal \bm{X}_{\tilde S}) = \min_{\bm{v}\in\mathbbm{R}^{|\tilde S|}:\bm{v}\neq \bm{0}}\frac{\bm{v}^\intercal\bm{X}_{\tilde S}^\intercal \bm{X}_{\tilde S}\bm{v}}{\|\bm{v}\|^2}\geq \lambda(K)\|\bm{X}\|_{o}^2 .
\end{equation}
The last inequality comes from $|\tilde S\backslash S_\star\cup S_\star|\leq (K+1)s_\star$. 
Then combine \eqref{eq:Tr-bound1} and \eqref{eq:Tr-bound2}, we have:
\begin{equation}
\label{eq:trace-inequality}
    Tr(\bm{\Sigma}_{\tilde S})\leq \frac{\sigma_\star^2|\tilde S|p_{\max}}{\lambda(K)\|\bm{X}\|_o^2}.
\end{equation}

We then bound the first term in \eqref{eq:ineq-l2-bound}.
\begin{equation}\nonumber
    \|\bm{\mu}_{\tilde S}-\bm{\theta}^{\star}_{\tilde S}\|\leq \|(\bm{X}_{\tilde S}^\intercal\bm{X}_{\tilde S})^{-1}\bm{X}_{\tilde S}^\intercal \bm{X}_{\tilde S^c}\bm{\theta}^{\star}_{ \tilde S^c}\|+\sigma_\star\|(\bm{X}_{\tilde S}^\intercal\bm{X}_{\tilde S})^{-1}\bm{X}_{\tilde S}^\intercal \bm{\epsilon}\|: = I+II.
\end{equation}

For $I$, we first calculate $\|\bm{X}_{\tilde S}^\intercal\bm{X}_{\tilde S^c}\bm{\theta}^{\star}_{\tilde S^c}\|^2$
\begin{equation}\nonumber
\begin{aligned}
    \|\bm{X}_{\tilde S}^\intercal\bm{X}_{\tilde S^c}\bm{\theta}^{\star}_{\tilde S^c}\|^2 &= \sum_{g\in\tilde S}\|\bm{X}_g^\intercal \bm{X}_{\tilde S^c}\bm{\theta}^{\star}_{\tilde S^c}\|^2\\
    &=\sum_{g\in\tilde S}\|\sum_{j\in \tilde S^c\cap S_{\star}}\bm{X}_g^\intercal\bm{X}_j\bm{\theta}^\star_{j}\|^2\\
    &\leq \sum_{g\in \tilde S}|\tilde S^c\cap S_{\star}| \sum_{j\in \tilde S^c\cap S_{\star}}\|\bm{X}_g^\intercal\bm{X}_j\bm{\theta}^{\star}_{j}\|^2\\
    & \leq \sum_{g\in \tilde S}|\tilde S^c\cap S_{\star}| \sum_{j\in \tilde S^c\cap S_{\star}}\Lambda_{\max}(\bm{X}_g\bm{X}_g^\intercal)\Lambda_{\max}(\bm{X}_j^\intercal\bm{X}_j)\|\bm{\theta}^{\star}_{j}\|^2\\
    &\leq \sum_{g\in \tilde S}|\tilde S^c\cap S_{\star}| \sum_{j\in \tilde S^c\cap S_{\star}}\|\bm{X}\|_o^4\|\bm{\theta}^{\star}_{j}\|^2\\
    & \leq \|\bm{X}\|_o^4|\tilde S|s_\star \|\bm{\theta}^\star_{\tilde S^c}\|^2
\end{aligned}
\end{equation}
where the first inequality follows from Cauchy inequality and the second to last inequality follows from the fact that $\Lambda_{\max}(\bm{X}_g\bm{X}_g^\intercal) = \Lambda_{\max}(\bm{X}_g^\intercal\bm{X}_g)\leq \|\bm{X}\|_o^2$.
Then we can upper bound I by:
\begin{equation}
\label{eq:bound-I}
    I\leq \Lambda_{\max}((\bm{X}_{\tilde S}^\intercal\bm{X}_{\tilde S})^{-1})\|\bm{X}_{\tilde S}^\intercal\bm{X}_{\tilde S^c}\bm{\theta}^{\star}_{\tilde S^c}\|\leq \frac{\|\bm{X}\|_o^2|\tilde S|^{1/2}s_\star^{1/2} \|\bm{\theta}^{\star}_{ S^c}\|}{\lambda(K)\|\bm{X}\|_o^2} \leq \frac{ |\tilde S|^{1/2}s_\star^{1/2}\|\bm{\theta}^{\star}_{S^c}\|}{\lambda(K)}.
\end{equation}
Then it remains to upper bound II. On $\Gamma$, we have:
\begin{equation}
\label{eq:bound-II}
    II\leq \frac{\sigma_\star\|\bm{X}_{\tilde S}^\intercal \bm{\epsilon}\|}{\lambda(K)\|\bm{X}\|_o^2} = \frac{\sigma_\star(\sum_{g\in\tilde S}\|\bm{X}_g^\intercal\bm{\epsilon}\|^2)^{1/2}}{\lambda(K)\|\bm{X}\|_o^2}\leq \frac{4\sigma_\star|\tilde S|^{1/2}\sqrt{p_{\max}\log G}}{\lambda(K)\|\bm{X}\|_o}.
\end{equation}
Combining \eqref{eq:bound-I} and \eqref{eq:bound-II}, we get the upper bound for $\|\bm{\mu}_{\tilde S}-\bm{\theta}_{\tilde S}^\star\|$, which, together with \eqref{eq:trace-inequality}, gives the upper bound for the first term in \eqref{eq:expectation-and-difference}.

For the second term in \eqref{eq:expectation-and-difference}, we have:
\begin{equation}
\label{eq:estimation-error-predecompose}
\begin{aligned}
    \|\bm{Y}-\bm{X}_{\tilde S}\bm{\theta}_{\tilde S}\|^2 -v&= \|\bm{Y}-\bm{X}\bm{\bar\theta_{\tilde S}}\|^2 -v= \|\bm{Y}-\bm{X}\bm{\theta}^\star+\bm{X}(\bm{\theta}^\star-\bm{\bar\theta}_{\tilde S})\|^2-v\\
    &= \sigma_\star^2\|\bm{\epsilon}\|^2+2\sigma_\star\bm{\epsilon}^\intercal\bm{X}(\bm{\theta}^\star-\bm{\bar{\theta}}_{\tilde S})+\|\bm{X}(\bm{\theta}^\star-\bm{\bar\theta}_{\tilde S})\|^2-\|(\bm{I}-\bm{P}_{\tilde S})\bm{Y}\|^2.
\end{aligned}
\end{equation}
For $\|(\bm{I}-\bm{P}_{\tilde S})\bm{Y}\|^2$, we have:
\begin{equation}\nonumber
    \|(\bm{I}-\bm{P}_{\tilde S})\bm{Y}\|^2 = \|(\bm{I}-\bm{P}_{\tilde S})(\bm{X}_{\tilde S}\bm{\theta}^\star_{\tilde S}+\bm{X}_{\tilde S^c}\bm{\theta}^\star_{\tilde S^c}+\sigma_\star\bm{\epsilon})\|^2 = \|(\bm{I}-\bm{P}_{\tilde S})(\bm{X}_{\tilde S^c}\bm{\theta}^\star_{\tilde S^c}+\sigma_\star\bm{\epsilon})\|^2.
\end{equation}
Thus we can decompose \eqref{eq:estimation-error-predecompose} into the summation of three terms as follows:
\begin{equation}\nonumber
\begin{aligned}
    &I = \sigma_\star^2(\|\bm{\epsilon}\|^2-\|(\bm{I}-\bm{P}_{\tilde S})\bm{\epsilon}\|^2),\\
    & II = 2\sigma_\star\bm{\epsilon}^\intercal\bm{X}(\bm{\theta}^\star- \bm{\bar\theta}_{\tilde S}
    )-2\sigma_\star\bm{\epsilon}^\intercal(\bm{I}-\bm{P}_{\tilde S})\bm{X}_{\tilde S^c}\bm{\theta}_{\tilde S^c}^\star,\\
    & III = \|\bm{X}(\bm{\theta}^\star- \bm{\bar\theta}_{\tilde S}
    )\|^2 - \|(\bm{I}-\bm{P}_{\tilde S})\bm{X}_{\tilde S^c}\bm{\theta}^\star_{\tilde S^c}\|^2.
\end{aligned}
\end{equation}
For I, we have under $\Gamma$:
\begin{equation}\nonumber
  I = \sigma_\star^2\|\bm{P}_{\tilde S}\bm{\epsilon}\|^2\preceq n\epsilon_n^2.
\end{equation}

For the second term, we have:
\begin{equation}\nonumber
\begin{aligned}
\bm{\epsilon}^\intercal\bm{X}(\bm{\theta}^\star-\bm{\bar{\theta}}_{\tilde S})-\bm{\epsilon}^\intercal(\bm{I}-\bm{P}_{\tilde S})\bm{X}_{\tilde S^c}\bm{\theta}_{\tilde S^c}^\star& = \bm{\epsilon}^\intercal\bm{X}_{\tilde S}(\bm{\theta}^\star_{\tilde S}-\bm{\theta}_{\tilde S})+\bm{\epsilon}^\intercal\bm{X}_{\tilde S^c\cap S_\star}\bm{\theta}^\star_{\tilde S^c}- \bm{\epsilon}^\intercal(\bm{I}-\bm{P}_{\tilde S})\bm{X}_{\tilde S^c\cap S_{\star}}\bm{\theta}_{\tilde S^c}^\star\\
& = \bm{\epsilon}^\intercal\bm{X}_{\tilde S}(\bm{\theta}^\star_{\tilde S}-\bm{\theta}_{\tilde S})+\bm{\epsilon}^\intercal\bm{P}_{\tilde S}\bm{X}_{\tilde S^c\cap S_\star}\bm{\theta}^\star_{\tilde S^c}\\
&\leq \|\bm{\epsilon}^\intercal\bm{X}_{\tilde S}\|\|\bm{\theta}^\star_{\tilde S}-\bm{\theta}_{\tilde S}\|+\|\bm{\epsilon}^\intercal\bm{P}_{\tilde S}\bm{X}_{\tilde S^c\cap S_\star}\|\|\bm{\theta}^\star_{\tilde S^c}\|\\
&= \|\bm{X}_{\tilde S}^\intercal \bm{P}_{\tilde S}\bm{\epsilon}\|\|\bm{\theta}^\star_{\tilde S}-\bm{\theta}_{\tilde S}\|+\|\bm{X}^\intercal_{\tilde S^c\cap S_\star}\bm{P}_{\tilde S}\bm{\epsilon}\|\|\bm{\theta}^\star_{\tilde S^c}\|\\
&\preceq (|\tilde S|^{1/2}+s_\star^{1/2})\|\bm{X}\|_o \sqrt{n}\epsilon_n\vartheta.
\end{aligned}
\end{equation}

For the last term, under $\Gamma$, we have:
\begin{equation}\nonumber
\begin{aligned}
\|\bm{X}(\bm{\theta}^{\star}-\bm{\bar\theta}_{\tilde S})\|^{2} - \|(\bm{I}-\bm{P}_{\tilde S})\bm{X}_{\tilde S^c}\bm{\theta}^\star_{\tilde S^c}\|^2& = \|\bm{X}_{\tilde S}(\bm{\theta}^{\star}_{\tilde S}-\bm{\theta}_{\tilde S})+\bm{X}_{\tilde S^{c}}\bm{\theta}^{\star}_{\tilde S^c}\|^{2} - \|(\bm{I}-\bm{P}_{\tilde S})\bm{X}_{\tilde S^c}\bm{\theta}^\star_{\tilde S^c}\|^2\\
&\leq 2\|\bm{X}_{\tilde S}(\bm{\theta}^{\star}_{\tilde S}-\bm{\theta}_{\tilde S})\|^{2}+2\|\bm{X}_{\tilde S^{c}}\bm{\theta}^{\star}_{\tilde S^c}\|^{2}-\|(\bm{I}-\bm{P}_{\tilde S})\bm{X}_{\tilde S^c}\bm{\theta}^\star_{\tilde S^c}\|^2\\
&= 2\|\bm{X}_{\tilde S}(\bm{\theta}^{\star}_{\tilde S}-\bm{\theta}_{\tilde S})\|^{2}+\|\bm{X}_{\tilde S^{c}}\bm{\theta}^{\star}_{\tilde S^c}\|^{2}+\|\bm{P}_{\tilde S}\bm{X}_{\tilde S^c}\bm{\theta}^\star_{\tilde S^c}\|^2\\
&\leq 2|\tilde S|\|\bm{X}\|^2_o\|\bm{\theta}^{\star}_{\tilde S}-\bm{\theta}_{\tilde S}\|^{2}+2s_\star\|\bm{X}\|^2_o\|\bm{\theta}^{\star}_{\tilde S^c}\|^{2}\\
&\leq 4\|\bm{X}\|_o^2(|\tilde S|+s_\star)\vartheta^2.
\end{aligned}
\end{equation}
Combining these three pieces together, we obtain an upper bound for the second term in \eqref{eq:expectation-and-difference}. Then we can upper bound $\text{KL}(\widehat Q\|\Pi(\cdot\mid\bm{X}, \bm{Y}))$ as
\begin{equation}\nonumber
\begin{aligned}
\text{KL}(\widehat Q\|\Pi(\cdot\mid\bm{X}, \bm{Y}))&\preceq (K+1)s_\star\log G+p_{\tilde S}+L|\tilde S|\left(\frac{\sqrt{p_{\max}}}{\sqrt{\lambda(K)}\|\bm{X}\|_o}+\frac{s_\star^{1/2}\vartheta}{\lambda(K)}+\frac{\sqrt{p_{\max}\log G}}{\lambda(K)\|\bm{X}\|_o}\right)\\
& +\epsilon_n\left( n\epsilon_n^2+(|\tilde S|^{1/2}+s_\star^{1/2})\|X\|_o\sqrt{n}\epsilon_n\vartheta+\|\bm{X}\|_o^2(|\tilde S|+s_\star)\vartheta^2\right)+L|\tilde S|^{1/2}\vartheta.
\end{aligned}
\end{equation}

Note $\lambda(K)<\sqrt{\lambda(K)}$, $p_{\tilde S}\leq p_{\max}|\tilde S|$ and $|\tilde S|\asymp s_\star$, we accomplished the proof.
\ 

We define a sub-family of $\mathcal{P}_{MF}^\prime$ as 
\begin{equation}\nonumber
    \mathcal{Q}_{MF} = \{P_S = q(\sigma^2)\prod_{j\in S}q(\bm{\theta}_j)\otimes \delta_{S^c}\},
\end{equation}
where $q(\bm{\theta}_j) = N(\bm{\mu}_j, \bm{D}_j)$. 
Note that any distribution in $\mathcal{Q}_{MF}$ can be obtained by restricting $\bm{\Sigma}_S$ in $\mathcal{P}_{MF}^\prime$ to be block-wise diagonal, i.e., for any $i,j\in S$ and not in the same group, their correlation equals to $0$. We then upper bound the KL divergence between the VB posterior arising from the family $\mathcal{Q}_{MF}$ and the true posterior. Let $\bm{D}_S$ be the block-wise diagonal matrix where the diagonal terms are $\bm{D}_j,$ $j\in S$.

\begin{lemma}
\label{lemma:KL-bound-Q-MF}
For the VB posterior $\widetilde Q$ arising from the family $\mathcal{Q}_{MF}$, it satisfies:
\begin{equation}\nonumber
\begin{aligned}
    \text{KL}(\widetilde Q\|\Pi(\cdot\mid\bm{X}, \bm{Y}))\mathbbm{1}_{\Gamma}&\preceq (K+1)s_\star\log G+p_{\tilde S}+\epsilon_n\left( |\tilde S|^{1/2}\|X\|_o\sqrt{n}\epsilon_n\vartheta+\|\bm{X}\|_o^2|\tilde S|\vartheta^2\right)\\
    &+L |\tilde S|\left( \frac{s_\star^{1/2}\epsilon}{\lambda(K)}+\frac{4\sqrt{p_{\max}\log G}}{\|\bm{X}\|_o\lambda(K)} \right)+|\tilde S|p_{\max}\log(1/\lambda(K)).
\end{aligned}
\end{equation}
\end{lemma}

\textit{Proof of Lemma \ref{lemma:KL-bound-Q-MF}.\text{}}
Similar as in the proof of Lemma \ref{lemma:KL-bound-Q}, it suffices to upper bound:
\begin{equation}
\label{eq:KL-Q-MF}
    \mathbbm{E}_{\bm{\mu}_{\tilde S}, \bm{D}_{\tilde S}, v} \log\frac{dN(\bm{\mu}_{\tilde S}, \bm{D}_{\tilde S})\otimes q(\sigma^2)}{dN(\bm{\mu}_{\tilde S}, \bm{\Sigma}_{\tilde S})\otimes q(\sigma^2)}+ \mathbbm{E}_{\bm{\mu}_{\tilde S}, \bm{D}_{\tilde S}, v}\log\frac{dN(\bm{\mu}_{\tilde S}, \bm{\Sigma}_{\tilde S})\otimes q(\sigma^2)}{d\Pi_{\tilde S}(\cdot|Y)}
\end{equation}
where $\tilde S$ satisfies \eqref{eq:condition-tilde-S}.

We first deal with the first term in \eqref{eq:KL-Q-MF}, which is just the KL divergence between two multivariate Gaussian random variables and can thus be calculated as:
\begin{equation}\nonumber
   \frac{1}{2}\left(\log(|\bm{\Sigma}_{\tilde S}\|\bm{D}^{-1}_{\tilde S}|)-p_{\tilde S}+Tr(\bm{\Sigma}_{\tilde S}^{-1}\bm{D}_{\tilde S})\right).
\end{equation}
Now define the $i$-{th} diagonal block of $D_{\tilde S}$ as $\sigma_\star^2(\bm{X}_i^\intercal\bm{X}_i)^{-1}$ for any $i\in\tilde S$, then we have: $Tr(\bm{\Sigma}_{\tilde S}^{-1}\bm{D}_{\tilde S}) = p_{\tilde S}$. Thus it remains to upper bound the first term: $\log (|\bm{\Sigma}_{\tilde S}\|\bm{D}_{\tilde S}^{-1}|)$.
\begin{equation}
\label{eq:det-bound}
\begin{aligned}
    &|\bm{D}_{\tilde S}^{-1}| = \prod_{j\in \tilde S}|\bm{X}_{j}^\intercal\bm{X}_{j}|\leq \prod_{j\in \tilde S} \Lambda_{\max}(\bm{X}_{j}^\intercal\bm{X}_{j})^{p_j}\leq\|\bm{X}\|_o^{2p_{\tilde S}}\\
    &|\bm{\Sigma}_{\tilde S}|\leq \Lambda_{\max}(\bm{\Sigma}_{\tilde S})^{p_{\tilde S}}\leq (\|\bm{X}\|_o\sqrt{\lambda(K)})^{-2p_{\tilde S}}.
\end{aligned}
\end{equation}
Note $p_{\tilde S}\leq |\tilde S|p_{\max}$, then the first term in \eqref{eq:KL-Q-MF} can be upper bounded by: $|\tilde S|p_{\max}\log (1/\lambda(K))/2$.

We can upper bound the second term in \eqref{eq:KL-Q-MF} similarly as in Lemma \ref{lemma:KL-bound-Q}. The only difference is now we need to bound $Tr(\bm{D}_{\tilde S})$ instead of $Tr(\bm{\Sigma}_{\tilde S})$ and the rest of the proof is the same. 
\begin{equation}
\label{eq:Tr-bound-D}
    Tr(\bm{D}_{\tilde S}) = \sum_{j\in \tilde S}Tr((\bm{X}_{j}^\intercal \bm{X}_{j})^{-1})\leq \sum_{j\in \tilde S}\frac{p_j}{\Lambda_{\min}(\bm{X}_j^\intercal\bm{X}_j)}\leq \sum_{j\in\tilde S}\frac{p_j}{\lambda(1)\|\bm{X}\|_o^2}  = \frac{p_{\tilde S}}{\lambda(1)\|\bm{X}\|_o^2}
\end{equation}

Combining \eqref{eq:KL-Q-MF}, \eqref{eq:det-bound} and \eqref{eq:Tr-bound-D}, we have:
\begin{equation}\nonumber
\begin{aligned}
    \text{KL}(\widetilde Q\|\Pi(\cdot\mid\bm{X}, \bm{Y}))&\preceq |\tilde S|\log G+p_{\tilde S}+L |\tilde S|^{1/2}\vartheta+\epsilon_n\left( n\epsilon_n^2+(|\tilde S|^{1/2}+s_\star^{1/2})\|\bm{X}\|_o\sqrt{n}\epsilon_n\vartheta+\|\bm{X}\|_o^2(|\tilde S|+s_\star)\vartheta^2\right)\\
    &+L |\tilde S|\left( \frac{s_\star^{1/2}\epsilon}{\lambda(K)}+\frac{4\sqrt{p_{\max}\log G}}{\|\bm{X}\|_o\lambda(K)}+\frac{\sqrt{p_{\max}}}{\sqrt{\lambda(1)}\|\bm{X}\|_0} \right)+|\tilde S|p_{\max}\log(1/\lambda(K)).
\end{aligned}
\end{equation}
Note that $\lambda(1)\geq \lambda(K)$, we have accomplished the proof.

\begin{corollary}
\label{cor:KL-bound-P}
The variational posterior $\widetilde\Pi$ arising from the family $\mathcal{P}_{MF}$ satisfies:
\begin{equation}\nonumber
\begin{aligned}
    \text{KL}(\widetilde \Pi\|\Pi(\cdot\mid\bm{X}, \bm{Y}))\mathbbm{1}_{\Gamma}&\preceq (K+1)s_\star\log G+p_{\tilde S}+\epsilon_n\left( |\tilde S|^{1/2}\|\bm{X}\|_o\sqrt{n}\epsilon_n\vartheta+\|\bm{X}\|_o^2|\tilde S|\vartheta^2\right)\\
    &+L |\tilde S|\left( \frac{s_\star^{1/2}\epsilon}{\lambda(K)}+\frac{4\sqrt{p_{\max}\log G}}{\|\bm{X}\|_o\lambda(K)} \right)+|\tilde S|p_{\max}\log(1/\lambda(K)).
\end{aligned}
\end{equation}
\end{corollary}
The proof of this Corollary can be shown easily by noting that $\mathcal{Q}_{MF}$ is a subclass of $\mathcal{P}_{MF}$.

Then we are ready to present the proof of Theorem \ref{thm:VB-contraction}.

\textit{Proof of Theorem \ref{thm:VB-contraction}.} Let $\mathbbm{E}_\star$ be the expectation under the true data generating process $\bm{Y} = \bm{X}\bm{\theta}^\star+\sigma_\star\bm{\epsilon}$. Note that $\mathbbm{E}_{\star}\Pi(\widehat\Omega_{\rho_n}^c)\leq \mathbbm{E}_{\star}\Pi(\widehat\Omega_{\rho_n}^c)\mathbbm{1}_{\Gamma}+o(1) $ and 
\begin{equation}\nonumber
\mathbbm{E}_{\star}\Pi(\widehat\Omega_{\rho_n}^c)\mathbbm{1}_{\Gamma}\preceq e^{-C_2\rho_nKs_\star\log G}.
\end{equation}
Using Theorem 5 in \citet{2019rayvilinear}, we have:
\begin{equation}\nonumber   \mathbbm{E}_{\star}\widetilde\Pi(\widehat\Omega_{\rho_n}^c)\mathbbm{1}_{\Gamma}\leq \frac{2}{C_2\rho_nKs_\star\log G}\text{KL}(\widetilde\Pi\|\Pi(\cdot\mid\bm{X},\bm{Y}))+o(1)
\end{equation}

By Corollary \ref{cor:KL-bound-P}, we have:
\begin{equation}\nonumber
\mathbbm{E}_{\star}\widetilde\Pi(\widehat\Omega_{\rho_n}^c)\mathbbm{1}_{\Gamma}\preceq \frac{1}{\rho_n}\left\{1+\sqrt{\frac{s_\star}{K\lambda(K^\prime)}}\epsilon_n+\frac{s_\star\epsilon_n}{\lambda(K^\prime)K}+L\bigg(\frac{s_\star}{\sqrt{n\log G}\lambda(K)}+\frac{1}{\|\bm{X}_o\|\sqrt{\log n}\lambda(K)}\bigg)\right\}.
\end{equation}
By Assumption \eqref{eq:assump-constant-group} and $\rho_n\gg \max\{s_\star\epsilon_n, 1\}$ we have completed the proof for general slab functions. 

We then state the proof for hierarchical slab functions. The proof is basically the same as the above reasoning. As an example, we upper bounded the KL divergence between the VB posterior $\widehat Q$, which arises from the augmented mean field family defined in \eqref{eq:mean-field-augment} and $\Pi_{\tilde S}(\cdot\mid\bm{X}, \bm{Y})$. For any $h(\bm{\theta}_i)$ which can be rewritten hierarchically as a scale mixture of multivariate normal distributions, we write it as: $$\bm{\theta}_i\mid\alpha_i^2\sim N(\bm{0}, \alpha_i^{-2} \bm{I}_{p_i}),\quad \alpha_i^2\sim \tilde h(\alpha_i^2).$$ Then the posterior density $\Pi_{\tilde S}(\cdot\mid \bm{X}, \bm{Y})$ can be calculated as:
\begin{equation}\nonumber
    \frac{1}{D_\pi}e^{-\frac{1}{2\sigma^2}\|\bm{Y}-\bm{X}_{\tilde S}\bm{\theta}_{\tilde S}\|^2+\frac{v}{2\sigma_\star^2}-\frac{v}{2\sigma_\star^2}}(\sigma^2)^{-n/2}g(\sigma^2)\prod_{j\in\tilde S}(\alpha_j^2)^{p_j/2}e^{-\frac{\alpha_j^2\|\bm{\theta}_j\|^2}{2}}\tilde h(
    \alpha_j^2),
\end{equation}
where $D_{\pi}$ is the normalizing constant.
The density of VB posterior  can be written as
\begin{equation}\nonumber
    \frac{1}{D_{N}}e^{-\frac{1}{2\sigma_\star^2}\|\bm{Y}-\bm{X}_{\tilde S}\bm{\theta}_{\tilde S }\|^2+\frac{1}{2\sigma_\star^2}\|(\bm{I}-\bm{P}_{\tilde S})\bm{Y}\|^2-\frac{v}{2\sigma^2}}(\sigma^2)^{-n/2}g(\sigma^2)\prod_{j\in \tilde S}(\alpha_j^2)^{p_j/2}e^{-\frac{\alpha_j^2\kappa_j^2}{2}}\tilde h(
    \alpha_j^2),
\end{equation}
where $D_{N}$ is the normalizing constant. 

Choose $v = \|(\bm{I}-\bm{P}_{\tilde S})\bm{Y}\|^2$ and $\kappa_j^2 = \mathbbm{E}_{\bm{\mu}_{\tilde S}, \bm{\Sigma}_{\tilde S}}\|\bm{\theta}_j\|^2$, then we have:
\begin{equation}\nonumber
\begin{aligned}
    &\ \ \ \ \mathbbm{E}_{\bm{\mu}_{\tilde S}, \bm{\Sigma}_{\tilde S}, v}\frac{\log dN(\bm{\mu}_{\tilde S}, \bm{\Sigma}_{\tilde S})\otimes \pi(\sigma^2)}{d\Pi_{\tilde S}(\cdot|\bm{X},\bm{Y})}\\
    &= \log\frac{D_\pi}{D_N}+\mathbbm{E}_{v}\bigg[\frac{\sigma_\star^2}{\sigma^2}-1\bigg]\bigg(\frac{\mathbbm{E}_{\bm{\mu}_{\tilde S}, \bm{\Sigma}_{\tilde S}}\|\bm{Y}-\bm{X}_{\tilde S}\bm{\theta}_{\tilde S}\|^2-v}{2\sigma_\star^2}\bigg)\\
    &+\frac{\|(\bm{I}-\bm{P}_{\tilde S})\bm{Y}\|^2-v}{2\sigma_\star^2}+\sum_{j\in \tilde S}\frac{\mathbbm{E}_{\kappa_i}\alpha_j^2}{2}\mathbbm{E}_{\bm{\mu}_{\tilde S}, \bm{\Sigma}_{\tilde S}}\left[\|\bm{\theta}_j\|^2-\kappa_j^2\right]\\
    & = \log\frac{D_\pi}{D_N}+\mathbbm{E}_{v}\bigg[\frac{\sigma_\star^2}{\sigma^2}-1\bigg]\bigg(\frac{ \|(\bm{I}-\bm{P}_{\tilde S})\bm{Y}\|^2+tr(\bm{P}_{\tilde S})-v}{2\sigma_\star^2}\bigg)+\frac{\|(\bm{I}-\bm{P}_{\tilde S})\bm{Y}\|^2-v}{2\sigma_\star^2}\\
    &=\log \frac{D_\pi}{D_N}+\mathbbm{E}_{v}\bigg[\frac{\sigma_\star^2}{\sigma^2}-1\bigg]\frac{p_{\tilde S}}{2\sigma_\star^2}.
    \end{aligned}
\end{equation}
Thus it suffices to upper bound the first term. Note that if the slab function $h(\bm{\theta}_j)$ can be written as a scale mixture of multivariate Gaussian random variables, then the density function $h$ only depends on the $\ell_2$ norm of the $\bm{\theta}_j$. In this case,  with a little abuse of notation, we denote the slab function for $\bm{\theta_j}$ as $h(\|\bm{\theta}_j\|)$. 

By integrading out $\alpha_i^2$, we can  calculate $ D_{\pi}/D_{N}$ as 
\begin{equation}\nonumber
\frac{D_\pi}{D_N} = \frac{\int\int \exp\left(-\frac{1}{2\sigma^2}\|\bm{Y}-\bm{X}_{\tilde S}\bm{\theta}_{\tilde S}\|^2+\frac{v}{2\sigma_\star^2}-\frac{v}{2\sigma_\star^2}\right)(\sigma^2)^{-(n/2)}g(\sigma^2)\prod_{j\in\tilde S} h(\|\bm{\theta}_j\|)d\sigma^2d\bm{\theta}_{\tilde S}}{\int\int \exp\left(-\frac{1}{2\sigma_\star^2}\|\bm{Y}-\bm{X}_{\tilde S}\bm{\theta}_{\tilde S}\|^2+\frac{1}{2\sigma_\star^2}\|(\bm{I}-\bm{P}_{\tilde S})\bm{Y}\|^2-\frac{v}{2\sigma^2}\right)(\sigma^2)^{-(n/2)}g(\sigma^2)\prod_{j\in\tilde S}h(\kappa_j)d\sigma^2d\bm{\theta}_{\tilde S}}.
\end{equation}

Therefore on $\Gamma$, we can upper bound $\log\frac{D_\pi}{D_N}$ as 
\begin{equation}\nonumber
\begin{aligned}
\sup_{(\bm{\theta}_{\tilde S},\ \sigma^2)\in B_{\tilde S}} \bigg(1-\frac{\sigma_\star^2}{\sigma^2}\bigg)\frac{\|\bm{Y}-\bm{X}_{\tilde S}\bm{\theta}_{\tilde S}\|^2-v}{2\sigma_\star^2}+L\sum_{j\in \tilde S}\left|\|\bm{\theta}_j\|-\kappa_j\right|+\log 2.
\end{aligned}
\end{equation}
For $\bm{\theta}_{\tilde S}$ in $B_{\tilde S}$, we have: $\|\bm{\theta}_{\tilde S}-\bm{\theta}^\star_{\tilde S}\|<2\vartheta$.
\begin{equation}\nonumber
\begin{aligned}
    \sup_{\bm{\theta}_{\tilde S}\in B_{\tilde S}}\sum_{j\in\tilde S}\left|\|\bm{\theta}_j\|-\kappa_j\right| = \sup_{\bm{\theta}_{\tilde S}\in B_{\tilde S}} \sum_{j\in\tilde S}|\|\bm{\theta}_j\|-\mathbbm{E}_{\bm{\mu}_{\tilde S}, \bm{\Sigma}_{\tilde S}}\left[\|\bm{\theta}_j\|\right]|\leq |\tilde S|^{1/2}\mathbbm{E}_{\bm{\mu}_{\tilde S}, \bm{\Sigma}_{\tilde S}}\left[\|\bm{\theta}_j-\bm{\theta}_j^{\star}\|\right]+2|\tilde S|^{1/2}\vartheta.
\end{aligned}
\end{equation}
The first term has already been upper-bounded in the proof of Theorem \ref{thm:VB-contraction}.

\subsection{Proof of Theorem \ref{thm:VB-contraction-additive}}
The proof of Theorem \ref{thm:VB-contraction-additive} is rather similar as the proof of Theorem \ref{thm:VB-contraction}. The first step is to obtain the contraction rates for the true posterior. It suffices to validate Lemma \ref{lemma:posterior-key}. We only emphasize the main difference from the proof of Theorem \ref{thm:VB-contraction} here. To make better comparisons between this proof and the proof of Theorem \ref{thm:VB-contraction}, we still use $\bm{X}$ to denote the basis expanded design matrix $\bm{\widetilde X}$.

\textbf{Difference 1:} When we want to prove Lemma \ref{lemma:posterior-step-b1} for additive models, $\phi_1$ is now equivalent to
\begin{equation}\nonumber
\phi_1 = \mathbbm{1}\left\{\max_{S\in\mathcal{F}:|S\backslash S_{\star}|\leq Ks_\star}\bigg|\frac{(\bm{\epsilon}+\bm{\delta})^\intercal(\bm{I}-\bm{P}_{S\cup S_\star})(\bm{\epsilon}+\bm{\delta})}{n}-1\bigg|>M_1\epsilon_n\right\}.
\end{equation}
Expanding the numerator of the test function, we have:
\begin{equation}\nonumber
(\bm{\epsilon}+\bm{\delta})^\intercal (\bm{I}-\bm{P}_{S\cup S_{\star}})(\bm{\epsilon}+\bm{\delta}) = \|(\bm{I}-\bm{P}_{S\cup S_{\star}})\bm{\epsilon}\|^2+2\bm{\epsilon}^\intercal(\bm{I}-\bm{P}_{S\cup S_\star})\bm{\delta}+\|(\bm{I}-\bm{P}_{S\cup S_{\star}})\bm{\delta}\|^2.
\end{equation}
Note that:
\begin{equation}\nonumber
-\|(\bm{I}-\bm{P}_{S\cup S_{\star}})\bm{\epsilon}\|\|(\bm{I}-\bm{P}_{S\cup S_{\star}})\bm{\delta}\|\leq \bm{\epsilon}^\intercal(\bm{I}-\bm{P}_{S\cup S_\star})\bm{\delta}\leq \|(\bm{I}-\bm{P}_{S\cup S_{\star}})\bm{\epsilon}\|\|(\bm{I}-\bm{P}_{S\cup S_{\star}})\bm{\delta}\|,
\end{equation}
we have:
\begin{equation}\nonumber
    (\|(\bm{I}-\bm{P}_{S\cup S_{\star}})\bm{\epsilon}\|-\|(\bm{I}-\bm{P}_{S\cup S_{\star}})\bm{\delta}\|)^2\leq (\bm{\epsilon}+\bm{\delta})^\intercal (\bm{I}-\bm{P}_{S\cup S_{\star}})(\bm{\epsilon}+\bm{\delta})\leq (\|(\bm{I}-\bm{P}_{S\cup S_{\star}})\bm{\epsilon}\|+\|(\bm{I}-\bm{P}_{S\cup S_{\star}})\bm{\delta}\|)^2.
\end{equation}
Then using the above inequality, we can repeat the proof of Lemma \ref{lemma:posterior-step-b1} for additive models.

\textbf{Difference 2:} We then  prove Lemma \ref{lemma:posterior-step-b2} for additive models. 
Under the null, 
\begin{equation}\nonumber
    \frac{\|\bm{X}_{S\cup S_{\star}}^\dagger\bm{Y}-\bm{\theta}_{S\cup S_{\star}}^\star\|^2}{\sigma_\star^2} =  (\bm{\epsilon}+\bm{\delta})^\intercal\bm{X}_{S\cup S_{\star}}(\bm{X}_{S\cup S_{\star}}^\intercal \bm{X}_{S\cup S_{\star}})^{-2}\bm{X}_{S\cup S_{\star}}(\bm{\epsilon}+\bm{\delta}).
\end{equation}
Thus we have under the null:
\begin{equation}\nonumber
\begin{aligned}
    \phi_2&\leq \mathbbm{1}\left\{\max_{S\in\mathcal{F}: |S\backslash S_\star|\leq Ks_\star}(\bm{\epsilon}+\bm{\delta})^\intercal \bm{P}_{S\cup S_\star}(\bm{\epsilon}+\bm{\delta})>M_2^2n\epsilon_n^2/4\right\}\\
    &\leq \mathbbm{1}\left\{\max_{S\in\mathcal{F}: |S\backslash S_\star|\leq Ks_\star}(\|\bm{\epsilon}^\intercal \bm{P}_{S\cup S_\star}\bm{\epsilon}\|+\|\bm{\delta}^\intercal \bm{P}_{S\cup S_\star}\bm{\delta}\|)^2>M_2^2n\epsilon_n^2/4\right\}.
\end{aligned}
\end{equation}
Note that $\bm{\delta}^\intercal \bm{P}_{S\cup S_\star}\bm{\delta}\leq  \|\bm{\delta}\|^2\leq s_\star nd^{-2\kappa}\asymp s_\star d = o(n\epsilon_n^2)$. Thus 
\begin{equation}\nonumber
\begin{aligned}
    \mathbbm{E}_{\star}\phi_2\leq \binom{p}{Ks_\star}\times \mathbbm{P}\left(\chi^2_{(K+1)s_\star d}>(M_2^2/4-o(1))n\epsilon_n^2\right).
\end{aligned}
\end{equation}
Furthermore, under the alternative, we have:
\begin{equation}\nonumber
\begin{aligned}
\|\bm{X}^\dagger_{S\cup S_{\star}}\bm{Y}-\bm{\theta}^\star_{S\cup S_\star}\| \geq \|\bm{\theta}_{S\cup S_{\star}}- \bm{\theta}_{S\cup S_{\star}}^\star\|- \sigma\| \bm{X}^\dagger_{S\cup S_{\star}}\bm{\epsilon}\|-\| \bm{X}^\dagger_{S\cup S_{\star}}\bm{\delta}\|.
\end{aligned}
\end{equation}
Note that $\| \bm{X}^\dagger_{S\cup S_{\star}}\bm{\delta}\| = o(\sqrt{n}\epsilon_n/(\|\bm{X}\|_o\sqrt{\lambda}))$,
the rest of the proof follows similarly as the proof of Lemma \ref{lemma:posterior-step-b2}.

\textbf{Difference 3:} When we repeat the proof of Lemma \ref{lemma:posterior-step-c} for the additive models, the only difference is in part (3). We here prove that if: 
\begin{equation}\nonumber
\|\bm{P}_{S_\star}\bm{\epsilon}\|^2\leq Cn\epsilon_n^2,\quad\|\bm{\epsilon}\|^2< 2Cn,\quad\textnormal{and}\quad \bm{\epsilon}^\intercal\bm{\delta}<\eta_4n\epsilon_n^2
\end{equation}
then for any small $\eta_3>0$:
\begin{equation}\nonumber
    \inf_{(\sigma, S, \bm{\theta})\in\widehat \Omega}\frac{N(\bm{Y}\mid \bm{X\theta}, \sigma^2\bm{I})}{N(\bm{Y}\mid \bm{X\theta}^\star, \sigma_\star^2\bm{I})}\gtrsim e^{-[C\eta_1+\eta_2^2/2+\sqrt{C}\eta_2+\eta_3+\eta_4+o(1)]n\epsilon_n^2}.
\end{equation}
Note that $\eta_4$ can be arbitrarily small, since  for any $\eta_4$, we have:
\begin{equation}\nonumber
    \mathbbm{P}(\bm{\epsilon}^\intercal\bm{\delta}>\eta_4n\epsilon_n^2) = \mathbbm{P}(\|\bm{\delta}\|Z>\eta_4n\epsilon_n^2)\leq e^{-\eta_4^2n^2\epsilon_n^4/\|\bm{\delta}\|^2} = o(e^{-n\epsilon_n^2})
\end{equation}
where $Z$ is a standard normal random variable. Note that 
for additive models, we have:
\begin{equation}\nonumber
\begin{aligned}
-2\log \Delta & = \|\sigma_\star\bm{\epsilon}+\bm{\delta}+\bm{X}(\bm{\theta}^\star-\bm{\theta})\|^2/\sigma^2-\|\bm{\epsilon}\|^2+2n\log(\sigma^2/\sigma_\star^2)\\
& = (\sigma_\star^2/\sigma^2-1)\|\bm{\epsilon}\|^2+\|\bm{X}(\bm{\theta}^\star-\bm{\theta})/\sigma\|^2+2\sigma_\star\bm{\epsilon}^\intercal\bm{X}(\bm{\theta}^\star-\bm{\theta})/\sigma^2+2n\log(\sigma^2/\sigma_\star^2)\\
&+\|\bm{\delta}\|^2/\sigma^2+2\sigma_\star\bm{\epsilon}^\intercal\bm{\delta}/\sigma^2+2\bm{\delta}^\intercal\bm{X}(\bm{\theta}^\star-\bm{\theta})/\sigma^2.
\end{aligned}
\end{equation}
Then the proof is completed by noting that $\|\bm{\delta}\|^2 = o(n\epsilon_n^2)$ and $\bm{\delta}^\intercal\bm{X}(\bm{\theta}^\star-\bm{\theta})\leq \|\bm{\delta}\|\|\bm{X}(\bm{\theta}^\star-\bm{\theta})\| = o(n\epsilon_n^2)$.

After establishing the contraction rate for the true posterior, we then complete the proof by bounding the KL divergence between the VB posterior and the true posterior. The process is also rather similar to that in the proof of Theorem \ref{thm:VB-contraction} and we only mention key differences here.

\textbf{Difference 4:} We define $\bm{\mu}_{\tilde S}$ similarly as \eqref{eq:define-mu-Sigma} and then 
\begin{equation}\nonumber
    \|\bm{\mu}_{\tilde S}-\bm{\theta}^{\star}_{\tilde S}\|\leq \|(\bm{X}_{\tilde S}^\intercal\bm{X}_{\tilde S})^{-1}\bm{X}_{\tilde S}^\intercal \bm{X}_{\tilde S^c}\bm{\theta}^{\star}_{ \tilde S^c}\|+\sigma_\star\|(\bm{X}_{\tilde S}^\intercal\bm{X}_{\tilde S})^{-1}\bm{X}_{\tilde S}^\intercal \bm{\epsilon}\|+\|(\bm{X}_{\tilde S}^\intercal\bm{X}_{\tilde S})^{-1}\bm{X}_{\tilde S}^\intercal \bm{\delta}\|: = I+II+III.
\end{equation}
The upper bound for I and II can be derived similarly as in \eqref{eq:bound-I} and \eqref{eq:bound-II}. For III, we have:
\begin{equation}\nonumber
    III = \sqrt{\bm{\delta}^\intercal \bm{X}_{\tilde S}(\bm{X}_{\tilde S}^\intercal\bm{X}_{\tilde S})^{-2}\bm{X}_{\tilde S}\bm{\delta}}\leq \frac{\sqrt{\bm{\delta}^\intercal\bm{P}_{\tilde S}\bm{\delta}}}{\|\bm{X}\|_o\sqrt{\lambda}}\leq \frac{\sqrt{s_\star n}d^{-\kappa}}{\|\bm{X}\|_o\sqrt{\lambda}}.
\end{equation}
We then upper bound $\|\bm{Y}-\bm{X}_{\tilde S}\bm{\theta}_{\tilde S}\|^2-v$. Note that
\begin{equation}\nonumber
    \|\bm{Y}-\bm{X}_{\tilde S}\bm{\theta}_{\tilde S}\|^2-v = I+II+III,
\end{equation}
where
\begin{equation}\nonumber
\begin{aligned}
    &I = \|\sigma_\star\bm{\epsilon}+\bm{\delta}\|^2-\|(\bm{I}-\bm{P}_{\tilde S})(\sigma_\star\bm{\epsilon}+\bm{\delta})\|^2,\\
    & II = 2(\sigma_\star\bm{\epsilon}+\bm{\delta})^\intercal\bm{X}(\bm{\theta}^\star- \bm{\bar\theta}_{\tilde S}
    )-2(\sigma_\star\bm{\epsilon}+\bm{\delta})^\intercal(\bm{I}-\bm{P}_{\tilde S})\bm{X}_{\tilde S^c}\bm{\theta}_{\tilde S^c}^\star,\\
    & III = \|\bm{X}(\bm{\theta}^\star- \bm{\bar\theta}_{\tilde S}
    )\|^2 - \|(\bm{I}-\bm{P}_{\tilde S})\bm{X}_{\tilde S^c}\bm{\theta}^\star_{\tilde S^c}\|^2.
\end{aligned}
\end{equation}
Note that 
\begin{equation}\nonumber
    I = \|\bm{P}_{\tilde S}(\sigma_\star\bm{\epsilon}+\bm{\delta})\|^2\leq2\sigma_\star^2\|\bm{P}_{\tilde S}\bm{\epsilon}\|^2+2\|\bm{\delta}\|^2\preceq n\epsilon_n^2.
\end{equation}
For $II$, we only have to take care of $\bm{\delta}^\intercal \bm{X}(\bm{\theta}^\star-\bm{\bar{\theta}}_{\tilde S})$ and $\bm{\delta}^\intercal(\bm{I}-\bm{P}_{\tilde S})\bm{X}_{\tilde S^c}\bm{\theta}^\star_{\tilde S^c}$ since the rest of the terms have been dealt with in the proof of Lemma \ref{lemma:KL-bound-Q}.
For both terms, we have:
\begin{equation}\nonumber
\begin{aligned}
    &\bm{\delta}^\intercal \bm{X}(\bm{\theta}^\star-\bm{\bar{\theta}}_{\tilde S})\leq \|\bm{\delta}\|\|\bm{X}(\bm{\theta}^\star-\bm{\bar{\theta}}_{\tilde S})\|\leq 2\sqrt{s_\star n}d^{-\kappa} \|\bm{X}\|_o\sqrt{|\tilde S|+s_\star}\vartheta,\\
    & \bm{\delta}^\intercal(\bm{I}-\bm{P}_{\tilde S})\bm{X}_{\tilde S^c}\bm{\theta}^\star_{\tilde S^c}\leq \|(\bm{I}-\bm{P}_{\tilde S})\bm{\delta}\|\|\bm{X}_{\tilde S^c}\bm{\theta}^\star_{\tilde S^c}\| \leq \sqrt{s_\star n}d^{-\kappa} \|\bm{X}\|_o\sqrt{s_\star}\vartheta
\end{aligned}
\end{equation}
The rest of the proof  can be completed by repeating the proof of Theorem \ref{thm:VB-contraction}.

\section{Methodological Details}\label{append:compute}
In this section, we provide a comprehensive derivation of the CAVI updates, which encompass Section \ref{subsec:CAVI} and Section \ref{subsec:hierarchical-sas}, in addition to the EM updates discussed in Section \ref{subsec:hyperparameter}. For CAVI, the formulas are derived by minimizing the KL divergence with respect to one parameter while keeping others constant. The EM updates are rooted in the principle of maximizing the Evidence Lower Bound (ELBO).

\subsection{Proofs for CAVI updates with general slab functions}\label{sec:CAVI-proof}

Proof of \eqref{eq:mu-sigma-update-general}: Since the discrete component of $\text{KL}({P_{\bm{\mu}, \bm{\Sigma}, \bm{\gamma}, v}\|\Pi(\cdot\mid \bm{Y})})$ is independent of $\bm{\mu}$ and $\bm{\Sigma}$, it suffices to compute the KL divergence between the continuous part of the variational family and the posterior. By expanding and simplifying the formula for $\text{KL}({P_{\bm{\mu}, \bm{\Sigma}, \bm{\gamma}, v\mid z_i=1}\|\Pi(\cdot\mid \bm{Y})})$ with respect to the $i$-th group, we have:
\begin{equation}
    \begin{aligned}
     h_i(\bm{\mu},\bm{\Sigma}):&=\text{KL}(P_{\bm{\mu}, \bm{\Sigma}, \bm{\gamma}, v|z_i=1}\Vert\Pi(\cdot\mid \bm{Y}))\\
&=\mathbbm{E}_{\bm{\mu}, \bm{\Sigma}, \bm{\gamma}, v|z_i=1}
	    \left[\log P_{\bm{\mu}, \bm{\Sigma}, \bm{\gamma}, v|z_i=1}-\log\frac{\Pi(\bm{Y}|\cdot)\Pi(\cdot)}{\int \Pi(\bm{Y}|\cdot)\Pi(\cdot)}\right]\\
&=\mathbbm{E}_{\bm{\mu}, \bm{\Sigma}, \bm{\gamma}, v|z_i=1}
	    \left[\log\frac{P_{\bm{\mu}, \bm{\Sigma}, \bm{\gamma}, v|z_i=1}}{\Pi(\cdot)}-\log\Pi(\bm{Y}|\cdot)\right]+C,
    \end{aligned}
    \label{eq:KL-init-Gaussian}
\end{equation}
where $C$ is a constant independent of $\bm{\mu}$ and $\bm{\Sigma}$, $\Pi(\cdot)$ is the prior and $\Pi(\bm{Y}|\cdot)$ is the likelihood function. 

Recall that $P_{\bm{\mu}, \bm{\Sigma}, \bm{\gamma}, v|z_i=1}$ takes the mean-field form \eqref{eq:mean-field-general}, it can be rewritten as 
\begin{equation}\nonumber
  P_{\bm{\mu}, \bm{\Sigma}, \bm{\gamma}, v|z_i=1}=  P_{\bm{\mu}_{-i},\bm{\Sigma}_{-i},\bm{\gamma}_{-i}, v}\times N(\bm{\mu}_i,\bm{\Sigma}_i),
\end{equation}
where $\bm{\mu}_{-i},\bm{\Sigma}_{-i}$ and $\gamma_{-i}$ are all parameters other than those in the $i$th group. Furthermore, the prior $\Pi(\bm{\theta})=h(\bm{\theta}_i)\times\Pi_{-i}$ is also an independent product of marginals. Note here that we update our parameters $(\bm{\mu}_i,\bm{\Sigma}_i)_{i=1}^G$ in a coordinate ascent manner. Therefore, by rewriting $h_i(\bm{\mu},\bm{\Sigma})$ as $\tilde{h}_i(\bm{\mu}_i,\bm{\Sigma}_i)$ for any fixed $\bm{\mu}_{-i}$, $\bm{\Sigma}_{-i}$ and plugging in the log-likelihood of linear regression, \eqref{eq:KL-init-Gaussian} can be computed as
\begin{equation}
\begin{aligned}
\tilde{h}_i(\bm{\mu}_i,\bm{\Sigma}_i)&=\mathbbm{E}_{\bm{\mu}, \bm{\Sigma}, \bm{\gamma}, v|z_i=1}\left[\log\frac{N(\bm{\mu}_i,\bm{\Sigma}_i)}{h(\bm{\theta}_i) }+\frac{1}{2\sigma^2}\Vert Y-\sum_{i=1}^{G}\bm{X}_i\bm{\theta}_i\Vert_2^2\right]+C'\\
&=\frac{1}{2\tilde\sigma^2}\bigg[Tr(\bm{X}_i^\intercal\bm{X}_i\bm{\Sigma}_i)+\bm{\mu}_i^\intercal\bm{X}_i^\intercal\bm{X}_i\bm{\mu}_i-2\bm{Y}^\intercal\bm{X}_i\bm{\mu}_i+2\sum_{j\neq i}\gamma_j\bm{\mu}_i^\intercal\bm{X}_i^\intercal\bm{X}_j\bm{\mu}_j\bigg]\\
& -\frac{1}{2}\log(|\bm{\Sigma}_i|)-\mathbbm{E}_{\bm{\mu}_i,\bm{\Sigma}_i}\log h(\bm{\theta}_i)+C'.
\end{aligned}
\label{eq:KL-computed-Gaussian}
\end{equation}
Here $C'$ is independent of $\bm{\mu}_i$ and $\bm{\Sigma}_i$, and $1/\tilde\sigma^2$ is the expectation of $1/\sigma^2$ under the mean-field family and will be provided later. Generally, a direct optimization of \eqref{eq:KL-computed-Gaussian} is not feasible. Here we also adopt the idea of separate optimization. By fixing $\bm{\Sigma}_i$ (or $\bm{\mu}_i$), we get the objective functions $f_i$ (or $g_i$) in \eqref{eq:mu-sigma-update-general}.

Proof of \eqref{eq:Gaussian-gamma}: This proof is similarly to the proof of \eqref{eq:mu-sigma-update-general}. First, notice that the parameter $\gamma_i$ controls the sparsity in the variational posterior. Instead of computing the conditional KL divergence, here we expand the full KL divergence $\text{KL}({P_{\bm{\mu}, \bm{\Sigma}, \bm{\gamma}, v}\|\Pi(\cdot\mid\bm{X}, \bm{Y})})$ with respect to the $i$-th group as:
\begin{equation}\nonumber
    \begin{aligned}
    &\quad \text{KL}(P_{\bm{\mu}, \bm{\Sigma}, \bm{\gamma}, v}\Vert \Pi(\cdot\mid\bm{X}, \bm{Y}))\\
&=\mathbbm{E}_{\bm{\mu}, \bm{\Sigma}, \bm{\gamma}, v}\left[\log\frac{P_{\bm{\mu}_{-i},\bm{\Sigma}_{-i},\bm{\gamma}_{-i}, v}\times [\gamma_iN(\bm{\mu}_i,\bm{\Sigma}_i)+(1-\gamma_i)\delta_{\bm{0}}]}{\Pi_{-i}\times [w h(\bm{\theta}_i)+(1-w)\delta_{\bm{0}}]}+\frac{1}{2\sigma^2}\Vert \bm{Y}-\sum_{i=1}^{G}\bm {X}_i\bm{\theta}_i\Vert_2^2\right]+C\\
&=\mathbbm{E}_{\bm{\mu}, \bm{\Sigma}, \bm{\gamma}, v}
	    \left[  1_{\{z_i=1\}}\log\frac{ \gamma_iN(\bm{\mu}_i,\bm{\Sigma}_i)}{w h(\bm{\theta}_i)}+1_{\{z_i=0\}}\log\frac{1-\gamma_i}{1-w}+ \frac{1}{2\sigma^2}\Vert \bm{Y}-\sum_{i=1}^{G}\bm {X}_i\bm{\theta}_i\Vert_2^2 \right]+C\\
& = \gamma_i\log(\gamma_i)+(1-\gamma_i)\log(1-\gamma_i)-\gamma_i\log\frac{w}{1-w}- \frac{\gamma_ip_i}{2}\log (2\pi)-\frac{\gamma_i p_i}{2}-\frac{\gamma_i}{2}\log|\bm{\Sigma}_i|\\
&-\gamma_i\mathbbm{E}_{\bm{\mu}_i,\bm{\Sigma}_i}\log h(\bm{\theta}_i)+\frac{1}{2\tilde\sigma^2}\mathbbm{E}_{\bm{\mu},\bm{\Sigma},\bm{\gamma}}[\|\bm{Y}-\sum_{g=1}^{G}\bm {X}_i\bm{\theta}_i\|^2]+C.
    \end{aligned}
\end{equation}
The constant $C$, which is independent of $\gamma_i$, may vary in different lines. Expanding the last two lines and taking its derivative to zero, we obtain the equation
\begin{equation}
\begin{aligned}
            \log\frac{\gamma_i}{1-\gamma_i}&-\left(\frac{1}{2\tilde\sigma^2}\left[2\bm{\mu}_i^\intercal\bm{X}_i^\intercal\bm{Y}-2\sum_{j\neq i}\gamma_j\bm{\mu}_i^\intercal\bm{X}_i^\intercal \bm{X}_j\bm{\mu}_j-Tr(\bm{X}_i^\intercal\bm{X}_i(\bm{\mu}_i\bm{\mu}_i^\intercal+\bm{\Sigma}_i))\right]\right.\\
        &\left.+\frac{p_i}{2}\log2\pi+\frac{1}{2}\log|\Sigma_i|+\frac{p_i}{2}+\mathbbm{E}_{\bm{\mu}_i, \bm{\Sigma}_i}\log h(\bm{\theta}_i)+\log\frac{w}{1-w}\right)=0,
\end{aligned}
\end{equation}
which gives the update rule \eqref{eq:Gaussian-gamma}.

Proof of \eqref{eq:Gaussian-sigma}: It's well known that the optimal form for $q(\sigma^2)$ will be the exponential of the expected log of the complete conditional, i.e.,
$$q(\sigma^2)\propto \exp\{\mathbbm{E}[\log p(\sigma^2|\cdot)]\},$$
where the expectation is taken with respect to the mean-field family.

For the linear model, when we impose a prior distribution $g(\sigma^2)$ on $\sigma^2$, the posterior distribution can be expressed as::
$$p(\sigma^2|\cdot) \propto (\sigma^2)^{-(n/2)}g(\sigma^2)\exp\left(-\frac{\|\bm{Y}-\sum_{i=1}^{G}\bm{X}_i\bm{\theta}_i\|_2^2}{2\sigma^2}\right).$$
 Thus it suffices to calculate $\mathbbm{E}[\|\bm{Y}-\sum_{g=1}^G\bm{X}_i\bm{\theta}_i\|_2^2]$, where the expectation is taken with respect to $\bm{\theta}$. This quantity has been previously computed in the proof of \eqref{eq:Gaussian-gamma}. We have:
\begin{equation}\nonumber
    \begin{aligned}
    &\quad\mathbbm{E}[\|\bm{Y}-\sum_{i=1}^{G}\bm{X}_i\bm{\theta}_i\|_2^2]\\
    &= \bm{Y}^\intercal \bm{Y}-2\sum_{i=1}^G\gamma_i \bm{Y}^\intercal\bm{X}_i\bm{\mu}_i+\sum_{i=1}^G \gamma_i Tr(\bm{X}_i^\intercal\bm{X}_i(\bm{\mu}_i\bm{\mu}_i^\intercal+\bm{\Sigma}_i))+\sum_{i=1}^G\sum_{j\neq i}\gamma_i\gamma_j\bm{\mu}_i^\intercal \bm{X}_i^\intercal \bm{X}_j\bm{\mu}_j.
    \end{aligned}
\end{equation}

Denote the last line of the above equation to be $v$, then we have:
$$q(\sigma^2)\propto (\sigma^2)^{-(n/2)}g(\sigma^2) \exp\left(-\frac{v}{2\sigma^2}\right).$$
If we set $g(\sigma^2)$ as  the Inverse-Gamma$(\alpha,\beta)$ distribution, then we know $q(\sigma^2)$ also follows an Inverse-Gamma distribution with shape parameter  $n/2+\alpha$ and scale parameter $v/2+\beta$. As a result, $1/{\tilde\sigma^2}=\mathbbm{E}[1/\sigma^2] = (n/2+\alpha)/(v/2+\beta)$ and $\tilde\sigma^2 = (v/2+\beta)/(n/2+\alpha)$.

\subsection{Proofs for CAVI updates with hierarchical slab functions}
The proof follows a similar approach as in the case of general slab functions, with addition calculations of variables $\alpha_i$.

Proof of \eqref{eq:mu-sigma-update-hierarchical}: 
Similar to the proof of \eqref{eq:mu-sigma-update-general}, we consider the KL divergence condition on $z_i=1$, with an extra variable $\alpha_i$:
\begin{equation}\nonumber
    \begin{aligned}
    &\text{KL}(P_{\bm{\mu,\Sigma,\gamma,\kappa,\beta}|z_i=1}\Vert\Pi(\cdot\mid\bm{Y}))\\
     &=\mathbbm{E}\left[\frac{1}{2\sigma^2}\|\bm{Y}-\sum_{i=1}^{G}\bm{X}_i\bm{\theta}_i\|_2^2+\log\frac{N(\bm{\mu}_i, \bm{\Sigma}_i)}{N(\bm{0}, \alpha_i^{-2} \bm{I}_{p_i})}\right]+C\\
    &= \mathbbm{E}_v\bigg[\frac{1}{2\sigma^2}\bigg]\bigg[-2\bm{Y}^\intercal\bm{X}_i\bm{\mu}_i+\bm{\mu}_i^\intercal\bm{X}_i^\intercal\bm{X}_i\bm{\mu}_i+Tr(\bm{X}_i^\intercal\bm{X}_i\bm{\Sigma}_i)+2\sum_{j\neq i}\gamma_j\bm{\mu}_i^\intercal\bm{X}_i^\intercal\bm{X}_i\bm{\mu}_j\bigg]\\
    &+\frac{1}{2}\bigg[-\log|\bm{\Sigma}_i|-p_i\bigg]+\frac{1}{2}\mathbbm{E}_{\kappa_i}\left[\alpha_i^2\right]\left[Tr(\bm{\Sigma}_i)+\bm{\mu}_i^\intercal\bm{\mu}_i\right]+C.
    \end{aligned}
\end{equation}
By fixing $\bm{\Sigma}_i$ and $\bm{\mu}_i$ separately, we can get the expression of $f_i$ and $g_i$ in \eqref{eq:mu-sigma-update-hierarchical}, and the proof is thus completed.

Proof of \eqref{eq:GM-gamma}: With the knowledge from the proof of \eqref{eq:Gaussian-gamma}, $\gamma_i$ should optimize:
\begin{equation}\nonumber
    \mathbbm{E}\left[\frac{1}{2\sigma^2}\|\bm{Y}-\sum_{i=1}^{G}\bm{X}_i\bm{\theta}_i\|_2^2+\log\frac{\gamma_iN(\bm{\mu}_i, \bm{\Sigma}_i)q(\alpha_i^2)+(1-\gamma_i)\delta_{\bm{0}}(\bm{\theta}_i)\tilde{h}(\alpha_i^2)}{wN(\bm{0}, \alpha_i^{-2} \bm{I}_{p_i})\tilde{h}(\alpha_i^2)+(1-w)\delta_{\bm{0}}(\bm{\theta}_i)\tilde{h}(\alpha_i^2)}\right].
\end{equation}
We have calculated the first term before, and it suffices to calculate the second term, which is:
\begin{equation}\nonumber
\begin{aligned}
&\mathbbm{E}\left[\log\frac{\gamma_iN(\bm{\mu}_i, \bm{\Sigma}_i)q(\alpha_i^2)+(1-\gamma_i)\delta_{\bm{0}}(\bm{\theta}_i)\tilde{h}(\alpha_i^2)}{wN(\bm{0}, \alpha_i^{-2} \bm{I}_{p_i})\tilde{h}(\alpha_i^2)+(1-w)\delta_{\bm{0}}(\bm{\theta}_i)\tilde{h}(\alpha_i^2)}\right]\\
    & = \gamma_i\log\frac{\gamma_i}{w}+(1-\gamma_i)\log\frac{1-\gamma_i}{1-w}+\gamma_i\mathbbm{E}\left[\log\frac{N(\bm{\mu}_i, \bm{\Sigma}_i)q(\alpha_i^2)}{N(\bm{0}, \alpha_i^{-2} \bm{I}_{p_i})\tilde{h}(\alpha_i^2)}\right].
\end{aligned}
\end{equation}
The last term is the conditional KL divergence between the variational posterior and the prior of $\bm{\theta}_i$ and $\alpha_i$. This can be further extended as
\begin{equation}\nonumber
\mathbbm{E}\left[\log\frac{N(\bm{\mu}_i, \bm{\Sigma}_i)q(\alpha_i^2)}{N(0, \alpha_i^{-2} \bm{I}_{p_i})\tilde{h}(\alpha_i^2)}\right]=\frac{1}{2}\left[\mathbbm{E}\alpha_i^2(\bm{\mu}_i^\intercal\bm{\mu}_i+Tr(\bm{\Sigma}_i))-\log|\bm{\Sigma}_i|-p_i\mathbbm{E}\log\alpha_i^2-p_i\right]+\text{KL}(q\|\tilde{h}).
\end{equation}
Combining all terms together, the update for $\gamma_i$ can be obtained as:
\begin{equation}
\gamma_i:=\textnormal{logit}^{-1}\left[\log\frac{w}{1-w}+\frac{1}{2}[p_i\mathbbm{E}_{\kappa_i}\log\alpha_i^2+\log|\Sigma_i|+\bm{\mu}_i^\intercal\bm{\Sigma}_i^{-1}\bm{\mu}_i]-\text{KL}(q\|\tilde h)\right].
\label{eq:gamma-before-replace}
\end{equation}
Recall that in mean-field family, $q(\alpha_j^2)=\frac{1}{C_i}(\alpha_j^2)^{p_j/2}e^{-\alpha_j^2\kappa_j/2}\tilde h(\alpha^2_j)$, where $C_i$ is the normalizing constant. Substituting this into equation (\ref{eq:gamma-before-replace}) yields the update rule \eqref{eq:GM-gamma}.

Proof of \eqref{eq:GM-kappa}: Because $\alpha_i$ is not related to the log-likelihood, the first term in the KL divergence remains constant fixing all parameters other than $\alpha_i$. Therefore, we only need to optimize the KL divergence between the variational posterior and the prior, which is defined by:
\begin{equation}\nonumber
   \ell(\kappa_i):= \text{KL}(N(\bm{\mu}_i,\bm{\Sigma}_i)q(\alpha_i^2)\|N(\bm{0},\alpha_i^2\bm{I}_{p_i})\tilde{h}(\alpha_i^2))=\mathbbm{E}\log\frac{N(\bm{\mu}_i,\bm{\Sigma}_i)q(\alpha_i^2)}{N(\bm{0},\alpha_i^{-2}\bm{I}_{p_i})\tilde{h}(\alpha_i^2)}.
\end{equation}
Plugging in the $q(\alpha_i^2)$ in \eqref{eq:hierarchical-variational} here, it can be computed as:
\begin{equation}\nonumber
\begin{aligned}    \ell(\kappa_i)&=\mathbbm{E}\log\frac{N(\bm{\mu}_i,\bm{\Sigma}_i)q(\alpha_i^2)}{N(\bm{0},\alpha_i^{-2}\bm{I}_{p_i})\tilde{h}(\alpha_i^2)}=\mathbbm{E}\left[\frac{\alpha_i^2}{2}\bm{\theta}_i^\intercal\bm{\theta}_i-\frac{\alpha_i^2}{2}\kappa_i\right]-\log C_i+C'\\
&=-\frac{1}{2}\left[(\kappa_i-\bm{\mu}_i^\intercal\bm{\mu}_i-Tr(\bm{\Sigma}_i))\mathbbm{E}_{\kappa_i}\alpha_i^2\right]-\log C_i+C',
\end{aligned}
\end{equation}
where $C_i$ is the normalizing constant of $q(\alpha_i^2)$, and $C'$ is a constant irrelevant to $\kappa_i$. Furthermore, it's easily proven by the chain rule that ${\partial\log C_i}/{\partial \kappa_i}=-\frac{1}{2}\mathbbm{E}_{\kappa_i}\alpha_i^2$. Take derivative of $\ell(\kappa_i)$, we have
\begin{equation}\nonumber
\begin{aligned}
    \frac{\partial \ell(\kappa_i)}{\partial \kappa_i}&=\frac{\partial}{\partial \kappa_i}\left(-\frac{1}{2}\left[(\kappa_i-\bm{\mu}_i^\intercal\bm{\mu}_i-Tr(\bm{\Sigma}_i))\mathbbm{E}_{\kappa_i}\alpha_i^2\right]-\log C_i\right)\\
    &=-\frac{1}{2}\mathbbm{E}_{\kappa_i}\alpha_i^2-\frac{1}{2}(\kappa_i-\bm{\mu}_i\bm{\mu}_i^\intercal-Tr(\bm{\Sigma}_i))\frac{\partial \mathbbm{E}_{\kappa_i}\alpha_i^2}{\partial \kappa_i}+\frac{1}{2}\mathbbm{E}_{\kappa_i}\alpha_i^2\\
    &=-\frac{1}{2}(\kappa_i-\bm{\mu}_i\bm{\mu}_i^\intercal-Tr(\bm{\Sigma}_i))\frac{\partial \mathbbm{E}_{\kappa_i}\alpha_i^2}{\partial \kappa_i}.
\end{aligned}
\end{equation}
By the Cauchy-Schwarz inequality,
\begin{equation}\nonumber
    \frac{\partial \mathbbm{E}_{\kappa_i}\alpha_i^2}{\partial \kappa_i}=-\frac{1}{2}\left(\mathbbm{E}_{\kappa_i}\alpha_i^4-(\mathbbm{E}_{\kappa_i}\alpha_i^2)^2\right)\leq 0,
\end{equation}
and the equality doesn't hold for a continuous probability distribution $q(\alpha_i^2)$ here. Therefore, $\ell(\kappa_i)$ has a unique minimizer $\kappa_i = \bm{\mu}_i\bm{\mu}_i^\intercal+Tr(\bm{\Sigma}_i)$.

\subsection{Proofs for EM updates}
In this section, we provide the proof of equations \eqref{eq:general-hyper} and \eqref{eq:GM-hyper}. We begin by obtaining the explicit formula for the Evidence Lower Bound (ELBO) as defined in \eqref{eq:elbo}, and then determine its maximizer with respect to the hyper-parameters. For a general slab function, the ELBO is given by
\begin{equation}\nonumber
   \begin{aligned}
    \mathcal{L}(P_{\bm{\mu}, \bm{\Sigma}, \bm{\gamma}, v})=&\mathbbm{E}_{\bm{z,\theta}}\log\frac{\Pi(\bm{Y,z,\theta}|\lambda^2,w)}{P_{\bm{\mu}, \bm{\Sigma}, \bm{\gamma}, v}(\bm{z,\theta})}=\mathbbm{E}_{\bm{z,\theta}}\log\frac{\Pi(\bm{Y}|\bm{z},\bm{\theta})\Pi(\bm{z}|w)\Pi(\bm{\theta}|\bm{z},\lambda^2)}{P_{\bm{\mu}, \bm{\Sigma}, \bm{\gamma}, v}(\bm{z,\theta})}\\
    =&\int_{\bm{z}}\int_{\bm{\theta}}\log\frac{\Pi(\bm{z}|w)\Pi(\bm{\theta}|\bm{z},\lambda^2)}{P_{\bm{\mu}, \bm{\Sigma}, \bm{\gamma}, v}(\bm{z},\bm{\theta})}dP_{\bm{\mu}, \bm{\Sigma}, \bm{\gamma}, v}+C\\
    =&\sum_{i=1}^G\left(\gamma_i\log\frac{w}{\gamma_i}+(1-\gamma_i)\log\frac{1-w}{1-\gamma_i}+\gamma_i\mathbbm{E}_{\bm{\mu}_i,\bm{\Sigma}_i}\log\frac{h_\lambda(\bm{\theta}_i)}{N(\bm{\mu}_i,\bm{\Sigma}_i)}\right)+C,
   \end{aligned}
   \label{eq:EM-general}
\end{equation}
where $C$ is the expected log-likelihood and is not related to the update of hyperparameters. Then by taking partial derivatives with respect to $w$ , we can easily get the equation $w=\sum_{i=1}^G\gamma_i/G$. The hyperparameter $\lambda$ is the maximizer of $\sum_{i=1}^G\gamma_i\mathbbm{E}_{\bm{\mu}_i,\bm{\Sigma}_i}\log h_{\lambda}(\bm{\theta}_i)$. 

For the hierarchical spike-and-slab prior, the ELBO is similar to \eqref{eq:EM-general}, except that we incorporate another set of parameters $(\alpha_i)_{i=1}^G$. Specifically, it has the following form:
\begin{equation}\nonumber
    \begin{aligned}
    \mathcal{L}(P_{\bm{\mu}, \bm{\Sigma}, \bm{\gamma},\bm{\kappa}, v})&=\int_{\bm{z}}\int_{\bm{\alpha}}\int_{\bm{\theta}}\log\frac{\Pi(\bm{z,\theta,\alpha}|\lambda^2,w)}{P_{\bm{\mu}, \bm{\Sigma}, \bm{\gamma},\bm{\kappa}, v}(\bm{z,\theta,\alpha})}dP_{\bm{\mu}, \bm{\Sigma}, \bm{\gamma},\bm{\kappa}, v}+C\\
&=\int_{\bm{z}}\int_{\bm{\alpha}}\int_{\bm{\theta}}\log\frac{\Pi(\bm{z}|w)\Pi(\bm{\alpha}|\lambda^2)}{P_{\bm{\mu}, \bm{\Sigma}, \bm{\gamma},\bm{\kappa}, v}(\bm{z,\theta,\alpha})}dP_{\bm{\mu}, \bm{\Sigma}, \bm{\gamma},\bm{\kappa}, v}+C\\
    &=\sum_{i=1}^G\left(\gamma_i\log\frac{w}{\gamma_i}+(1-\gamma_i)\log\frac{1-w}{1-\gamma_i}+\gamma_i\int_{\alpha_i^2}q(\alpha_i^2){\log \tilde{h}_\lambda(\alpha_i^2)}d\alpha_i^2\right)+C.
    \end{aligned}
\end{equation}
Here the constant $C$ may vary in different lines. The update formula for $w$ can be derived by taking derivatives, and $\lambda$ should be chosen to optimize $\sum_{i = 1}^ G\gamma_i\int_{\alpha_i^2}q(\alpha_i^2){\log \tilde{h}_\lambda(\alpha_i^2)}d\alpha_i^2$.
\end{document}